\documentclass[sigconf]{acmart}

\usepackage{tikz}
\usepackage{amsmath}
\usepackage{pdfpages}
\usepackage{xurl}
\usepackage{multirow}
\usepackage{subcaption}
\usepackage{graphicx}
\usepackage{algorithm}
\usepackage[noend]{algpseudocode}
\usepackage{float}
\usepackage{adjustbox}
\usepackage{enumitem}
\usepackage{booktabs}

\algnewcommand\algorithmicswitch{\textbf{switch}}
\algnewcommand\algorithmiccase{\textbf{case}}
\algnewcommand\algorithmicdefault{\textbf{default}}
\algdef{SE}[SWITCH]{Switch}{EndSwitch}[1]{\algorithmicswitch\ #1\ \algorithmicdo}{\algorithmicend\ \algorithmicswitch}
\algdef{SE}[CASE]{Case}{EndCase}[1]{\algorithmiccase\ #1}{\algorithmicend\ \algorithmiccase}
\algdef{SE}[DEFAULT]{Default}{EndDefault}{\algorithmicdefault}{\algorithmicend\ \algorithmicdefault}
\algtext*{EndSwitch}
\algtext*{EndCase}
\algtext*{EndDefault}
\algnewcommand{\LeftComment}[1]{\Statex \(\triangleright\) #1}

\newcommand{\roomi}[1]{\textcolor{black}{\noindent[Roomi: #1]}}
\newcommand{\changes}[1]{\textcolor{black}{\noindent#1}}
\newcommand{\IEEEchange}[1]{\textcolor{black}{\noindent#1}}

\newcommand{\ccschanges}[1]{\textcolor{black}{\noindent#1}}

\newcommand{\ccschangestwo}[1]{\textcolor{black}{\noindent#1}}
\newcommand{\FsPSS}{.}
\newcommand{\ScPSS}{!}
\newcommand{\TdPSS}{\_}
\newcommand{\FrPSS}{\#}
\newcommand{\FfPSS}{-}
\newcommand{\SxPSS}{@}

\copyrightyear{2023} 
\acmYear{2023} 
\setcopyright{rightsretained} 
\acmConference[CCS '23]{Proceedings of the 2023 ACM SIGSAC Conference on Computer and Communications Security}{November 26--30, 2023}{Copenhagen, Denmark}
\acmBooktitle{Proceedings of the 2023 ACM SIGSAC Conference on Computer and Communications Security (CCS '23), November 26--30, 2023, Copenhagen, Denmark}
\acmDOI{10.1145/3576915.3623156}
\acmISBN{979-8-4007-0050-7/23/11}

\begin{document}

\pagestyle{plain} % removes running headers

\title{Measuring Website Password Creation Policies At Scale}

\author{Suood Alroomi}
\orcid{0000-0003-0454-3742}
\affiliation{
  \institution{Georgia Institute of Technology}
   % \city{Atlanta}
 % \state{Georgia}
  %\country{USA}
}

\affiliation{
  \institution{Kuwait University}
   % \city{Kuwait City}
 % \country{Kuwait}
}
\email{roomi@gatech.edu}

\author{Frank Li}
\orcid{0000-0003-2242-048X}
\affiliation{%
  \institution{Georgia Institute of Technology}
    %\city{Atlanta}
 % \state{Georgia}
  %\country{USA}
  }
\email{frankli@gatech.edu}

\newcommand\blfootnote[1]{%
  \begingroup
  \renewcommand\thefootnote{}\footnote{#1}%
  \addtocounter{footnote}{-1}%
  \endgroup
}

\begin{abstract}
Researchers have extensively explored how password creation policies influence the security and usability of user-chosen passwords, producing evidence-based policy guidelines. However, for web authentication to improve in practice, websites must actually implement these recommendations. To date, there has been limited investigation into what password creation policies are actually deployed by sites. Existing works are mostly dated and all studies relied on manual evaluations, assessing a small set of sites (at most 150, skewed towards top sites). Thus, we lack a broad understanding of the password policies used today.
In this paper, we develop an automated technique for inferring a website’s password creation policy, and apply it at scale to measure the policies of over 20K sites, over two orders of magnitude ($\sim$135x) more sites than prior work. 
%Our evaluated population is  diverse, spanning different rankings.
Our findings identify the common policies deployed, potential causes of weak policies, and directions for improving authentication in practice. Ultimately, our study provides the first large-scale understanding of password creation policies on the web.

\blfootnote{ $\dagger$ This is an extended version of a conference publication~\cite{ourpaper} that
includes several appendices, which were omitted due to space constraints.}

\end{abstract}




\maketitle

\section{Introduction}

Passwords remain the de facto standard method for online authentication~\cite{fido}, and password-based web authentication will likely remain ubiquitous for the foreseeable future. 
As a consequence, the security of the web ecosystem is critically dependent on how both users and websites manage password authentication.
Over the years, researchers have extensively explored how users behave with passwords, particularly when constrained by password creation policies (e.g.,~\cite{inglesant2010true, komanduri2011passwords, shay2010encountering, shay2014can, shay2015spoonful, shay2016designing, ur2015added, ur2016users, wash2016understanding, lee2022password}).
%~\cite{ur2017design, egelman2013does, felten_management, florencio2007large, hanamsagar2018leveraging, honey_shrunk_keys, inglesant2010true, komanduri2011passwords, mazurek2013measuring, melicher2016usability, nudges_impact, segreti2017diversify, service_name_passwords, shay2010encountering, shay2014can, shay2015spoonful, shay2016designing, stobert2014password, stobert2015expert, tangled_web, ur2012does, ur2015added, ur2016users, wash2016understanding}. 
These efforts have produced insights into how authentication \emph{should} be handled by websites to promote password security and usability.
These user-centric efforts have helped drive significant updates to modern password guidelines, such as spurring the US National Institute of Standards and Technology (NIST) to release new online identity management guidelines~\cite{grassi2017digital} in 2017, its first since 2004~\cite{burr2004electronic}.

Ultimately though, websites are the entities that must implement recommended practices to improve authentication security and usability in reality. To date, there has been significantly less investigation into how website operators actually manage password authentication and what password creation policies they enforce. A handful of studies~\cite{furnell2007assessment, furnell2011assessing, mannan2008security, kuhn2009survey, florencio2010security, mayer2017second, wang2015emperor, bonneau2010password, preibusch2010password, seitz2017differences, lee2022password} have manually analyzed the password policies of top websites. However, due to the manual methods used in this prior work, the scale of investigation is heavily limited, with the largest entailing only 150 websites\ccschangestwo{~\cite{bonneau2010password,preibusch2010password}}. The considered website populations also skew towards highly-ranked sites\ccschangestwo{~\cite{furnell2007assessment, furnell2011assessing, bonneau2010password, preibusch2010password, lee2022password, seitz2017differences, mayer2017second}}, across a few countries (i.e., US, Germany, and China)\ccschangestwo{~\cite{wang2015emperor,seitz2017differences,mayer2017second}}. Furthermore, most studies were conducted over a decade ago\ccschangestwo{~\cite{furnell2007assessment,mannan2008security,kuhn2009survey,florencio2010security,bonneau2010password,preibusch2010password, furnell2011assessing}}, predating significant updates to password guidelines, including those by NIST\ccschangestwo{~\cite{grassi2017digital}} and Germany's Federal Office for Information Security (BSI)\ccschangestwo{~\cite{bsi2019, bsi2020}}. Thus, we lack a large-scale modern understanding of the password creation policies deployed by sites today, \ccschangestwo{the authentication security and usability implications of these policies, and the adoption rate of authentication recommendations}.

%with no study more recent than 2017 (which notably pre-dates significant updates to widely used password guidelines, including those by NIST and Germany's Federal Office for Information Security (BSI)). 

In this work, we seek to close this gap. Given the incredible diversity of the web, doing so is challenging\ccschangestwo{~\cite{wang2015emperor,lee2022password}}, as websites and their password authentication are implemented in a myriad of ways, and password policy information is often not explicitly published. %\footnote{While some websites may display password guidance or requirement information during account creation, this too is also implemented in various ways and lacks consistent formats. \roomi{maybe we don't need this anymore with our new section discussing this?}}. 
In this work, we develop a web measurement method that automatically infers password creation policies in a blackbox fashion. Our method entails testing specifically-chosen passwords in a carefully constructed order during a site's account signup, identifying which passwords are accepted or rejected to infer the site's password creation policy. We \changes{construct} our inference method to \changes{reduce} its footprint on evaluated sites. We apply our technique to successfully infer the password creation policies of over 20K websites across the Tranco top 1M, evaluating a diverse population over two orders of magnitude ($\sim$135x) larger than any prior study. %Our population is also diverse, spanning different rankings.

Our analysis reveals how often websites employ certain creation policy parameters, such as acceptable characters, character composition requirements, disallowed password structures, and breached password blocking. 
%
\iffalse %\ country
We also assess the extent to which sites adhere to password guidelines and standards published by various organizations, including NIST, BSI, and \IEEEchange {the Open Web Application Security Project} (OWASP), particularly for sites within an organization's country (e.g., the US for NIST, Germany for BSI). Furthermore, we analyze how the password policies differ across rankings. 
\fi
%
We find that the most common policies today enforce few requirements on passwords, aligning with recent policy recommendations (e.g., NIST's 2017 guidelines~\cite{grassi2017digital}). However, counter to modern standards, acceptance of short passwords is widespread, with over half of sites allowing passwords of six characters or shorter, and an unexpected 12\% lacking any minimum length requirements. Furthermore, 30\% of sites do not support certain recommended characters in passwords, including spaces and special characters. We also observe only about 12\% of sites using password blocklists, resulting in the majority of sites being vulnerable to password spraying attacks~\cite{spraying,wang2015emperor}. Overall, only a minority of sites fully adhere to common guidelines, with most sites adhering to more dated guidelines. % (e.g., NIST's 2004 recommendations, rather than its 2017 guidelines). 
We also observe that top-ranked sites tend to support stronger policy parameters.
\iffalse %\ country  or category
and sites for different countries and categories exhibit notable policy differences. 
\fi
\changes{Through case studies of weak policy parameters, we identify how web frameworks and default configurations may be driving factors.}

%\roomi{should some of the above paragraph's results be added to our abstract?}

%\roomi{It is also important to note not only our data volume is high, but we have the largest non-US dataset analyzed (3,321 domains, 16.5\%) I could also calculate the number of non-English domains}

Ultimately, our study illuminates the state of modern password creation policies at scale for the first time, while also highlighting \ccschangestwo{authentication security and usability} problems requiring attention and identifying directions for improving authentication in practice.

\section{Related Work}
\label{sec:related}

Here we summarize prior work measuring real-world password policies and studies that relied upon automated account creation.

\subsection{Password Policy Measurements}

Over the past 15 years, multiple studies have manually investigated the password policies used by real-world websites.
Several initial studies~\cite{furnell2007assessment, furnell2011assessing, mannan2008security} were very limited in scale (considering up to 10 sites).
%
%S. Furnell conducted one of the earliest studies on website password policies in 2007, manually characterizing the policies of the top 10 most popular domains at that time~\cite{furnell2007assessment}. They found that policies varied considerably from site to site, although the sample size was small. They replicated the study four years later, in 2011, observing the wider adoption of password strength meters and two-factor authentication, and fewer cases of storing plaintext passwords~\cite{furnell2011assessing}.  
%They found only one website (YouTube) with dictionary check and several examples of domains that could be saving the password in plaintext or are reversibly encrypted (Friendster and MySpace). 
%
%Similarly, Mannan and Oorschot examined the security and usability of five Canadian banking sites in 2007~\cite{mannan2008security}, finding wide acceptance of weak passwords. 
%They found that the password restrictions for online banking at that time were very weak; most banks accepted passwords of lengths 5 and 6, some online banks allowed digit-only passwords, and one restricted the use of special characters. The authors were also able to register with a very weak passwords such as ‘123456’ and ‘111111’.
% 
At a larger scale, Kuhn et al. manually surveyed the password policies of 69 domains
%, which spanned eight different industries,
in 2007 and then again in 2009~\cite{kuhn2009survey}. The authors noted that 45\% of the websites changed their password policy in the two-year span. These changes included more widely imposing password complexity and length requirements, although policies on many sites remained weak.
%, although half of the sites still allowed passwords shorter than six characters during the second measurement. 
%
Similarly, in 2010, Florencio et al. explored the factors that influence the password policies employed by websites~\cite{florencio2010security}. The authors manually characterized the password policies of 75 US websites, 
%across different website categories and traffic rankings, 
%finding that factors related to security and asset valuation had no strong correlation with the strength of the password policy. Instead,
finding that factors related to monetization seemingly correlated inversely with policy strength. The study was replicated seven years later in 2017 by Mayer et al., using the same set of websites along with 67 additional German websites~\cite{mayer2017second}. This work replicated the earlier observations, and observed that overall, password policies on US websites had increased in strength over time, and were stronger than those on German sites.
In 2015, Wang et al. also compared the password policies between 30 Chinese websites and 20 English-language sites~\cite{wang2015emperor}. They observed several Chinese websites requiring digit-only passwords, and policies on English sites were overall more stringent.

At the largest scale, in 2010, Bonneau et al. conducted an extensive manual evaluation of the password policies on 150 domains chosen from the Alexa Top 500 sites~\cite{bonneau2010password}. %Their data set consisted of 150 domains chosen from Alexa’s top 500.
They found that half of the websites enforced a minimum password length of 6, and 18\% had no length restrictions. Furthermore, few sites disallowed common dictionary words for passwords. 
Due to password reuse by users across websites, the authors also highlighted the potential negative externalities caused by websites with weaker password policies, impacting the passwords chosen by users even on sites employing more secure policies. This concept was empirically explored further by Preibusch and Bonneau through a game-theoretic model using the same dataset~\cite{preibusch2010password}.  In 2017, Seitz et al. also characterized the potential for password reuse across sites by contrasting the password policies across 100 German sites~\cite{seitz2017differences}, finding that the policies were not diverse enough to mitigate the risk of password reuse. They were able to construct passwords that could be accepted across 99\% of the sites. 
%as they were able to come up with a password composition that can be accepted in 99\% of the websites, a 9-10 characters password with non-English word, lower and uppercase ASCII characters and digits (e.g., DenCHI2017). 
%
%Most recently, in 2020, Beno et al. examined the password policies of 14 popular websites and then conducted a user study where participants created passwords under each policy~\cite{beno2020hacking}. Using traditional password cracking tools, the researchers were able to crack a quarter of the passwords chosen on average. They observed that participants often created passwords longer than the required minimum, and that weak password policies correlated with weaker chosen passwords.
%
%Dudheria examined the password policies of 50 leading mobile applications of different categories~\cite{dudheria2019assessing}. They found that like websites, the password policies were diverse and did not follow any standard. Their results show that a minimum length of 6 was the most common among mobile applications, and a large percentage of sensitive applications accepted dictionary words.
Most recently, Lee et al.~\cite{lee2022password} manually investigated 120 top English websites, finding that
%only 19\%  employed password meters and 
over half did not blocklist common passwords. Overall, less than a quarter of the sites followed security and usability password policy recommendations.

A primary limitation of these studies is that they manually analyzed website password policies. As a consequence, the studies were small-scale, with the largest involving only 150 sites, and the characterized sites heavily skewed towards top sites (summarized in Table~\ref{tab:datasize_short} of Appendix~\ref{app:keywords}). Furthermore, most studies were over a decade ago, making their observations dated. The web has expanded significantly since then, and our understanding of secure password policies has also substantially evolved (including updates to modern authentication recommendations, such as NIST's latest password policy guidance released in 2017~\cite{grassi2017digital}). Thus, a more modern view of website password policies is needed. 
Our study leverages automation to provide the largest-scale picture of web password creation policies today, encapsulating a diverse population of websites across different rankings.

\subsection{Account Creation Studies}

Several studies have used automated account creation for different measurements.
DeBlasio et al. automatically created honey accounts on websites to detect potential credential theft~\cite{deblasio2017tripwire}. 
%They automatically created a honey account on a website where the chosen password is identical to the one used for a unique test email account. Successful attempts to log into that email account indicated that the website with the honey account might have been breached. 
They successfully created accounts across 2.3K websites, detecting 19 potential cases of website credential compromise. %To create the accounts, their web crawler tries to access different links within the domain, seeking out a registration page and then trying to fill it with accurate information. Heuristics are then used to verify the success of registration. Overall, they were able to register 2300 accounts for their study and later on detect 19 possible compromises through the email access attempts
Recently, Drakonakis  et al. investigated how websites handle cookies during authentication workflows~\cite{drakonakis2020cookie}. They attempted automated account creation and login across 1.5M domains, successfully creating accounts on 25K domains in total. They found half of the domains vulnerable to cookie-hijacking attacks. 
While our automated account creation process shares similarities with the prior work, we designed our method from the ground up, as our end-to-end empirical method required overcoming distinct challenges, such as more extensive account creation activity and inferring password policies.

%While our registration crawler shares many elements with the abovementioned studies, we have relied on a more effective approach throughout our stages as we deal with a large dataset that requires multiple attempts. In each stage of the framework, we relied on constructing data-driven ML models for classification and NLP for keywords and phrases extraction, which we found to be superior to using simple heuristics. Further, we did not exclude domains with captcha but tried to solve them instead.

\section{Method and Implementation}
\label{sec:Methodology}

\begin{figure}[t]
    \centering
    \includegraphics[scale=0.75]{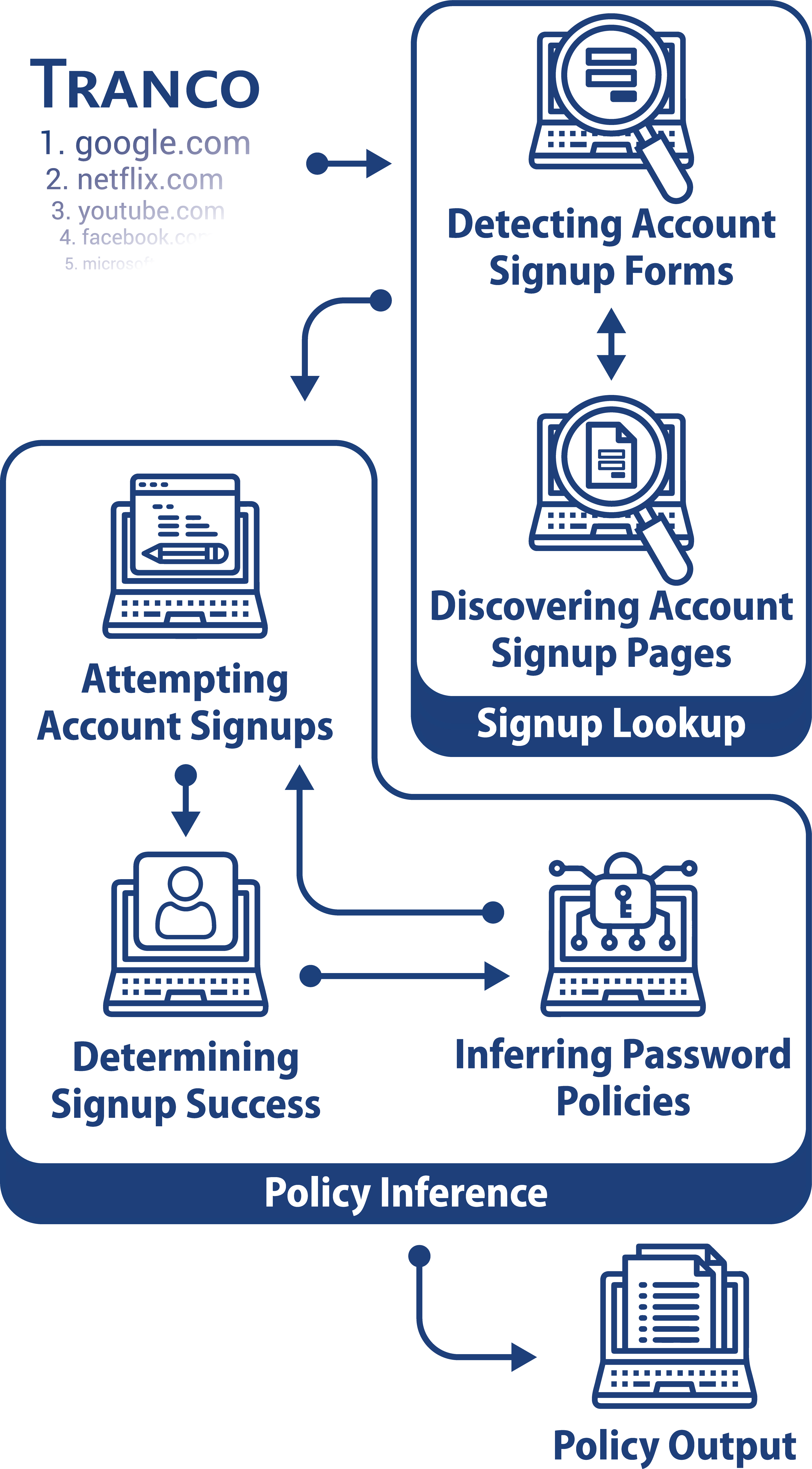}

    \caption{\ccschangestwo{Illustration of the stages of our password policy measurement method.}
    }

    \label{fig:methodology}
\end{figure}

Here, we describe our method for automatically inferring password policies. At a high level, we attempt multiple account signups on a website using different passwords, observing which accounts are successfully created to identify password policy parameters. 
\ccschangestwo{As shown in Figure~\ref{fig:methodology}, we first discover a website's account signup workflow. To do so, we search for account signup forms (Section~\ref{sec:SignupFormDetection}) across a website's pages to detect an account signup page (Section~\ref{sec:SignupLookup}). Then, we execute our policy inference process, which attempts multiple account signups with different passwords (Section~\ref{sec:FormFilling}) while evaluating whether the signup is successful (Section~\ref{sec:SignupSuccessVerification}). Based on which signup attempts (and the associated passwords) succeed, we infer the password policy parameters (Section~\ref{sec:PasswordPolicyInferenceAlgorithm}).}
\ccschangestwo{To conduct our measurements, we train two machine learning classifiers, one for signup form detection (Section~\ref{sec:SignupFormDetection}) and another for classifying signup attempt success (Section~\ref{sec:SignupSuccessVerification}). Other components of our method rely on keyword-based heuristics (Section~\ref{sec:KeywordsExtraction} and Appendix~\ref{app:keywords}), particularly for identifying potential account signup URLs and form fields.} We will share our measurement data and code to vetted researchers upon request, as otherwise these could potentially be used in online abuse.

\subsection{Ground-Truth Analysis}
\label{sec:KeywordsExtraction}
Modern websites and their authentication workflows are diverse, in both design and implementation. As a consequence, we require heuristics throughout our method for discovering and analyzing website account creation (as have prior work conducting similar automated account creation~\cite{deblasio2017tripwire,drakonakis2020cookie}).
These heuristics include keywords for classifying webpages and HTML elements. We additionally train machine learning classifiers for complex labeling tasks.

To identify keywords for our specific method in a systematic, \changes{language-agnostic}, and data-driven fashion, as well as to train our classifiers, we manually analyzed 2800 domains randomly sampled from the Tranco Top 1M~\cite{pochat2018tranco} (from June 6, 2021). We identified whether each domain supports account creation (26\% did), and if so, we analyzed the characteristics of its account signup workflow (including the location of its signup pages and forms). We refer to this dataset as our \textit{ground-truth data}.
For extracting relevant keywords, we applied keyword ranking algorithms to identify the top keywords prevalent in positive cases but uncommon in negative cases, \changes{agnostic to any specific language} (details in Appendix~\ref{app:keywords}). We discuss training our classifiers in the following sections.
%We selected all top ranked keywords that covered more than a few percent of sites in our random sample.
%We note that we also assessed our end-to-end method on a distinct random sample of \frank{FILL IN} domains, manually analyzing false positive and negative findings to identify if our keywords needed adjustment. We did not uncover refinements needed to our chosen keywords.

%we rely on the existence of a specific set of keywords as part of our prediction. This is needed when we classify links as signup or login, their forms, the field types in forms, and others. Our initial version of the framework used common set of keywords that we found to be appropriate for the targeted prediction, which is similar to what the prior works applied . However, we found many false negatives when we compared our classification to manual testing. Further, it was not clear how to empirically include relevant phrases without introducing too much noise. To improve our classifications, we applied several statistical NLP techniques that attempt to extract the most relevant and important keywords and phrases in a document or a corpus. 

\subsection{Detecting Account Signup Forms}
\label{sec:SignupFormDetection}
To assess a site's password policies, we first identify its signup page and form.
To distinguish account signup forms from others (e.g.,~login, newsletter), we use a binary SVM classifier.
For its features, we use the presence of signup-related keywords (chosen from our ground-truth data, discussed in Appendix~\ref{app:keywords}) in the HTML form's title, ID, class, and action, %(and any child HTML elements), 
as well as the numbers of form inputs in total and password-type inputs.
For training data, we manually labeled the HTML forms in our ground-truth data. We trained our model using Python's sklearn~\cite{Scikit}, selecting hyperparameters using grid search. Evaluating our model with 10-fold cross-validation, we observe an average accuracy of 94.7\% (errors discussed in Appendix~\ref{appendix:Challenges}). Note that while false negatives will cause us to skip evaluating sites, false positives will result in unsuccessful attempts to evaluate them (which we detect and filter out).

\subsection{Discovering Account Signup Pages}
\label{sec:SignupLookup}
Given a domain, our method starts by searching for its signup page, identified by the presence of a signup form (from Section~\ref{sec:SignupFormDetection}). This process proceeds as follows until a signup page is found.
\begin{enumerate}[nosep, leftmargin=*]

\item We search for a signup form on the domain's landing page.

\item We next crawl URL links found on the landing page that contain common keywords for account signup or login URLs. \ccschangestwo{We call these candidate URLs as they likely contain an account signup or login form}. Keywords are selected using ground-truth data (see Appendix~\ref{app:keywords}), with separate keywords for signup and login URLs. We use login URLs as they often contain links to a signup page (for users without an account). On login URLs, we attempt to detect a signup form, otherwise we collect further candidate signup URLs (now ignoring candidate login URLs). For each page, we visit at most four candidate URLs to avoid excessively crawling a domain. (In our ground-truth data, we observed that this threshold was sufficient for discovering signup URLs, as most pages had few, if any, candidate signup or login URLs.)

%If no signup form found at the landing page, our next step is to start looking for both signup and login candidate links by parsing the keywords in each link from the landing page against the extracted ones. Candidate links search also considers the length of the link and the number of slashes in them. The reason why we consider login pages and forms in our search is that we empirically found that they are usually a pathway to the signup ones. 

%The flagged candidate login pages from the landing page are first visited and a signup form is checked for there. If none found, we collect any signup candidate link to add to the list of candidates we have. When all login candidates are consumed with no signup form found we start processing all the collected candidate signup links. Each link is visited and checked for the existence of a signup form. Note that although it may look like we are visiting many links in our process, we do limit the number of candidates per page to be at most 4 which was empirically found to have high coverage. 

\item Finally, we query the Google search engine for the domain's account signup pages (using ScraperAPI~\cite{scraperapi}).
Our search query includes the domain along with ``account OR register OR sign+up OR create'', constructed using the most frequent keywords in the HTML titles of real signup pages in our ground-truth data. Given the search results, we again consider candidate signup and login URLs, crawling up to 4 candidate URLs in search of a signup page/form (using the same method for identifying candidate URLs and processing them as done with URL links on the domain's landing page). (We observed that this crawling threshold was sufficient on our ground-truth dataset.)

%If no signup found yet, the search continues, and at this stage we rely on using google search as an avenue. We extracted four keywords from processing the page titles of the dataset of signup pages to form our search query. The query is ``account OR register OR sign+up OR create'' along with the domain name. From there we look for the login and signup candidates of the first page results and processed them similar to what we did with earlier candidates. 

\item Here, we record the domain as lacking a  signup page.%If no signup page has yet been found, we record the domain as lacking a  signup page.
%At this point the module concludes and if no signup page found the next domain is processed. 

\end{enumerate}

We note that our crawler is non-interactive and does not simulate user actions on a page. Some sites require an action for the signup form to fully appear (e.g., clicking a ``signup'' button, or clicking through multi-page forms). However, in our ground-truth data, this behavior is not widespread, and automating it would be challenging.
%Note that we initially considered links that required clicking action to show but found them to both slow down the whole process and not to add much gain overall.   

\subsection{Attempting Account Signups}
\label{sec:FormFilling}
With a domain's signup page, we next fill out and submit the signup form. By testing different passwords across multiple signup attempts, we will infer the domain's password policy (discussed in Section~\ref{sec:PasswordPolicyInferenceAlgorithm}). Automatically filling and submitting a signup form encounters two key challenges.

First, we must identify signup form fields and provide acceptable values/actions.
We classify them based on the HTML input element's name, class, and ID, using relevant keywords identified in our ground-truth data (see Appendix~\ref{app:keywords}).
For common form fields (e.g., name, email), we use either pre-selected values (not real user data) or the Faker Python library~\cite{faker} to generate synthetic data. We handle the password field specifically, as discussed in Section~\ref{sec:PasswordPolicyInferenceAlgorithm}. For unrecognized fields, we generate a random string as a last resort. Some forms offer multiple button elements (e.g., signup and single sign-on buttons). We identify the account signup button using keywords derived from our ground-truth data  (see Appendix~\ref{app:keywords}).
%Then we fill them using a combination of predefined data, Faker library \cite{faker}, and auto fills by type as a last resort. 

%A captcha field is similarly recognized and AZcaptcha is used to solve it. We had to first recognize the type of captcha (text, image(s), ReCaptcha V2 or V3, or Hcaptcha) and then use the appropriate calls to the solver. For password field(s), our password inference algorithm, described in Section~\ref{sec:PasswordPolicyInferenceAlgorithm}, is responsible for feeding this module with the appropriate password to fill. 

A second challenge is that many signup workflows require completing a CAPTCHA. In our ground-truth data, we identified CAPTCHAs on at least 49\% of signup forms. We aimed to overcome CAPTCHAs to significantly increase our likelihood of successfully assessing sites. Given our measurement's scale and ethical concerns\footnote{Prior automated account creation work skipped sites with CAPTCHAs~\cite{drakonakis2020cookie} or used human CAPTCHA-solvers at a small scale~\cite{deblasio2017tripwire}.} with human-driven CAPTCHA solvers (discussed in Section~\ref{sec:Ethics}), we opted to rely on an automated CAPTCHA solver, AZcaptcha\footnote{AZcaptcha~\cite{azcaptcha} advertised an automated OCR-based method. We note that AZcaptcha's price point is also significantly lower than human-driven CAPTCHA solvers, reinforcing AZcaptcha's automation claims.}~\cite{azcaptcha}. 
%, and identified that it was able to effectively solve CAPTCHAs \frank{add description of AZcatpcha performance} \roomi {added}. The solver was able to break (94\%) of the captchas, with the errors mostly from not recognizing the captcha type or not being able to solve them correctly. The average response time of the solver is around 75.13 seconds, with reCAPTCHA and hCaptcha taking more than a minute and a half to solve (94.9 seconds). 
%
We identify CAPTCHAs during the signup process through fingerprinting the HTML/JavaScript code used by the CAPTCHA implementations supported by AZcaptcha, and pass the extracted CAPTCHAs to AZcaptcha to solve. (During our full measurement, AZcaptcha correctly solved 94\% of all CAPTCHAs we encountered, with failure cases discussed in Appendix~\ref{appendix:Challenges}.)

%Captcha is one of the bigger blocks against automation here, which is either skipped \cite{drakonakis2020cookie} or some solver is used \cite{deblasio2017tripwire}. While prior works used human solvers services \cite{deblasio2017tripwire}, we opted to use an automated online solver instead \cite{azcaptcha}. Surprisingly, the majority of captcha solvers services are human and AZcaptcha \cite{azcaptcha} was the few we found that is automated. 

%Another unexpected challenge  was the submission button that required some sort of an interaction by scrolling the page all the way down or clicking first on agreement for terms and services. Some forms also had other buttons such as reset and next for multi-page forms, which required us to design specific classifier for the submission button.

\subsection{Determining Signup Success}
\label{sec:SignupSuccessVerification}

Websites vary widely in response to submitting an account signup form, and behavior differs depending on the signup success. For example, some sites redirect to another page, while others display a message. To determine if a signup attempt is successful, we develop an ensemble decision tree classifier that operates on features of the webpage returned upon form submission. 
We collected training data from signup attempts on 160 domains in our ground-truth data. Our features include the presence of a signup form (detected as in Section~\ref{sec:SignupFormDetection}), keywords in the page and URL, and the similarity of the page and its URL with those before form submission. We then trained an XGBoost decision tree ensemble model with 100 trees, selecting hyperparameters using grid search. Evaluated using 4-fold cross-validation, we observe a 91.3\% accuracy. Note that classification errors primarily result in consistent successes or failures across all attempts for a domain, which we detect and filter out. %\frank{Provide a bit more parameter info, and FPR/FNR if possible.}

\subsection{Inferring Password Policies}
\label{sec:PasswordPolicyInferenceAlgorithm}

The prior sections discussed our method for finding signup pages, as well as completing, submitting, and determining the submission outcome for the signup forms. To infer the password policy, we perform multiple signup attempts where we provide consistent signup information except we vary the passwords provided systematically, allowing us to determine the password policy parameters based on which passwords are accepted or rejected. We determine whether a password is accepted based on the form submission outcome. However, form submission may fail due to other information we provide, rather than just the passwords. In such cases, as we provide consistent signup information across signup attempts, we will observe consistent signup failures for a domain, independent of the passwords tested, and we can subsequently filter out such domains from our analysis.
Also, a successful account signup results in a created account. To minimize the account-related resources we require of domains, we \changes{constructed} our method to reduce the number of accounts created, as discussed further in Section~\ref{sec:Ethics}.

%We are given a registration system with an unknown password policy. Such a system can be enquired by passing a string (password). This is called an \emph{\bf Attempt}. The registration system will then respond with either a $YES$ (a \emph{\bf Success}) or a $NO$ (a \emph{\bf Failure}) based on its password policy. The goal is to figure out the policy based on the responses given. Determining the policy consists of determining its behavior  with respect to a set of fixed characteristics, called \emph{parameters}. The main goal is to determine the parameters of a given unknown password policy, with a minimal amount of Attempts (inquiries to the registration system).

\subsubsection{Password Policy Parameters.}
\label{PasswordPolicyParameters}
We evaluate the following password creation policy parameters, which encapsulate all policy parameters investigated by prior work~\cite{bonneau2010password, wang2015emperor, seitz2017differences, lee2022password}, which fall into three classes.
The first class involves password \textbf{lengths}:

\begin{itemize}[nosep, leftmargin=*]

    \item \emph {\bf Length} (\(L_{min}\), \(L_{max}\)): %The length \(L\) of the password \newline
    % \centerline{$L_{min} \leq L \leq L_{max}$} \newline
      The minimum and maximum password lengths allowed, respectively. We conservatively consider \(L_{min}\in[0,32]\) and \(L_{max}\in[6, 128]\). 
     
\end{itemize}

The second class of parameters is \textbf{restrictive}, as they require that all passwords exhibit certain character structure.

\begin{itemize}[nosep, leftmargin=*]
     
      \item \emph {\bf Digits} (\(DIG_{min}\)): The minimum number of digits required. We consider \(DIG_{min} \in [0, 2]\).
    
    \item \emph {\bf Uppercase Letters} (\(UPP_{min}\)): The minimum number of uppercase letters required. We consider \(UPP_{min} \in [0, 2]\).
    
    \item \emph {\bf Lowercase Letters} (\(LOW_{min}\)): The minimum number of lowercase letters required. We consider \(LOW_{min} \in [0, 2]\).
    
    \item \emph {\bf Special Symbols} (\(SPS_{min}\)): The minimum number of special symbols required. We consider \(SPS_{min} \in [0, 2]\).
    
    \item  \emph {\bf Combination, 3 out of 4} (\(R_{cmb34}\)): Passwords must exhibit 3 out of 4 character classes (\emph{Digits}, \emph{Uppercase Letters}, \emph{Lowercase Letters}, \emph{Special Symbols}).
    %The parameter checks if it is required to have any combination of $3$ of those $4$ classes in the password. Note that it is possible to have none of the classes required and still require to have any $3$ out of $4$ to be present.
     \item  \emph {\bf Combination, 2 out of 4} (\(R_{cmb24}\)): Passwords must exhibit 2 out of 4 classes (same classes as \(R_{cmb34}\)).
    
    \item  \emph {\bf Combination, 2 out of 3} (\(R_{cmb23}\)): Passwords must exhibit 2 out of 3 classes (\emph{Digits}, \emph{Letters}, \emph{Special Symbols}).

    \item \emph {\bf Combination of Words} (\(R_{2word}\)): Passwords must have multiple words, where a word is defined as a string of three or more letters (any case), delimited by digits or special symbols\footnote{We assume that if $R_{2word}=True$, then $L_{min} \geq 10$ (whereas in theory, $L_{min}$ could be between 7 and 9). We argue that this is a reasonable assumption as requiring such word structure without allowing longer passwords would overly constrain user password selection, especially as the average word is 4.7 characters~\cite{Norvig}.}.
    
    %\roomi{, and is recommended for high-security environments, which emphasize security over usability~\cite{shay2015creating}.)} %and found in leaked datasets~\cite{johnson2019lost}}
    %An example acceptable password is \texttt{WHv201PeJ5}.

    %\frank{how do we select the two words to test, and what delimiter?} \roomi{here we simply test by moving the middle seperator of the safe bet to make a single word  replace(password, "MxT7zcS", "MxTzcS7"). Note here we don't use a dictionary word, but a word as defined above a string of letters, this definition is taken from prior works.}
    %

	\item  \emph {\bf No Arbitrary Special Symbols} (\(R_{no\_a\_sps}\)): Passwords cannot have arbitrary special characters (considering less popular special characters not accounted for by parameters $P_{spn1} - P_{spn4}$).
	 
	\item \emph {\bf Letter Start} (\(R_{lstart}\)): 
A password must start with a letter.
%The possibility of having a non-letter character at the beginning of the password. 
(Prior work observed such positional restrictions~\cite{florencio2010security}.)
%\changes{We test a password with a non-letter character at its start that satisfies the other restrictive policy parameters.}

%\roomi{we either take any existing non-letter and move it at start or we add zero if no non-letter exist}
	
%	\frank{We should revisit at the end whether we want to do some sort of merging of this with the permissive special character parameters.}

\end{itemize}

A final class of parameters is \textbf{permissive}, allowing certain password characteristics without requiring them.
%The structure of an admissible password is altered to include the tested permissive character (e.g. space) or substring (e.g. sequential) without violating the concluded restrictive parameters, and then an attempt is carried (see Appendix~\ref{Appendix:PasswordInferenceDetails} for details). 
\begin{itemize}[nosep, leftmargin=*]
    \item \emph {\bf Dictionary Words} (\(P_{dict}\)): Common dictionary words (e.g., \texttt{apple}) are permitted within the password, where a word is at least 3 letters. 
    %\changes{We identify the longest word (up to 8 characters) permitted in a password as constrained by other policy parameters, and test the inclusion of the most common English word~\cite{oxford}.}
    %, where a word is 3+ letters. 
    %\\frank{Is this anywhere in the password, or the whole password is a dictionary word?} \roomi{a substring depending on the size available in the password. "one", "time", "world", "number", "company", "sequence" these were inspired from your recommendation to use a common words in English https://en.wikipedia.org/wiki/Most_common_words_in_English}
    
    \item \emph {\bf Sequential Characters} (\(P_{seq}\)): Logical sequences of 3+ characters (e.g., \texttt{123}, \texttt{abc}) are permitted in the password. %\changes{We test a password including \texttt{abc}, \texttt{123}, or \texttt{ABC} (as permitted by other parameters).} %\frank{Which sequence(s) do we test? We can add them here}
    %\roomi{we have three options, depends on the minimum required characters which could limit our choirces in the space avaiable in the password, we use 'abc' or 'ABC' or '123', for example 123 is used when we can only fill numbers due to say Digmin =2}
    
    \item \emph {\bf Repeated Characters} (\(P_{rep}\)): 3+ consecutively repeated characters are permitted in the password. %Our tested password include a 3-character repeating substring of a class permitted by other parameters.
    
    %We test a password including either \texttt{111}, \texttt{aaa}, or \texttt{AAA}, selecting one permitted by other policy parameters.} %For example "111anf\$53" and "anfzzz\$53" are valid.

\item \emph {\bf Long-Digit Passwords} (\(P_{longd}\)): All-digit max-length passwords are permitted (observed before on Chinese websites~\cite{wang2015emperor}.)

%\changes{tested with such a password without sequential or consecutively repeating digits.} 

\item \emph {\bf Short-Digit Passwords} (\(P_{shortd}\)): 
All-digit min-length passwords are permitted (used along with \(P_{longd}\) to determine length's role in accepting digit-only passwords)

%All-digit minimum-length passwords are permitted, \changes{tested with such a password without sequential or consecutively repeating digits.}  (used along with \(P_{longd}\) to determine length's role in accepting digit-only passwords).

\item \emph {\bf Personal Information/Identifiers} (\(P_{id}\)): Personal information (e.g., username) is permitted in the password. %When creating an account, we use a username that is a 3-letter name followed by a random 5-digit string (e.g., ``joe31426''). We test a password with the 3-letter portion of the password (e.g., ``joe'').

\item \emph {\bf Space} (\(P_{space}\)): Whitespaces are permitted in the password. %\changes{We test a password with a space character in the middle.}  %The possibility of adding a space (ASCII code $33$) to a password. I.e. adding a space to a string, which may otherwise be accepted by the policy. \footnote{Since some policies truncate spaces from the proposed passwords, we require $L_{min} < L_{max}$.}  

\item \emph {\bf Emojis} (\(P_{emoji}\)): 
Emojis are permitted in the password. %\changes{We test a password with an arbitrary Emoji character at the end.} %(see Appendix~\ref{Appendix} for implementation details)

\item \emph {\bf Unicode Letters} (\(P_{unicd}\)): 
Unicode characters (e.g., accented characters) are permitted in the password. %\changes{We test a password with an arbitrary Unicode character at the end.}
%The possibility of having characters in the password that are letters different than the standard Latin letters. E.g., accented letters, or letters with umlauts or graves, or letters from a different alphabet, like Cyrillic, Greek, Arabic, or Chinese.

\item \emph {\bf Popular Special Symbols} (\(P_{spn1} - P_{spn4}\)): The four most popular special symbols (``.'', ``!'', ``\_'', and  ``\#'', respectively) are permitted in the password. We derive this list of top special symbols by analyzing 10M passwords in a popular password dataset~\cite{SecLists}. %\changes{We test passwords with each of these special character inserted at the end.}
%Those four parameters reflect whether some popular special symbols are allowed or are blocked. For instance, $P_{spn1}$ is $true$ if the most popular special symbol is allowed, and is $false$ if it's blocked. Similarly, $P_{spn2}$, $P_{spn3}$, and $P_{spn4}$ correspond to the second most popular special symbol, the third, and respectively the fourth one. \footnote{We processed 10 million passwords list of Seclist \cite{SecLists} to find the top four most popular special symbols to be . ! \_ and \#.}

\item \emph {\bf Breached Passwords} (\(P_{br}\)): A common password from a known password leak is permitted. %\changes{We test the highest-ranked breached password that conforms to other policy parameters, ranked based on a popular password breach~\cite{xato}.} %\ccschanges{As an example, if an all-digit password is accepted and $L_{min} \geq 6$, then we attempt the password ``123456'', as it is the highest-ranked breach password overall. If all-digit passwords are not allowed, we would try ``password'' instead, the second highest-ranked password.}

\end{itemize}

%There are $33$ possible values for $L_{min}$, $123$ for $L_{max}$, each of the parameters $LOW_{min}$, $UPP_{min}$, $DIG_{min}$, and $SPS_{min}$ has $3$ values, and the rest are Boolean parameters and have two possible values. Thus, the total password policy parameter space encapsulates nearly 345B possible distinct policies. Thus, a systematic method must be developed for determine policy parameters with a limited number of tested passwords.

%{$33 \times 123 \times 3^4 \times 2^{20} = 344,749,768,704$} 
%The result is a staggering total of nearly 345 billion possible combinations of the parameter values. Not all of those combinations are consistent, but even if the invalid configurations are discarded, the domain of consistent and possible password policies is still at least several billions in size.

\subsubsection{Inference Algorithm}
\label{sec:inference_alg_method}

With many parameters to infer, we require an efficient algorithm that evaluates a limited number of test passwords. We describe our algorithm here, with further details (including correctness and efficiency) \ccschanges{and an example} in Appendix~\ref{Appendix:PasswordInferenceDetails} %\ccschanges{and Appendix~\ref{app:example} shows our algorithm applied to a specific policy.}
%Further details on this algorithm, its correctness, and its efficiency is provided in }.
%Parameters $R_{2word}$ and $R_{nosps}$ will be called \emph{\bf pivotal}, because a lot of junctions in the logic of the current solution will depend on their values.
%\item Parameters $LOW_{min}$, $UPP_{min}$, $DIG_{min}$, $SPS_{min}$, $R_{cmb34}$, $R_{cmb23}$, and $R_{cmb24}$ will be called \emph{\bf restrictive}, since they impose general (usually independent of the length) restrictions on which passwords are accepted by the policy. Restrictive parameters describe necessities and affect all admissible passwords.
%\item Parameters $P_{dict}$, $P_{seq}$, $P_{rep}$, $P_{longd}$, $P_{shortd}$, $P_{id}$, $P_{space}$, $P_{unicd}$, $P_{emoji}$, $P_{nlstart}$, $P_{spn1}$, $P_{spn2}$, $P_{spn3}$, $P_{spn4}$, and $P_{br}$ will be called \emph{\bf permissive}, since they describe only specific cases for passwords that may be accepted by the policy, but are not always required.

\textbf{Algorithm Steps.} At a high level, our inference algorithm operates by first finding one acceptable password (chosen in a specific fashion). Then, we evaluate each policy parameter one by one, testing passwords that are modifications of the original admissible password where only the specific parameter's dimension is changed, to determine that parameter's value. The order of parameter evaluation is specifically chosen to isolate the impact of just that parameter and minimize the number of successful account signups. Concretely, our algorithm operates in five steps.

\textbf{Step 1. Admissible Password:} First we must find an admissible password to seed our exploration, \ccschangestwo{which satisfies the restrictive parameters (e.g.,~minimum class requirements) and all permissive parameters (e.g.,~avoiding the relevant password characteristic such as repeated letter and number sequences.)}

For a given length $l$, we identify that there exists only a small set of passwords (which we call the \emph{safe set}) for which one password will satisfy any possible parameter combination.  If a website accepts passwords of length $l$, then the safe set must contain at least one acceptable password.

While we consider a variety of parameters, the safe set is small because a password can satisfy multiple restrictive parameters simultaneously (e.g., contain multiple characters of all classes, satisfying all minimum class and class combination parameters), and also satisfy all permissive parameters by avoiding the relevant password characteristic (i.e., avoiding certain characters and sequences).

\begin{table}[t]
\centering
\begin{adjustbox}{width=0.47\textwidth}
\footnotesize

\begin{tabular}{|c|c|c|c|c|c|c|}
\hline
Password & $L$  & \textbf{$R_{no\_a\_sps}$} & \textbf{$LOW$} & \textbf{$UPP$} & \textbf{$DIG$} & \textbf{$SPS$} \\ \hline
M-7c4@     & \multirow{10}{*}{6} &   & 1 & 1 & 2 & 2 \\ \cline{1-1} \cline{3-7} 
M-7cS@     &                     &   & 1 & 2 & 1 & 2 \\ \cline{1-1} \cline{3-7} 
Mx-7c@     &                     &   & 2 & 1 & 1 & 2 \\ \cline{1-1} \cline{3-7} 
Mx7c4@     &                     &   & 2 & 1 & 2 & 1 \\ \cline{1-1} \cline{3-7} 
Mx7cS@     &                     &   & 2 & 2 & 1 & 1 \\ \cline{1-1} \cline{3-7} 
M-7cS4     &                     &   & 1 & 2 & 2 & 1 \\ \cline{1-1} \cline{3-7} 
M-7S4@     &                     &   & 0 & 2 & 2 & 2 \\ \cline{1-1} \cline{3-7} 
x-7c4@     &                     &   & 2 & 0 & 2 & 2 \\ \cline{1-1} \cline{3-7} 
Mx-cS@     &                     &   & 2 & 2 & 0 & 2 \\ \cline{1-1} \cline{3-7} 
Mx7cS4     &                     & T & 2 & 2 & 2 & 0 \\ \hline
M7-cS4@    & \multirow{5}{*}{7}  &   & 1 & 2 & 2 & 2 \\ \cline{1-1} \cline{3-7} 
Mx7-c4@    &                     &   & 2 & 1 & 2 & 2 \\ \cline{1-1} \cline{3-7} 
Mx-cS4@    &                     &   & 2 & 2 & 1 & 2 \\ \cline{1-1} \cline{3-7} 
Mx7-cS4    &                     &   & 2 & 2 & 2 & 1 \\ \cline{1-1} \cline{3-7} 
Mx7zcS4    &                     & T & 2 & 2 & 2 & 0 \\ \hline
Mx7-cS4@   & \multirow{2}{*}{8}  &   & 2 & 2 & 2 & 2 \\ \cline{1-1} \cline{3-7} 
MxT7zcS4   &                     & T & 2 & 2 & 2 & 0 \\ \hline
Mx7-cS4@y  & \multirow{2}{*}{9}  &   & 2 & 2 & 2 & 2 \\ \cline{1-1} \cline{3-7} 
MxT7zcS4t  &                     & T & 2 & 2 & 2 & 0 \\ \hline
MxT7zcS4-@ & \multirow{2}{*}{10} &   & 2 & 2 & 2 & 2 \\ \cline{1-1} \cline{3-7} 
MxT7zcS4t1 &                     & T & 2 & 2 & 2 & 0 \\ \hline

\end{tabular}
\end{adjustbox}

\caption{The safe set of passwords for different lengths $L$. For each password, we indicate which restrictive parameter configurations are satisfied. Note that all passwords satisfy the class combination parameters, $R_{lstart}$, and $R_{2word}$ (if $L\ge 10$). Permissive parameters are also all inherently satisfied. For $L > 10$, the safe set is identical as with $L=10$, except with passwords padded with arbitrary letters and digits to length.
\vspace{-5mm}
}

\label{tab:safesets}
\end{table}

We manually construct the safe sets for lengths $l\in [6,32]$, shown in Table~\ref{tab:safesets}, covering the range of lengths that we conservatively assume a site must accept (based on our $L_{min}$ and $L_{max}$ assumptions). As seen in Table~\ref{tab:safesets}, the safe set for a given length contains passwords covering all restrictive parameter combinations, while also satisfying all permissive parameters. %For example, the password ``Mx-cS@'' would be admissible for policies that allow passwords of length 6 and requiring the presence of two lower characters, two upper characters, and two special characters but not digit class. Here, the password “Mx-cS@” would also satisfy any lesser restrictive policy requiring zero to one lower, upper, and special characters, as long as digits are not required.
Note that for short lengths, fewer restrictive parameters can be concurrently satisfied, so the safe set is larger.
The largest safe set contains 10 passwords (for $l=6$), while for lengths 8 or larger, the safe set consists of only two passwords (with and without special characters).

We search for an admissible password through the safe sets in increasing length order, first testing passwords with special characters within each safe set. Whether the admissible password found contains a special character already determines our first restrictive parameter $R_{no\_a\_sps}$ (if arbitrary special characters are disallowed).
In subsequent steps, we modify this admissible password along a single parameter's dimension and identify whether the modified password remains accepted, revealing the parameter's value. 

\textbf{Step 2. Restrictive Parameters:} With an admissible password of length $l$ (and $R_{no\_a\_sps}$ determined, \ccschangestwo{which indicates whether arbitrary special characters are allowed}), we then evaluate the restrictive parameters first, as determining these reveal the constraints enforced on any further tests. %Here, we modify the admissible password while preserving the password length. 
\ccschangestwo{To determine the value of a restrictive parameter, we modify the admissible password to only violate that parameter, observing whether the modified password is accepted. If so, then the restrictive parameter is in effect.}

\emph{1) \ccschangestwo{Combination of Words ($R_{2word}$):}} If $R_{2word}=True$, the admissible password must contain a two-word structure, delimited by a non-letter character (if not, then we already know $R_{2word}=False$). To test $R_{2word}$, we modify the admissible password by moving the non-letter delimiter to the password end, eliminating the two-word structure \ccschanges{(e.g., Admissible Password: MxT7zcS4-@, Modified Password: MxTzcS4-@7)}. If this modified password is no longer accepted, $R_{2word}=True$, otherwise $False$.
This modification does not affect other parameters as the length and character composition remain identical, and there are no other positional restrictions on middle-of-password characters. Permissive parameters are also not affected as the modification does not introduce a character sequence related to a permissive parameter (e.g., sequential/repeated characters, dictionary word).

\emph{2) \ccschangestwo{Letter Start ($R_{lstart}$)}:} All our admissible passwords begin with a letter. To assess $R_{lstart}$, we move the first non-letter character in the admissible password to the start \ccschanges{(e.g., Admissible Password: Mx7-cS4@, Modified Password: 7Mx-cS4@)}. If accepted, $R_{lstart}=False$, otherwise $True$. If $R_{2word}=True$, we take care to avoid moving the two-word delimiter \ccschanges{(e.g., Admissible Password: MxT7zcS4t1, Modified Password: 4MxT7zcSt1)}, as all admissible passwords have multiple non-letter characters (see Table~\ref{tab:safesets}). This modification does not affect other parameters as the length and character composition remain identical, and the only other positional restriction %($R_{2word}$) 
remains satisfied. Also, moving the non-letter characters does not introduce a character sequence affecting a permissive parameter.

\emph{3) Character Class Minimums \ccschangestwo{($DIG_{min}$, $UPP_{min}$, $LOW_{min}$, $SPS_{min}$)}:} To find the character class minimum for class $C$ %(where $C \in [DIG, UPP, LOW, SPS]$),
\ccschangestwo{(where $C$ is either digits, uppercase letters, lowercase letters, or special symbols}), we modify the admissible password to contain no $C$ characters, by replacing $C$ characters with characters of other classes \ccschanges{(e.g., if $C=LOW$, Admissible Password: Mx7-cS4@, Modified Passwords: MX7-CS4@)}. If accepted, $C_{min}=0$. Otherwise, we modify the admissible password to contain only one $C$ character \ccschanges{(e.g., if $C=LOW$, Admissible Password: Mx7-cS4@, Modified Passwords: MX7-cS4@)}. If accepted, $C_{min}=1$, otherwise $C_{min}=2$.

To avoid conflicting with other restrictive parameters, our default replacement policy is to swap between lowercase and uppercase characters (to not impact $R_{2word}$ and $R_{lstart}$), and between digits and special symbols (to not affect $R_{2word}$).
If $R_{no\_a\_sps} = True$ (no special characters allowed), digits are instead replaced with any letters (note here that if $R_{2word}=True$, then $DIG_{min} \geq 1$).

In most cases, all class combination parameters ($R_{cmb23}$, $R_{cmb24}$, $R_{cmb34}$) remain satisfied without further consideration. As seen in Table~\ref{tab:safesets}, most admissible passwords already have four character classes, so three classes remain after eliminating one class in the admissible password. A few admissible passwords have only three character classes (none have fewer classes), either because they are short (specifically, $l=6$) or because $R_{no\_a\_sps}=True$ (so only three classes are allowed).
For $l=6$ admissible passwords, there are two characters of each class, and we can replace the second $C$ character with one from the missing class, following the default replacement policy for the first character \ccschanges{(e.g., if $C=UPP$, Admissible Password: Mx-cS@, Modified Password: mx-c1@)}. This preserves $R_{lstart}$ while maintaining 3 distinct classes. 
%(note $R_{2word}$ is not relevant for such short passwords).
When $R_{no\_a\_sps}=True$, the class combination parameters either implicitly imply class minimums which we will correctly infer (e.g., $R_{cmb34}=True$ means there needs to be one character of each class), or will remain satisfied (the modified password still has two classes).

%\textcolor{purple}{The replacing candidate is chosen based on the values of the evaluated restrictive parameter and also avoids violating future ones such as the combinational restrictions and letter start. For example, $R_{2word}$ being true restricts the replacing candidate for lowercase to uppercase and special to number (and visa versa). Further, all admissible passwords in the safe set are structured to have at least three character classes. For those with exactly three present classes (2 general),  the replacing candidates are the ones missing in the admissible password, thus satisfying all combinational restrictions. For the starting letter restriction, the first character is not considered when the replacing candidate is a non-letter. Further, a special character cannot be selected as a candidate when $R_{no\_a\_sps}$ is true}

\emph{4) Combinations Requirements ($R_{cmb23}$, $R_{cmb24}$, $R_{cmb34}$):} To evaluate the final set of restrictive parameters, the class combination requirements, we modify the admissible password to have fewer classes and test for acceptance.

We start by identifying required character classes based on the other restrictive parameters.
$R_{2word}$ and $R_{lstart}$ both require letters; we select the required case based on class minimums, selecting lowercase letters by default. Similarly, $R_{2word}$ requires either digits or special characters; we select which based on class minimums and $R_{no\_a\_sps}$, selecting digits by default.

For modifying our admissible password, we replace all characters of non-required classes with those of a required class (replacing with lowercase letters if no class is required). If letters are required at certain positions, we replace any letters of a non-required class with letters of the required class (likewise between digits and special characters). This modified password has the minimum number of classes while adhering to other restrictive parameters, without impacting length or permissive parameters \ccschanges{(e.g., if $UPP_{min} \geq 1$, Admissible Password: Mx7-cS4@, Modified Password: MXZNCSZA)}.
If the modified password is accepted, we can determine the class combination parameters given the required classes in the password \ccschanges{(in the prior example, there are no class combination requirements)}.

However, if not accepted, then an explicit class combination requirement is in effect. We determine its configurations based on the properties of the rejected modified password, as follows:
\begin{itemize}[nosep, leftmargin=*]

    \item \emph{All non-letters of one class \ccschanges{(e.g., if $DIG_{min} \geq 1$, $R_{lstart} = False$, Admissible Password: Mx7-cS4@, Rejected Modified Password: 32729041)}}. Here, the other restrictive parameters require a single non-letter class. We test a new modification of the admissible password with only that non-letter class and letters of one case, using lowercase  by default \ccschanges{(e.g., New Modified Password: a2729041)}. If this new password is accepted, $R_{cmb23}=R_{cmb24}=True$ (and $R_{cmb34}=False$), otherwise only $R_{cmb34}=True$.
    
    \item \emph{All non-letters of both classes (\ccschanges{e.g., if $DIG_{min} \geq 1$, $SPS{min} \geq 1$, $R_{lstart} = False$, Admissible Password: Mx7-cS4@, Rejected Modified Password: 157-824@)}}. Here, we can immediately infer that only $R_{cmb34}=True$ as a two-class password was rejected.
    
    \item \emph{All letters of one class/case \ccschanges{(e.g., if $UPP_{min} \geq 1$, Admissible Password: Mx7-cS4@, Rejected Modified Password: MXZNCSZA)}}. We test a new modified password with letters of both cases (\ccschanges{e.g., New Modified Password: MxZNCSZA}). If accepted, only $R_{cmb24}=True$. If not, move to the following case.
    
    \item \emph{All letters of both classes/cases \ccschanges{(e.g., if $UPP_{min} \geq 1$, Admissible Password: Mx7-cS4@, Rejected Modified Password: MxZNCSZA)}}. If both letter cases are required, we know $R_{cmb23}=R_{cmb34}=True$. Otherwise, we test a new modified password with letters of only one case (whichever is required, defaulting to lowercase letters) and digits (\ccschanges{e.g., New Modified Password: M3ZNCSZA}). If accepted, only $R_{cmb23}=True$, otherwise only $R_{cmb34}=True$.
    
    \item \emph{Contains one non-letter class and one letter-class \ccschanges{(e.g., if $UPP_{min} \geq 1 $, $DIG_{min} \geq 1$, Admissible Password: Mx7-cS4@, Rejected Modified Password: MX71CS41)}}. 
    Here, we can immediately infer that only $R_{cmb34}=True$ as the two-class password was rejected.

\end{itemize}

\textbf{Step 3. Length Parameters:} Having now determined the restrictive parameter values that constrain password structure, we can construct passwords of different lengths that satisfy the restrictive parameters (while implicitly satisfying all permissive parameters by avoiding associated characters and sequences).
We can then determine the password length minimum and maximum through using binary search to test the acceptance of passwords of varying length (within the ranges $L_{min} \in [0,32]$ and $L_{max} \in [6, 128]$).  
\ccschangestwo{For example, to evaluate $L_{max}$, we first construct and test a password of length 67 (halfway point of our range). If accepted, we recursively explore $L_{max}$ within the upper half $[68, 128]$, otherwise we explore the lower half $[6, 66]$), following the logic of binary search.}

We detail our password construction algorithm in Appendix~\ref{Appendix:PasswordInferenceDetails}. At a high-level, the restrictive parameters provide a set of required characters and positional constraints,
and we satisfy these constraints first before adding additional characters to construct a password of an evaluated length $l$.
We start constructing a password using characters required by the class minimums, then using characters of other not-yet-used classes to satisfy class combination requirements (adhering to $R_{no\_a\_sps}$).
If $R_{lstart}$ and/or $R_{2word}$ are true, we satisfy these positional constraints at the start of the password, again first using allowed characters of classes required by the class minimums and combination requirements (and any remaining required characters are added after the positional constraints).
At this point, our partially-constructed password is the shortest that satisfies all restrictive parameters. If its length already exceeds the evaluated length $l$, we consider $l$ an unacceptable length. Otherwise, we pad the password with arbitrary letters and digits to length $l$ \ccschanges{(e.g., if $UPP_{min}=DIG_{min}=1$ and other restrictive parameters are false, Constructed Length-11 Password: M7ak3jCbE43)}.

%we detail how we construct a satisfying password for a given length $l$. The method attempts to generate a password of length $l$ that satisfies all restrictive requirements $R$. It starts with a queue of all required classes ($C_{min} > 0$) and arbitrary fills them while considering positional and structural restrictions of $R_{2word}$ and $R_{lstart}$. Class count of the constructed password is then compared with the values of the required combinations to append classes of the missing class (discarding special if $R_{no\_a\_sps} =True$). If the length already exceeds $l$, we deem $l$ as infeasible. If it is shorter than $l$, then we pad the password to length $l$ with arbitrary alternating letters and digits. 

%Step 4 focuses on finding the minimum and maximum length of the password using binary search. Expanding and reducing the password length must ensure that we do not violate any of the earlier deduced parameters or the permissive parameters that are not tested. E.g., we can’t expand by adding ‘aaaaa’ or ‘12345’, not to violate the repetition and sequence checks done later.

\textbf{Step 4. Permissive Parameters:} Next, we determine the permissive parameters (i.e., what is allowed in passwords).
\ccschangestwo{To do so, we inject the character(s) associated with a permissive parameter (e.g., emoji, dictionary word) into an admissible password, while still satisfying restrictive, length, and other permissive parameters, and test if the modified password is accepted. If so, then the permissive parameter is true, and the associated characters are permitted.}

\emph{1) Permitted Characters ($P_{space}$, $P_{unicd}$, $P_{emoji}$, $P_{spn1}$ - $P_{spn4}$):} We first generate an admissible password of maximum length (described in Step 3). We then test a modified password where a non-essential character (i.e., one not used to satisfy a restrictive parameter) is replaced with the evaluated character (if not possible, then the parameter value is inherently false) \ccschanges{(e.g., for $P_{spn4}$, Generated Password: Mx7-a1p5b2, Modified Password: Mx7-a1p5b\#)}. For $P_{space}$, we require that the whitespace character is not at the start or end of the password. %\ccschanges{(e.g., Generated Password from Step 3: Mx7-a1p5b2 when $L_{max}$ is 10 and exactly one of each class is required to be present), Modified password: Mx7- a1p5b to test for $P_{space}$}.
This modified password remains adherent to restrictive, length, and other permissive parameters. If accepted, the permissive parameter value is true. 

\emph{2) Permitted Sequences ($P_{rep}$, $P_{seq}$ , $P_{dict}$, $P_{id}$):} 
\ccschanges{ Here, we construct a password with the evaluated sequence and test for acceptance. For \ccschangestwo{repeated characters ($P_{rep}$)} the sequence is three repeating consecutive characters (e.g., \texttt{111}, \texttt{aaa}, or \texttt{AAA}), and for \ccschangestwo{sequential characters ($P_{seq}$)} it is \texttt{abc}, \texttt{123}, or \texttt{ABC}. For both parameters, we select one as permitted by other policy parameters. }

For \ccschangestwo{dictionary words ($P_{dict}$)},  we identify the longest word (up to 8 characters) permitted in a password as constrained by other policy parameters, and test the inclusion of the most common English word~\cite{oxford}. For \ccschangestwo{personal identifiers ($P_{id}$)}, the evaluated sequence is a subset of the username used during account creation. We choose our usernames to be a 3-letter names followed by 5 random digits, and the sequence is the 3-letter portion of the username (e.g. if the registered username is joe31426, we evaluate the acceptance of the sequence "joe" in the password)

We first construct the shortest password $P$ that satisfies the restrictive requirements (as done in Step 3).
If the evaluated sequence can be added to the end of $P$ while remaining within $L_{max}$, we simply test this augmented password, padding if necessary to reach $L_{min}$ \ccschanges{(e.g., to test for $P_{seq}$ if $L_{min}=6$, $L_{max}=64$ and the shortest password satisfying restrictive parameters is: AQ16-@, Modified Password: AQ16-@abc)}.  This augmentation does not affect restrictive parameters (nor length and other permissive constraints).

However, it is possible that appending the sequence to $P$ does not fit within $L_{max}$. In such cases, $P$ must already be near $L_{max}$-length (as we only require appending 3 characters). Instead, we must construct the evaluated sequence using characters already existing in $P$. We find the most common class $C$ in $P$ amongst lowercase letters, uppercase letters, and digits (for $P_{dict}$ and $P_{id}$, we only consider the two letter classes). We then rearrange the characters in $P$ to cluster $C$ characters together. If three (or more) $C$ characters are consecutive, we replace them with the evaluated sequence. Otherwise, we add the $C$ characters necessary to form a 3-$C$-character substring, again replacing this with the evaluated sequence. By using the most common class, we minimize the additional characters that may need to be added \ccschanges{(e.g., to test for $P_{seq}$ if $L_{max}=8$ and the shortest password satisfying restrictive parameters is: AQ16-@, Modified Password: ABC16-@)}. If the password cannot be constructed within length $L_{max}$, it is inherently false.

If restrictive parameters do not specify positional constraints, the rearrangement of $P$'s characters does not violate any restrictive parameters (nor length or other permissive parameters). 
If $R_{lstart}$ or $R_{2word}$ specify positional constraints, we handle each specifically. We ensure that the rearranged password starts with a letter if $R_{lstart}=True$. If $R_{2word}=True$, then $P$ contains a two-word structure, which must have at least 3 characters of one letter class. We cluster three letters of this class as one of the 3-letter words, and replace it with the evaluated sequence. % Thus, we can construct and test a password with the evaluated sequence (if possible within length).

%The password is then reordered so that characters of the same class are consecutive (starting with a letter character). Finally, the class with the most number of relevant character class count for the tested parameter is padded to the maximum allowed length $L_{max}$ (non-sequential and non-repetitive). If such a class is more than or equal to three after padding (or the length of the minimum allowed ID for $P_{id}$), we replace it with the tested substring (e.g., "abc"). We note that when $R_{2word}=True$, we can only reorder the letters of the two-words for the first seven characters and the non-letter delimiter. The second word (or after it) can then be used to test the substring.

%The constructing of a satisfying password used here adheres to all restrictions by design. The re-ordering and additional padding of a specific class count does not violate restrictive parameter beside $R_{2word}$ which is accounted for by only reordering letters of the two words and characters after the second word.

\emph{3) \ccschangestwo{Long and Short Digit Passwords ($P_{longd}$, $P_{shortd}$:)}} We generate digit-only passwords of lengths $L_{max}$ and $L_{min}$, respectively (without sequences/repetition). Here, restrictive parameters are ignored to explore exceptions for all-digit passwords \ccschanges{(e.g., if $L_{min}=6$, Attempted Password: 147036)}.

\emph{4) \ccschangestwo{Breached Passwords ($P_{br}$)}:} With other parameters determined, we test the highest-ranked breached password~\cite{xato} satisfying them \ccschanges{(e.g. if $DIG_{min} \geq 1$, $R_{lstart} = False$, $L_{min} \geq 6$, all other restrictive parameters are false and permissive parameters are true, Attempted Password: 123456, the most popular password in~\cite{xato} satisfying policy parameters)}. If accepted, $P_{br}=False$, otherwise true. 

\textbf{Step 5. Sanity Check:} Given an inferred policy, we test one final password that should not succeed (e.g., too short, violates restrictive parameters), as a sanity check. %\ccschanges{This is done by randomly sampling a password from~\cite{xato} and attempting it if it violates one or more of the policy parameters.} 
A detected success indicates a policy inference error, which we can filter out. (We also filter out other errors, where all attempts are successes or failures, and those where trailing attempts all fail, as discussed later.)

%\paragraph{Effectiveness of the Solution}

%Direct approaches to determining the parameters, like exhaustive or ``brute force'' searches, will be extremely wasteful on the attempts. This is mainly due to the fact that there are multiple (mostly independent) parameters. As previously stated, this results in at least several billion feasible policies, and this is even when only the consistent policies are considered (i.e. when an ``intelligent'' search is applied). The direct ``brute force'' approach will, in its worst case, have to loop through all 345 billion theoretically possible policies.

%Several features of the current solution allow this explosion in complexity to be avoided. First, the safe sets are significant optimization, as they allow the system to figure out the value for a singled out parameter with attempts that are ``safe'' for the other parameters (thus, applying a ``divide and conquer'' strategy). Also, the division and prioritization between the parameter types allows the system to figure out the most restrictive values first, and then this reduces the attempts for the rest of the parameters. Furthermore, the use of binary search techniques lowers the complexity of Step 4 from linear to logarithmic. Finally, the algorithm is integrated with a lot of logic about the dependencies between some parameters, which clears out a lot of complicated cases and saves further attempts.

\textbf{Algorithm Efficiency.}
Our algorithm systematically evaluates a website's password policy in an efficient fashion that avoids brute-force guessing passwords. As we can pre-compute the safe sets for our full range of explored lengths, and all policy parameters have a limited range of values (including length, which is efficiently investigated through binary search), we can determine the bounds on the number of passwords tested, as well as the bounds on the number of successful passwords accepted by a website.
Table~\ref{tab:algo_summary} depicts these bounds for each step of our inference algorithm, as well as for the entire algorithm.
In the worst-case, our method will create up to 37 accounts on a website, with at most 105 account signup attempts (in most cases, the number of attempts and accounts created is significantly lower).
%\frank{Can we make the prior sentence more concrete with the mean number of attempts + successes across our actual measurement?}
We note that we \changes{prioritized} fewer accounts created, as the impact of a failed account signup attempt on a website is much lower. Also, there is precedence in the research community for creating test accounts for measurement purposes; existing studies on password policies also created multiple accounts to evaluate policy parameters, but did so manually~\cite{florencio2010security, bonneau2010password, preibusch2010password, seitz2017differences, mayer2017second, wang2015emperor,lee2022password}. % (e.g., Seitz et al.~\cite{seitz2017differences} created up to 15 accounts per site).

\changes{
\textbf{Algorithm Correctness.}
Appendix~\ref{Appendix:PasswordInferenceDetails} describes how each parameter is correctly evaluated in isolation. To further ensure correctness, we tested our inference algorithm on a thousand randomly-generated valid policies, observing only correct inferences.
}

%%% Attempts median: 33 ; Success medians: 24

\begin{table}[t]
\begin{center}
\begin{tabular}{l c c}
\toprule
Algorithm Step & \# Attempts & \# Successes \\ 
\midrule
%& $[\;\stackrel{\text{best}}{\text{\scriptsize case}}, %\stackrel{\text{worst}}{\text{\scriptsize case}}\;]$ & %$[\;\stackrel{\text{best}}{\text{\scriptsize case}}, %\stackrel{\text{worst}}{\text{\scriptsize case}}\;]$ \\ \hline
%& $[\;\text{\scriptsize best}, \text{\scriptsize worst}\;]$ & %$[\;\text{\scriptsize best}, \text{\scriptsize worst}\;]$ \\ 
Step 1: Admissible password & [1, 65] & [1, 1] \\ 
%Step 2: Pivotal parameters & [0, 1] & [0, 1] \\ 
Step 2: Restrictive parameters & [4, 13] & [0, 9] \\ 
Step 3: Length parameters & [11, 12] & [0, 12] \\ 
Step 4: Permissive parameters & [2, 14] & [0, 14] \\ 
Step 5: Sanity check & [1, 1] & [0, 1] \\
\midrule
Total: Whole algorithm & [19, 105] & [1, 37] \\ 
\bottomrule
\end{tabular}
\end{center}
\caption{Bounds on the number of account signup attempts and successes required by our method, per domain.
}

\label{tab:algo_summary}

\end{table}

\subsection{Measurement Implementation}

We implement our measurement method using Selenium browser automation~\cite{selenium} with headless Chrome instances\footnote{When crawling with a headless browser, websites may detect and block such a crawler. However, when debugging our method, we tested full browser instances and did not observe higher crawling success, likely because many sites either do not block crawlers or apply anti-bot techniques that are similarly effective on full browsers.} %Meanwhile, using full browser instances requires more compute resources and execution time.}.
To minimize the computational load we induce on websites, as well as avoid triggering anti-bot detection, we rate limit our crawling of a domain to at most one page load every 30 seconds, and at most one account signup attempt every 30 minutes. We also use a pool of 14 proxies, switching to a new proxy for each signup attempt to provide IP diversity. Given the rate limiting, we highly parallelize our analysis across sites, such that sites are assessed in a round-robin fashion. %(i.e., other domains are analyzed while the next attempt on one domain is waiting).

%Another potential issue when testing an online policy is that the very quick programmatic succession of attempts may be recognized as a bot-like behaviour (especially when very similar and successive usernames are used). This is a risk both when a website's policy is being analyzed and the security system of the website considers the source of the inquiries, and also when an independent, third-party security service is monitoring the source (i.e. the IP address that runs the analysis of the website or websites). 

%Thus, it is needed to modify the behavior of the framework so that it rather resembles a human creating accounts on their own. For that purpose, we added a random waiting time in between making consecutive attempts upper-bounded by a specific fixed waiting time. Further, the inference job (consecutive attempts) for a domain is presented by a thread, and batches of domains are multi-threaded. This creates a round-robin effect in which threads take turns in querying the separate domains for every attempt. This enabled us to progress through our analysis of the domains while still having a large waiting time of at least 30 minutes between attempts for each investigated policy. IP blocking was also avoided by switching to a different proxy for every attempt on the domains.

%, which combines the four main popularity rankings (Alexa, Cisco Umbrella, Majestic and Quantcast) and filter out some domains to achieve a list with higher stability over time.

\subsection{Limitations}
\label{limitation}
Our measurement method is best-effort, relying on multiple heuristics. It can exhibit false negatives, missing some sites with account signups, such as those with complex workflows (e.g., multi-page forms), user verification (email or phone) prior to signup form submission, registration fees, or offline membership (details in Appendix~\ref{appendix:Challenges}). Furthermore, our evaluation may fail on sites that can detect our measurements (e.g., sites deploying anti-bot defenses) or where our machine learning models misclassify. However, as our method follows a consistent workflow for account signup attempts, we can filter out errors where all attempts are detected as successful or failures, which is infeasible, as well as those where trailing attempts are all failures (as this is highly unlikely, as discussed in Section~\ref{sec:Results}). Also, our final method step involves testing the inferred policy, further reducing the likelihood of false positives.

\changes{Our measurements also assume static policy parameters, rather than dynamic rules, such as if a site were to enforce password strength requirements. To evaluate whether password strength enforcement occurs at scale, we calculated the strength of all accepted passwords on successfully evaluated sites using password strength estimator \texttt{zxcvbn}~\cite{wheeler2016zxcvbn}.
We observe that for 94\% of sites, the weakest accepted password was rated 2 or lower (out of 4), which is considered a relatively weak password (ranging from ``too guessable'' to ``somewhat guessable''). Thus, it is unlikely that most sites are enforcing high password strength requirements. }
%Furthermore, any kind of anti-bot mechanism that we couldn’t overcome with proxy-rotation and long delays between attempts can result in measurement error. We also note that our developed ML techniques applied are subject to misclassification, which we best mitigate them in our final filtering step.}

%\roomi{Our measurement also assumes concrete set of policies rules employed instead of dynamic ones (e.g. through password meters or dynamically changing ones~\cite{yang2017dppg}). To evaluate the possible influence of password meter enforcing the requirements, we calculated the strength of all accepted password using password strength estimator zxcvbn~\cite{wheeler2016zxcvbn}. Extracting the weakest accepted password in each domain, we found that 94\% of the passwords were rated 2 or lower out of 4, indicating low possible influence of password meters.}

\ccschangestwo{Due to our method limitations, our evaluated sites may skew from domains with complex or unique workflows, as our analyzed domains use single-step account creation workflows, specific common keywords, and do not require verification or payment for signups. While our work does not comprehensively evaluate all sites (similar to all prior automated account creation works, including those investigating authentication~\cite{drakonakis2020cookie,deblasio2017tripwire}), our dataset (discussed in Section~\ref{sec:Results}) is still orders of magnitude larger and more diverse (including across rankings) than prior studies, serving as more generalizable empirical grounding. Furthermore, as detailed in Appendix~\ref{Generalizability}, we manually investigated the password policies of a random sample of domains that our method does not handle, and found that our study's core findings generalize to these domains.}

\subsection{\changes{Alternative Measurement Approaches}}
\label{sec:method_alt}

\changes{While our automated account creation process is similar to prior work~\cite{drakonakis2020cookie}, our task involves distinct challenges (e.g., password policy inference), so we designed our method in a data-driven fashion from scratch.  In comparison, while prior work applied rule-based heuristics for keyword selection, form detection, and verification, we applied machine learning techniques for such tasks. Our signup discovery process also uses search engine results to improve discovery. Our efforts resulted in effective account creation automation, even compared to prior work (see Appendix~\ref{appendix:comparisontocookiehunter}).}

\changes{We initially explored non-blackbox methods for assessing password policies, which could reduce website interactions. However, we manually evaluated a random sample of 200 signup websites and identified significant limitations.}

\textbf{Mining Textual Policy Descriptions:} Only 25\% of sampled sites provided policy descriptions (prior work observed 22\%~\cite{bonneau2010password}, as well as inconsistencies between policies and their descriptions~\cite{lee2022password}). Such descriptions are also diverse, often displayed only upon user action, and require natural language processing, yet often still do not  describe all policy parameters (e.g., password blocklisting).

\textbf{Inspecting Client-Side Policy Checks:} Only 10\% of sampled sites had client-side JavaScript password policy checks, which were custom implemented per site, inhibiting automated analysis.

\changes{\textbf{Analyzing Strength Meters:} Only 11\% of sampled sites displayed password meters (recent work found only 19\% on top English sites~\cite{lee2022password}). Prior work has also observed widespread custom meter designs~\cite{ur2012does}, inhibiting automated analysis. Furthermore, sites typically use password meters as nudges instead of enforcing strength requirements~\cite{wang2015emperor,furnell2011assessing}, and various policy facets (e.g., blacklisting, allowed characters) may not be factored into strength meters.}

%\roomi{These alternative methods are also implemented in inconsistent formats/styles (e.g., ~\cite{ur2012does} found 46 distinct meter designs on the top 100 sites), and many policy facets are not explicitly stated (e.g., blacklisting/allowing spaces/emojis). Further, they can be biased towards sites with better policies as their implementation are part of the recommended practices. }

\changes{\textbf{Using Password Resets:} One might assess password policies through password reset workflows. However, we did not log into accounts to avoid account activity (as discussed in Section~\ref{sec:Ethics}). Furthermore, many sites prevent choosing a new password similar to previous ones, which would interfere with policy inference. Finally, sites exhibit diverse password recovery workflows, often requiring user verification, complicating automated analysis.}

%We didn’t log into accounts to avoid any use of created accounts (see Ethics \ref{sec:Ethics}). Sites can also apply rate-limit on password change frequency or require significantly different passwords. Our inference algorithm relies on attempting subsequent passwords with a small change to test a single parameter requirements/acceptance in its working logic. Login also entails challenges verifying accounts (e.g., over email), custom password change workflow, and distinguishing login success.}

\subsection{Ethics}
\label{sec:Ethics}
%\roomi{general note: I have not added or edited this section, do we have to modify any parts after their initial comments?}
As our study involves evaluating a large number of websites, there are several important ethical considerations. It is impractical to obtain consent from all sites. Furthermore, obtaining consent could negatively impact the scientific validity of our study, as websites may opt-out in a biased manner, may change their policies in light of our investigation, or may specifically block our measurements. Thus, we do not seek consent from the studied sites, and must carefully design our measurement methods. \changes{We extensively explored various measurement methods (as detailed in Section~\ref{sec:method_alt}).} Here, we discuss the concerns with our resulting approach, the potential harm associated with our study, and our mitigations.

To assess the password policies on websites, we attempt multiple account signups in an automated fashion, succeeding for some attempts. 
Prior studies have performed similar automated account creation~\cite{deblasio2017tripwire, drakonakis2020cookie}, and we draw inspiration from their ethical considerations in designing our method.
The potential harm that this activity causes for websites includes the computational resources incurred by the website in processing our signup attempts and created accounts. To limit the resources that websites must expend due to our study, we \changes{constructed} our password inference algorithm to \changes{reduce} the number of attempts and successful accounts registered. For successfully created accounts, we never access, verify, or use those accounts. We also crawl websites and attempt account signups in a heavily rate-limited fashion, ensuring that a website receives at most one attempt every half hour (and in most cases, attempts occur even less frequently). We believe that for websites supporting account registrations, this rate of signup attempts and the number of accounts created requires a limited amount of storage and load on websites, and should not tax even small websites. Furthermore, there is precedence in the research community for creating small numbers of test accounts for measurement purposes; existing studies on password policies also created test accounts to evaluate policy parameters, but did so manually~\cite{florencio2010security, bonneau2010password, preibusch2010password, seitz2017differences, mayer2017second, wang2015emperor,lee2022password} (e.g., Seitz et al.~\cite{seitz2017differences} created up to 15 accounts per site).
As part of our account creation method, we solve CAPTCHAs using an automated CAPTCHA solver. We avoid human-driven CAPTCHA solvers due to ethical issues identified with such services~\cite{captcha_economics}.

%There is also the potential for reputational harm in disclosing websites with weak or unexpected password policies. Thus, we keep websites anonymous in this paper, and report aggregate statistics. However, we are attempting to contact sites with weak policies to inform them of our investigation.

From the legal perspective, we consulted our organization's general counsel, as our methods may be contrary to some websites' policies and terms of services, which we are unable to explicitly check for all sites in our study. General counsel reviewed this study and determined that the legal risk is minimal, with support from judicial precedence, and that there lacked damages incurred by websites. Our organization's administration also reviewed and approved this study. 
%Legal asepcts, consulted with general counsel, acted with the permission and knowledge of our administrators. Didn't check ToS. Fake info was provided. Legal risk is low, real damage to any party is absent, limitations of ToS enforceability, and outweighed by value of work
Finally, there are no human subjects concerns with this study (as such, we were not reviewed by our organization's Institutional Review Board). No real user data was used for this study, and our study did not interact with any individuals.

\section{Results}
\label{sec:Results}
Here, we apply our measurement method to evaluate the password policies of websites in the Tranco Top 1M. We analyze the top password policies, the values of the various policy parameters, adherence to modern guidelines, and differences across rankings. 

\subsection{Aggregate Measurement Results}
\label{results:agg}
We conducted our large-scale measurement in Dec. 2021, evaluating password policies across Tranco Top 1M (Dec. 13). Appendix Figure~\ref{fig:funnel} visualizes the site population at each method stage.

Out of the 1M domains, we find signup pages on 141K domains (14.12\%). While we could successfully submit one signup attempt (including CAPTCHA solving) on 59K domains, we were able to fully evaluate (across multiple attempts) 26K domains. Finally, we filter out domains where all signup attempts are reported as successes or failures (as this is not feasible, especially with our sanity check signup attempt), or where all trailing attempts are failures (we test permissive parameters last, and as discussed shortly, it is highly unlikely that any site truly does not permit all tested characters/structures). This filtering leaves us with 20,119 domains for which we successfully analyze password policies. We manually validated our results are accurate on a random sample of 100 evaluated sites. We note that this population is two orders of magnitude larger than prior work (as discussed in Section~\ref{sec:related}), providing large-scale data on password policies for the first time.

Our analyzed sites are also broadly distributed across rankings (unlike prior work's focus on top sites), with a slight skew towards lower-ranked sites, \ccschanges{as shown in Appendix Figure~\ref{fig:rankings}.}
Across each 100K ranking interval, our final dataset contains between 1.4K–3.7K sites (and between 12.1K-19.2K signup sites found).
\ccschangestwo{In the subsequent discussion of our results, we 
separately consider our evaluated sites that are within the top 10K, 100K, and 1M (full dataset). Here, our results for Top X sites represent only the domains that we evaluated within the Top X ranking, rather than all Top X sites (as we did not evaluate all sites).}

\subsection{Top Policies}
\label{sec:top_policies}

\begin{table}[t]
\centering
\begin{tabular}{|c|c|c|}
\hline
Rank & Policy & \%  \\
\hline
1 & $L_{min}=1$ & 8.3 \\
2 & $L_{min}=6$ & 7.1\\
3 & $L_{min}=5$, $L_{max}=40$ & 4.1\\
4 & $L_{min}=8$ & 3.4  \\
5 & $L_{min}=5$  & 2.9\\
6 & $L_{min}=12$ & 2.8\\
7 & $L_{min}=4$ & 1.2 \\
8 & $L_{min}=8$, $R_{cmb34}=T$ & 0.8 \\
9 & $L_{min}=8$, $L_{max}=72$ & 0.8 \\
10 & $L_{min}=7$ & 0.7 \\
11 & $L_{min}=4$, $L_{max}=40$ & 0.5 \\
12 & $L_{min}=8$, $P_{longd}=F$, $P_{shortd}=F$ & 0.5\\
13 & $L_{min}=8$, $LOW_{min}=UPP_{min}=DIG_{min}=1$ & 0.4  \\
14 &$L_{min}=4$, $L_{max}=20$  & 0.3\\
15 & $L_{min}=6$, $L_{max}=100$, $P_{emoji}=F$  & 0.3\\
\hline
\end{tabular}
\caption{Top 15 password policies for all evaluated sites. %We list the policy as well as the percent of sites exhibiting that policy. 
For each policy, unless specified otherwise, $L_{max}=128$, minimum required characters of a class is 0, restrictive parameters are false, and permissive parameters are true.}
\label{tab:top_policies}
\vspace{-4mm}
\end{table}

To start, we group websites with identical password policy configurations (across all policy parameters), and consider the top password policies observed among our websites.  Table~\ref{tab:top_policies} lists the top 15 policies observed across our 20K websites (spanning the Tranco top 1M sites), and the percent of sites using those policies. Among the top policies, the majority (11 of 15) are simple policies, only constraining the password length without further restrictions. Surprisingly, the most popular policy (8.3\% of sites) allowed passwords of any length without any constraints. Such a policy allows even single character passwords (we manually verify this behavior on a sample of sites), which are extremely weak passwords. Other top policies allow short passwords (e.g., 4, 5, and 6 characters).
In addition, 5 top policies also cap the password's length (including one that limits passwords to only 20 characters).
Other password constraints are less prominent in top policies, with only 4 of the top 15 policies applying any non-length constraints.% (requiring character class combinations, requiring minimum numbers of certain character classes, disallowing emojis, and preventing all digit passwords.

We find that policy popularity among sites exhibits a long-tail distribution. While the most popular policy was seen on 8.3\% of sites, the top 10 policies cover only 32.1\% of sites, with a total of 11,184~distinct policy configurations. Most policies appear on only one site, which highlights enormous diversity in the policies deployed (with implications for guidelines, password usability, and password managers, as will be discussed in Section~\ref{sec:lessons}).

%\textbf{Implications.} The latest password guidelines discourage applying password complexity constraints, and instead prioritize ensuring that passwords are suitably long (e.g., NIST 2017 requires 8 character minimums~\cite{grassi2017digital}). From the top policies we observed, we identify that a significant portion of websites already deploy simple policies that would align with these recommendation by increasing the minimum length accepted. We note though that besides password length, modern guidelines recommend checking passwords against breached password lists and dictionary words while permitting as many characters as possible (e.g., unicode, emojis). However, we did not observe any top policies employing such checks, and one top policy disallowing emojis.

%%%%%%%%%%%%%%%%%%%%%%%%%%%%%%%%%%%%%%%%%%%%%%%%%%%%%%%%%%%%%%%%%%

\subsection{Policy Parameters Values}
\label{sec:parameters}
Here, we evaluate individual password policy parameters. 
As the top 15 policies (Section~\ref{sec:top_policies}) capture only a third of our sites, their parameters do not necessarily reflect an aggregate perspective.

\begin{figure}[t]

    \centering
    \begin{subfigure}{0.37\textwidth}
        \includegraphics[width=\textwidth]{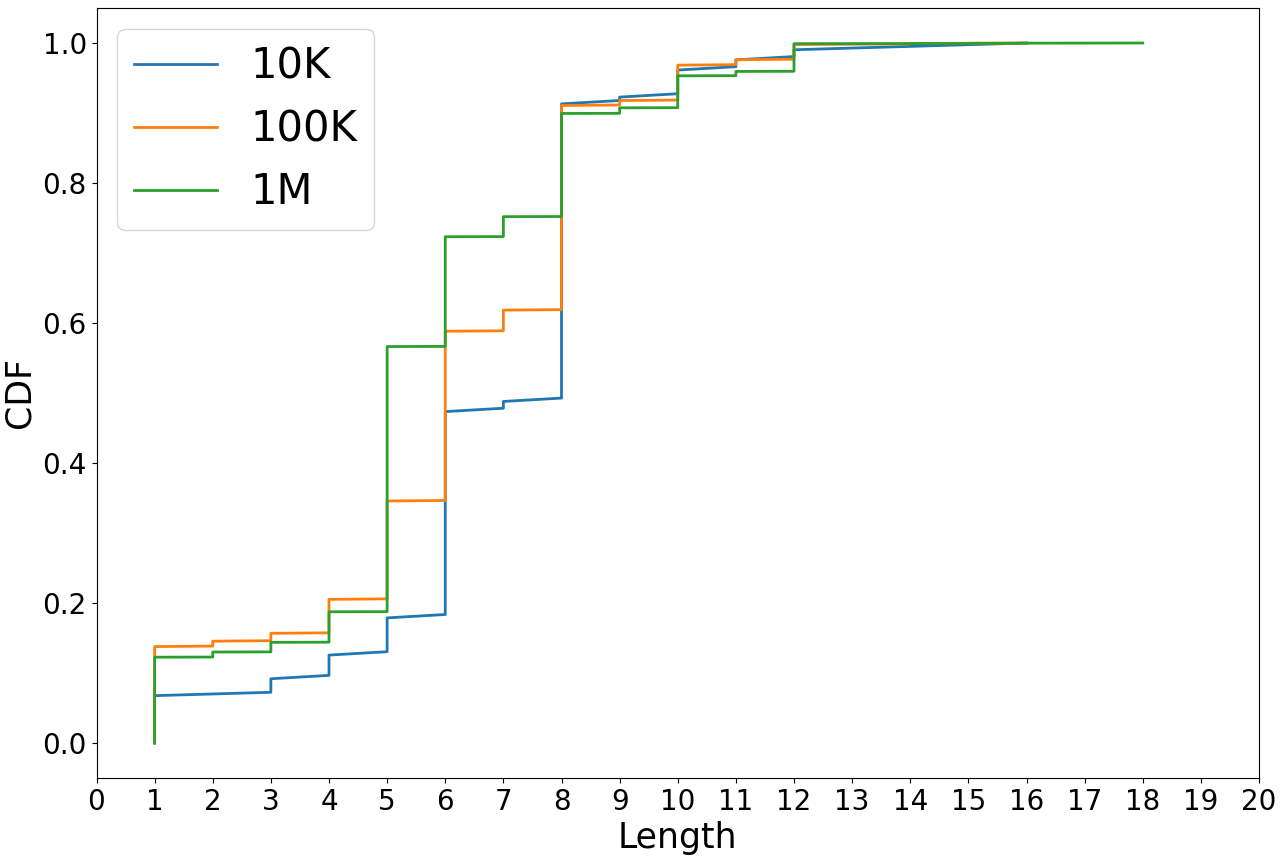}
        \caption{Minimum password lengths}
        \label{fig:minlength}    
    \end{subfigure}
    
    \begin{subfigure}{0.37\textwidth}
        \includegraphics[width=\textwidth]{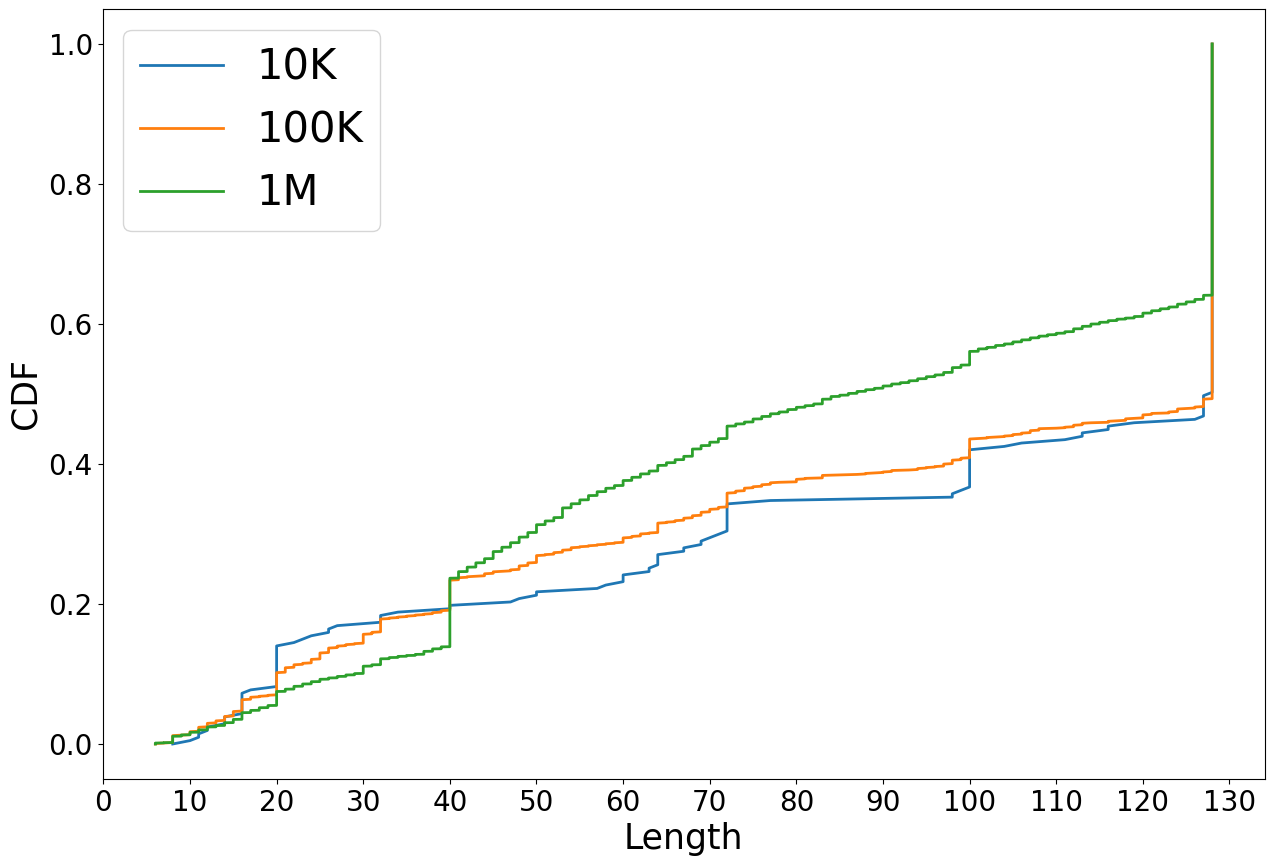}
        \caption{Maximum password lengths}
        \label{fig:maxlength}
    \end{subfigure}
    \caption{CDFs of password minimum and maximum length requirements, for all sites in our dataset (Top 1M) as well as those ranked in the top 10K and 100K.}
    \label{fig:length}

\end{figure}

\subsubsection{Length.} Figure~\ref{fig:minlength} plots the CDF of the minimum password lengths enforced by password policies across our websites (Top 1M). As also seen with top policies, we find that a non-trivial fraction of sites ($\sim$12\%) allow single-character passwords. The most prevalent minimum length is 5, seen at nearly 40\% of sites. Only 25\% of sites require passwords of length 8 or longer, as recommended by most modern guidelines~\cite{grassi2017digital, OWASP, NCSC2018, bsi2020, cert2009}, and $\sim$10\% require 10+ lengths.

Figure~\ref{fig:maxlength} similarly depicts the CDF of the maximum password lengths allowed by our websites. We observe that 36\% of sites do not cap the password length (or allow at least 128 characters). The most common cap was 40 characters, observed at about 10\% of sites. For other sites, the maximum length widely varied, although we notice prevalent use of lengths 20, 72, and 100. Overall, nearly 60\% of sites allowed passwords of at least 64 characters, as recommended by many current guidelines~\cite{grassi2017digital, OWASP, cert2009}.  We also find that a small portion of sites (1.7\%) do not allow passwords longer than 10 characters, which is shorter than some sites' \emph{minimum} lengths. %We plan to explore the domains behind these values in our future work.

\changes{
\textbf{Case Study: $L_{min}=1$.} We manually investigated 475 detected sites and verified the correctness of our measurements. Through analyzing the JavaScript libraries and links embedded on these sites, we identified that the common pattern exhibited was simply accepting any non-empty password field, without  applying password length logic. Interestingly, while this logic was customized for the majority of sites, we observed the prevalence of several web frameworks across these sites that we manually confirmed do not support password length constraints by default, such as WooCommerce (19\% of such sites) and XenForo (1\%).
}
%\roomi{important note: both WooCommerce and XenForo were confimred but Shopify is not outside of simply finding domains with size 1. Actually contacting the support team, they confirmed the default is 5 at the moment}

%\roomi{it maybe worth mentioning that the number of non-empty checks domains that are not using specific framework is 306, 64.3\%}

\changes{
\textbf{Case Study: $L_{min}=5$.} We investigated the most common minimum length of 5 (38\% of sites). Manually investigating a sample of 500 domains, we found 85\% using the Shopify platform. We confirmed with Shopify customer support that their default length minimum was 5, indicating the influence a platform can have.}

%\roomi{found this extremely interesting, how Shopify had a huge influence in the most common minimum we found = 5, I found this to be very interesting and significant, and also reaffirm our hypothesis of the strong influence of defaults on passwords implemented}

%requiring one-character password by extracting all link and script tags referenced in their signup pages (11,403 total links) along with manually sampling large of them, confirming they accept one character. We categized the links per page based on their services (e.g., CDN, framework, font-services,  jQuery, Google-services). Most of the domains within the list (64.3\%) were custom-built without any identifiable framework. These domains were found to implement a simple check to the password field not to be empty. Further, we found the presence of one popular e-commerce (WooCommerce) framework (18.5\%) and verified it does not enforce length requirements by default (and its meter is advisory). Another popular e-commerce framework (Shopify) was used in many domains (9\%) that we suspect its older (or custom) versions did not enforce length requirements (current default is . We also found examples of domains implementing a commercial forum software package (XenForo) that we confirmed did not enforce length requirements.}

\begin{table}[t]
\centering

\begin{tabular}{|cc|c|c|c|c|}
\hline
\multicolumn{2}{|l|}{}                          & \textbf{Lower} & \textbf{Upper} & \textbf{Digit} & \textbf{Special} \\ \hline
\multicolumn{1}{|c|}{\multirow{3}{*}{0}} & 10K  & 78.4           & 81.7           & 71.2           & 80.3             \\ \cline{2-6} 
\multicolumn{1}{|c|}{}                   & 100K & 79.2           & 79.4           & 76.3           & 82.5             \\ \cline{2-6} 
\multicolumn{1}{|c|}{}                   & 1M   & 84.1           & 83.7           & 82.0             & 86.3             \\ \hline
\multicolumn{1}{|c|}{\multirow{3}{*}{1}} & 10K  & 10.6           & 10.1           & 20.7          & 14.9             \\ \cline{2-6} 
\multicolumn{1}{|c|}{}                   & 100K & 11.6           & 10.9           & 14.5           & 9.8              \\ \cline{2-6} 
\multicolumn{1}{|c|}{}                   & 1M   & 8.3            & 8.7            & 10.0             & 7.0                \\ \hline
\multicolumn{1}{|c|}{\multirow{3}{*}{2}} & 10K  & 11.1           & 8.2            & 9.7            & 9.2              \\ \cline{2-6} 
\multicolumn{1}{|c|}{}                   & 100K & 9.2            & 9.7            & 9.2            & 7.7              \\ \cline{2-6} 
\multicolumn{1}{|c|}{}                   & 1M   & 7.5            & 7.6            & 8.0              & 6.8              \\ \hline
\end{tabular}

\caption{For different character classes, we list the percent of sites in the Tranco Top 10K, 100K, and 1M (full dataset) that require a certain number of characters of that class.}
\label{tab:restrictive}
\vspace{-3mm}
\end{table}

\subsubsection{Restrictive Parameters.} In Table~\ref{tab:restrictive}, we display the percent of sites requiring a minimum number of class characters, for each character class. We see that the vast majority of sites (82-86\%) do not enforce such requirements, with special characters being least likely to be required and digits being most likely. Of the remaining sites that do, approximately half require one character of a class, while another half require two (or more). We note that higher numbers of required characters of a class increase the complexity in creating passwords, which prior research has demonstrated can ultimately diminish the security and usability of passwords~\cite{shay2014can}, and is no longer recommended by many guidelines~\cite{grassi2017digital, OWASP, NCSC2018}.

%Similarly, Table~\ref{tab:parameter_vals} lists the prevalence of the remaining restrictive requirements. We observe a similar prevalence of character class combinations (15\% of sites) as with character class minimums (with 11\% of sites using both character class minimums and class combination requirements). We note that a non-trivial portion of sites (2.4\%) require word structure in passwords, while 2.9\% of sites require passwords to begin with a letter. Thus, many sites are not as permissive as recommended

\ccschangestwo{Similarly, Table~\ref{tab:parameter_vals} lists the prevalence of the remaining restrictive requirements. Derived from these results, we observed a similar prevalence of character class combinations (15\% of distinct sites have at least one required combination, considering all combination possibilities) as with character class minimums (with 11\% of sites using both character class minimums and class combination requirements). Furthermore, as seen in Table~\ref{tab:parameter_vals}, we note that a non-trivial portion of sites (2.4\%) require word structure in passwords, while 2.9\% of sites require passwords to begin with a letter. Thus, many sites are not as permissive as recommended~\cite{grassi2017digital, OWASP, NCSC2018}.}

\textbf{Case Study: Required Word Structure and Letter Start.} 
We manually investigated 100 domains requiring a two-word structure as well as domains enforcing letter start, confirming our inference. We did not identify common platforms or frameworks, but many sites used form validation JS libraries (e.g., jQuery Validation, FormCheck.js, Knockout Validation) to enforce a password regex.

%However, we note that this policy was seen for older versions of LinkedIn~\cite{linkedin2,johnson2019lost} as well as in leaked password datasets~\cite{johnson2019lost,melicher2016usability,shay2016designing,seitz2017differences}. We suspect this policy has been suggested for high security environments~\cite{shay2015creating}.}

%\roomi{note the 2.4 for above was written 7.2 by mistake in the prior versions of our writing. It was obvious mistake since we talked about 2word, but mention 7.2 instead of 2.4 shown in the table}

%\frank{I believe this parameter, R\_NoSPS doesn't include the 4 most common special characters. I think we should recompute consider sites that don't allow any special character, so are strictly alphanumeric.} \roomi{I just looked them up, almost all Rnosps =true are false in any of the 4 commons, with the exception of only 17 policies, recalculating the parameters did not change the \% with our rounding}

\subsubsection{Permissive Parameters.} 
\label{sec:results_permissive}
Finally, we evaluate the prevalence of permissive parameter values for our sites, as shown in Table~\ref{tab:parameter_vals}. 
Two widely recommended password policies~\cite{OWASP, bsi2020, NCSC2018, grassi2017digital, cert2009} are disallowing users to choose dictionary words and common breached passwords. We observe limited deployment of such password blocking though, as 72\% of sites permit dictionary words as passwords and 88\% allow breached passwords. Certain password structures are also often discouraged~\cite{grassi2017digital}, however we detect limited prevention of these patterns as well. Approximately 71\% of sites permit sequences, repeating characters, and personal identifiers (e.g., username) in passwords, and 78\% allow all-digit passwords.
Recent password guidelines~\cite{OWASP, grassi2017digital} also recommend allowing various types of characters. We observe  over 30\% of sites do not support spaces, Unicode, or emojis in passwords, and about 30\% disallow one of the four most popular special characters (``.'', ``!'', ``\_'', and ``\#''). 

\textbf{Case Study: Accepting Popular Passwords.} We assess whether sites accept popular passwords using the top four passwords in a password breach dataset~\cite{xato}. We list these passwords and their acceptance by sites across ranking ranges in Table~\ref{tab:breach}: 39\% of sites accepted the top password and nearly half accepted one of the top four passwords. These sites may be vulnerable to password spraying attacks~\cite{spraying,wang2015emperor} as their policies permit users to choose popular passwords. We note that most restrictive parameters and password blocklisting would disallow such passwords.

%\roomi{In Table~\ref{tab:breach}, we display the performance of policies against popular breached password attempts when it conforms to the evaluated policy (rejection is a direct indication of a deployed breach check). We note that more than a third of the policies conform and accept the most popular password “123456”, and at least one of the top 4 most popular breach passwords is accepted to around half of the policies in 10k, 100k, and 1m ranked domains. The combination of weak password policies along with no breach password checks makes these domains and their users vulnerable to both targeted~\cite{wang2015emperor} and password spraying attacks (login attempts to multiple users with a small list of high probable passwords)~\cite{spraying}}

\begin{table}[t]
\centering

\begin{tabular}{|c|c|c|c|}
\hline
                                 & \textbf{10K} & \textbf{100K} & \textbf{1M} \\ \hline
\textbf{Restrictive Parameters}    &              &               &             \\ \hline
Requires 2 Words                  & 4.8          & 3.9           & 2.4         \\ \hline
Requires No Arbitrary Special   & 4.3          & 3.0           & 1.8         \\ \hline
Any 3 of 4 Classes               & 6.7          & 6.4           & 7.4         \\ \hline
Any 2 of 4 Classes              & 14.9       &  11.0           &  9.1       \\ \hline
Any 2 of 3 General Classes      & 10.1         & 10.0          & 9.3        \\ \hline
Starting With a Letter              & 1.7         & 2.0          & 2.9        \\ \hline
\textbf{Permissive Parameters}    &              &               &             \\ \hline
Dictionary Words               & 83.7         & 80.1          & 72.0        \\ \hline
Sequential Characters           & 84.1         & 79.1          & 71.7        \\ \hline
Repeated Characters            & 82.2         & 79.8          & 71.1        \\ \hline
Short Digit-only               & 38.9         & 66.2          & 78.0        \\ \hline
Long Digit-Only                  & 57.2         & 69.4          & 78.2        \\ \hline
Personal Identifier              & 84.6         & 78.9          & 71.4        \\ \hline
Space                            & 75.5         & 73.3          & 69.0        \\ \hline
Unicode                          & 69.7         & 71.3          & 67.7        \\ \hline
Emoji                            & 59.6         & 65.8          & 64.4        \\ \hline
Breach Password                  & 84.1         & 84.8          & 88.2        \\ \hline
1st Popular Special = .          & 82.7         & 78.5          & 70.0        \\ \hline
2nd Popular Special = !         & 83.7         & 77.7          & 69.6        \\ \hline
3rd Popular Special = \_         & 84.1         & 78.3          & 69.7        \\ \hline
4th Popular Special = \#       & 82.2         & 76.4          & 69.4        \\ \hline

\end{tabular}

\caption{Policy parameter values for all sites within the Tranco Top 10K, 100K, and Top 1M (full population). For both restrictive and permissive parameters, we list the percent of sites where the parameter value is \emph{True}.}
\label{tab:parameter_vals}

\end{table}

\begin{table}[t]
\centering

\iffalse %%% REMOVING %%%%%
\begin{tabular}{|c|c|c|c|c|c|}
\hline
\textbf{Password} & \textbf{Rank} & \textbf{100} & \textbf{10K} & \multicolumn{1}{l|}{\textbf{100K}} & \textbf{1M} \\ \hline
n123456    & 1  &   11         & 21 & 39 & 39 \\ \hline
123456789 & 2   &    20       & 26 & 46 & 40 \\ \hline
qwerty    & 3    &     11     & 22 & 38 & 42 \\ \hline
password  & 4     &      19   & 27 & 49 & 48 \\ \hline
          & \textbf{Top 4} &20 & 27 & 51 & 53 \\ \hline
\end{tabular}
\fi  %%%%%%%% END OF REMOVED GRAPH %%%%%%%%%%%%

\begin{tabular}{|c|c|c|c|c|}
\hline
\textbf{Password} & \textbf{Rank}  & \textbf{10K} & \multicolumn{1}{l|}{\textbf{100K}} & \textbf{1M} \\ \hline
123456    & 1           & 21 & 39 & 39 \\ \hline
123456789 & 2        & 26 & 46 & 40 \\ \hline
qwerty    & 3        & 22 & 38 & 42 \\ \hline
password  & 4      & 27 & 49 & 48 \\ \hline
          & \textbf{Top 4} & 27 & 51 & 53 \\ \hline
\end{tabular}

\caption{\changes{Percentages of signup sites accepting the top four most popular passwords (based on a breach dataset~\cite{xato}).}}
\label{tab:breach}

\end{table}

\iffalse %%% REMOVING %%%%%
\begin{figure}
    \centering
    \includegraphics[width=8cm]{L_min_pdf.png}
    \caption{PDF - Minimum Required Password Length of 20,119 Domains
    }
    
    \label{fig:minlength}
\end{figure}

\begin{figure}
    \centering
    \includegraphics[width=8cm]{L_max_pdf.png}
    \caption{PDF - Maximum Required Password Length of 20,119 Domains
    }
    
    \label{fig:maxlength}
\end{figure}
\fi  %%%%%%%% END OF REMOVED GRAPH %%%%%%%%%%%%

%%%%%%%%%%%%%%%%%%%%%%%%%%%%%%%%%%%%%%%%%%%%%%%%%%%%%%%%%%%%%%%%%%

\begin{table}[]
\centering

\begin{tabular}{|c|ccc|}
\hline
\textbf{Standard Name}&  \textbf{1M} & \textbf{100K} & \textbf{10K} \\ \hline
NIST 1985 (Low)&  16.7            & 22.1 & 27.4    \\ \hline
NIST 1985 (Med)&  7.6             & 12.8 & 15.4    \\ \hline
NIST 1985 (High)&  3.8             & 7.4  & 9.1     \\ \hline
NIST 2004 (Lvl 1)&  42.1            & 65.3 & 77.9    \\ \hline
NIST 2004 (Lvl 2)&  5.5             & 7.5  & 6.7     \\ \hline
BSI 2005&  4.8             & 6.7  & 6.7     \\ \hline
US CERT 2009&  0.3             & 0.4  & 0.5     \\ \hline
DISA 2014 (Med)& 4.7             & 8.8  & 10.1    \\ \hline
DISA 2014 (High)&  0.1             & 0.1  & 0.5     \\ \hline
NIST 2017 (Should)&  30.8            & 40.8 & 34.6    \\ \hline
NIST 2017 (Shall)&  1.8             & 3.16    & 3.9     \\ \hline
NCSC 2018& 0.7             & 1.2  & 2.4     \\ \hline
BSI 2019&  14.6            & 22.3 & 32.7    \\ \hline
BSI 2020& 5.9             & 7.5  & 7.2     \\ \hline
OWASP& 1.3             & 2.9  & 4.3     \\ \hline
\end{tabular}

\caption{Percent of sites satisfying different guidelines, across the Tranco Top 10K, 100K, and 1M (full population).}
\label{tab:compliance}

\end{table}

\subsection{Adherence to Standards and Guidelines}
\label{sec:compliance}

Over time, various organizations have released password policy guidelines. Here, we assess the extent to which sites adhere to these guidelines. In Table~\ref{tab:compliance}, we list 9 prominent guidelines in order of publication year, including different security levels offered by some.  Appendix Table~\ref{tab:guideline_descriptions} summarizes these recommendations. While we can determine if a site's policy adheres to a standard, we do not know if the site's owners explicitly chose to follow the standard.
%Here we measure the adherence of our password policies against twelve national, international, and community-driven policies from 1985 to 2021, shown in Table~\ref{tab:compliance}. A summary of the password-policy related guidelines is provided in Table x of Appendix a.

We observe that NIST's 2004 guidelines have been most widely adopted, with \ccschangestwo{42.1\%} of sites adhering. Meanwhile, \ccschangestwo{30.8\%} of sites' policies satisfy NIST 2017's guidelines, although \ccschangestwo{16.7\%} of sites exhibit policies that follow NIST's old 1985 recommendation. These results indicate the staying power of recommendations, as old NIST guidelines are still observed on most sites, even more than 5 years after updated guidelines were released. Similarly, fewer websites adhere to  Germany BSI's latest guidelines compared to older ones. 

Across NIST and DISA guidelines, we also observe that stronger security levels are significantly less adopted. For example, only \ccschangestwo{5.5\%} of sites have policies satisfying NIST 2004 Level 2, compared to \ccschangestwo{42.1\%} for Level 1. We also see low adoption of stricter password guidelines, such as those of US CERT, NCSC, and OWASP. Notably, these guidelines and higher security levels generally required stricter length requirements (particularly $L_{min}=8$), and checks against dictionary words and breached passwords. This suggests incentives to adopt stronger policies are ineffective and the costs of deploying these strong policy parameters are non-trivial.

\subsection{Variation by Website Rankings}
\label{sec:website_characteristics}
Here we consider how password policies differ across websites ranked within the Tranco Top 10K, 100K, and 1M.% (our full site population). %These differences highlight how looking at only top sites can provide a different picture of password policies compared to a larger-scale measurement.

%\roomi{This subsection was initially called "Variation by Website Characteristics" but now the country and category content and reference are removed}

\iffalse %\ country  or category
Here we consider how password policies differ across different website subpopulations, specifically focusing on ranking differences, and relegating analysis of different site countries and categories to Appendices~\ref{sec:countries} and~\ref{sec:categories}, respectively (due to space). 
We compare websites ranked within the Tranco Top 10K, \changes{Top 100K}, and our full set of websites (Top 1M). 

\fi
\textbf{Length.} Figure~\ref{fig:length} shows the CDFs of minimum and maximum passwords lengths, respectively, for \changes{all three} groups. We observe that in all graphs, the CDFs for top-ranked sites skew towards longer lengths, which is recommended for stronger passwords. The median minimum password length for \changes{top 10K} sites is 8 characters, compared to 5 and 6 characters for \changes{the top 100K and} all sites, respectively. Similarly, while about 40\% of all sites allow long passwords that are at least 128 characters, \changes{50\% and 55\% of top 100K and top 10K sites do, respectively} (although a higher portion of top-ranked sites cap passwords at 20 or fewer characters than among all sites).

%Tables x and y show the restrictive and permissive parameters for both groups. We note that domains within the 1k ranking are less likely to accept digit-only passwords of any length (38.94\% for short-digit and 57.21\% for long-digit) compared to the one million group (78.04\% and 78.24\%, respectively). Both groups were found to allow sequential, repetitive, and accept dictionary word passwords at a high percentage range of 71-83\%. This contradicts with the general early guidelines from the literature [] and the industry [] that recommends complex passwords with these parameters checked. We also observed a high acceptance of Emoji (59.62\%, 64.44\%), Unicode (67.71\%), and space (75.48\%, 68.98\%), which inlined with the recent shift standards [] advocating for passphrases permitting such characters. 

%Approximately one fifth of the domains required one or more of a specific type of character to be present (lower, upper, digit special). The top 1k domains had higher requirements of one or more of a particular class being present (21.63\%, 18.27\%, 28.85\%, 19.71\% for lower, upper, digit and special, respectively) compared to the top one million (15.86\%, 16.29\%, 18.00\%, 13.72\%). 

\textbf{Restrictive and Permissive Parameters.} 
Table~\ref{tab:parameter_vals} depicts the parameter values for \changes{all three ranking ranges}, showing the percent of sites within each population where a parameter value is true. We observe that overall, top sites are more likely to enforce restrictions on the password (e.g.,~$R_{cmb24}$ is true for 15\% of top 10K sites, compared to 9\% of all sites).
Top sites are generally more permissive in which special characters they accept, including periods, exclamation marks, underscores, pound signs, space, and Unicode characters (although slightly fewer top sites accept emojis compared to all sites). Surprisingly, top sites also are more permissive of oft-discouraged password patterns, including dictionary words, sequential and repeated characters, and the inclusion of personal identifiers. However, top sites are significantly less likely to accept all-digit passwords, accepted by only 39-57\% of top 10K sites compared to 78\% of all sites. Top sites are also slightly less likely to allow breached passwords compared to all websites though (84\% of the top 10K versus 88\% for all).
Overall, top sites apply more password composition requirements but also permit more characters/structures (except all-digit passwords).
%When inspecting the top 15 policies on top 10K sites (shown in Table~\ref{tab:top10k_policies} of the Appendix), we observe that the top policies are also generally more complex than those of all sites, again suggesting that top sites skew towards enforcing more policy constraints.

\textbf{Adherence to Guidelines.} Table~\ref{tab:compliance} lists the adherence to common guidelines \changes{across ranking ranges}. We observe that across all guidelines, higher-ranked sites generally exhibit higher adherence, suggesting that they are more likely to follow recommendations. However, the most recent guidelines are still only adopted by a minority of sites across all three ranking ranges (see Section~\ref{sec:compliance}).

%%%% END OF TABLE REMOVAL %%%%%%%%

\section{Comparison with Prior Findings}
\label{sec:comparison}

%\frank{Discuss the high-level implications of our results, any next steps that can be done/future work, and concluding remarks. There's a movement in many papers these days to not have a separate discussion + conclusion, especially since most papers' conclusion is just a rewrite of the abstract/intro in past tense. So I think we'll do that here both to save space and be more direct about our concluding remarks.}
%In this study, we conducted a large-scale empirical measurement of password policies across the web. We automatically collected policies of over 20K sites, analyzing the top policies, policy configurations, and adherence to popular guidelines.
\iffalse %\ country  or category

and variation of policies across website rankings, countries, and categories.
\fi

Prior works on assessing website password policies are small-scale and largely dated\ccschangestwo{~\cite{furnell2007assessment,mannan2008security,kuhn2009survey,florencio2010security,bonneau2010password,preibusch2010password, furnell2011assessing}} (see Section~\ref{sec:related}). Here, we compare our results with prior findings, to understand how policies may have changed over time, and the insights afforded by a large-scale perspective.

\textbf{Top Policies and Parameter Values.}
Prior work assessed policy parameter values, rather than top policies, likely due to small sample sizes. 
In comparison, our large-scale study identified the top policies, most of which enforced only length constraints, as well as a long tail of policies which are mostly unique to a site.
%Seitz et al. found that 70\% of policies on the 83 sites investigated only enforced length constraints, similar to many of our top policies.

\emph{Length:} %Most prior work observed that a minimum length of 6 was most frequently enforced (at over half of evaluated sites~\cite{bonneau2010password, seitz2017differences, wang2015emperor}), with few sites  (e.g., \cite{bonneau2010password} only observed 2\%) requiring even longer passwords. 
A recent 2022 analysis of 120 top English sites observed that a minimum length of 8 was most frequently enforced, followed by lengths 6 and 5~\cite{lee2022password}. We observe the same for our top 10K sites, with 40\% of sites requiring length 8 passwords, 30\% requiring length 6, and 7\% requiring length 5. However, when considering the top 1M sites, length 5 was the most prevalent, on nearly 40\% of sites. Meanwhile, length 6 and length 8 passwords were required by approximately 15\% of sites each.
%However, over 40\% of the sites now require passwords longer than 6 characters. 
Further,~\cite{bonneau2010password, mayer2017second} observed few sites without length requirements, but at scale, we observed this policy at nearly a quarter of the sites. Thus, our large-scale measurement identified shorter password length minimums on most sites than reported by recent studies focused on top sites.

Prior work observed widespread use of length caps (note, \cite{lee2022password} did not investigate length maximums). Seitz et al.~\cite{seitz2017differences} observed an average max length of 43 characters, and Wang et al.~\cite{wang2015emperor} did not observe any max lengths greater than 64. In contrast, we observe over a third of all sites allowing 128+ character passwords, with a median length cap of 86 (with even fewer sites using length caps among top-ranked sites). As these prior studies are over a half decade ago and of limited scale, it seems likely that sites today have broadly shifted towards accepting longer passwords.

\emph{Restrictive and Permissive Parameters:} Few works systematically characterized restrictive and permissive parameters, with most highlighting case studies rather than comprehensive analysis. However, prior work~\cite{bonneau2010password, wang2015emperor,lee2022password} observed between 30-50\% of sites enforced several restrictive parameters. We observe a smaller fraction, with only 1.8-9.3\% of sites employing any given restrictive parameter, although top-ranked sites employed restrictive parameters more. Thus when considering websites at scale, restrictive parameters are less prevalent overall. Earlier work from 2010~\cite{bonneau2010password} also found few sites performing dictionary checks. However, we observed a modest rate today, at 28\% (\cite{lee2022password} observed 41\% on top English sites).

%Bonneau: 18\% didn't impose length. 6 was most common (half sites), with 5 and 4 next (covering 76\% of sites). 2\% of sites had more than length 6. (Kuhn, even fewer had length). Did observe few sites implemented other restrictions. 9\% do dictionary checks. 5\% had any character requirements. 

%Seitz: 70\% only specified length, primarily len 6. Remaining 30\% specified various complexity requirements. Half sites didn't specify max length, but max length when set was 43 characters on average. Seitz observe a number of blocked characters.

%Wang: 60\% lmin6, 30\% lmin8. 72 lmax<=64. Half enforce charset requirements. Observed acceptance of only long digit-only password.
%32\% used blacklist of popular passwords. Only 28\% blocked IDs.

\textbf{Adherence to Standards and Guidelines.}
Prior work mostly predates modern password guidelines\ccschangestwo{~\cite{furnell2007assessment,mannan2008security,kuhn2009survey,florencio2010security,bonneau2010password,preibusch2010password, furnell2011assessing}} (e.g., NIST 2017, BIS 2019), and did not identify comprehensive comparisons of password policies with the standards prevalent at a study's publication. %(even for recent studies). %However, we note that a number of studies identified that common policy parameter values appeared to adhere to the guidelines of the times (e.g., the frequency of 6-character minimum length passwords was attributed to NIST 2004 by prior work~\cite{bonneau2010password}).

\textbf{Variation by Website Ranking.}
Prior work~\cite{florencio2010security, mayer2017second} looked at several US university websites, and found that top-ranked sites had weaker policies than lower-ranked ones, although policies were evaluated using an entropy metric with notable limitations~\cite{weir2010testing, grassi2017digital}. In contrast, our site population is orders of magnitude larger and has substantially broader ranking coverage, and we observe stronger policy characteristics for top sites (e.g., longer length requirements, broader adherence to modern recommendations).

\section{Concluding Discussion}
\label{sec:lessons}

In this study, we conducted the largest evaluation of website
password creation policies to date, assessing over 20K sites ($\sim$135x more sites than prior work). Our results revealed the state
of modern web authentication, and identified insecure policies deployed (especially outside of the top sites). 
\ccschanges{Of note, we observed that 75\% of sites allow shorter passwords than the recommended 8 characters~\cite{OWASP,bsi2020, NCSC2018,grassi2017digital,cert2009} (with 12\% allowing single-character passwords) and 40\% cap password lengths below the 64 characters recommendation~\cite{OWASP, grassi2017digital, cert2009}.
Meanwhile, 15\% of sites enforce character constraints, which is no longer recommended~\cite{OWASP, NCSC2018, grassi2017digital}.
Only 12\%-28\% of sites employ password blocking, as widely advocated~\cite{OWASP,bsi2020,NCSC2018,grassi2017digital,cert2009}. Finally, a third of sites did not support certain password characters as suggested~\cite{OWASP,grassi2017digital}, including whitespaces needed for passphrases. Ultimately, only a minority of sites adhered to modern guidelines overall.}
Here, we synthesize our findings into
lessons for moving web authentication forward.

%\textbf{Weak policies remain widespread.} Our results reveal that despite recent emphasis on increasing password length requirements~\cite{grassi2017digital}, most sites still permit short passwords and many disallow long ones. Also, most sites do not block dictionary words or breached passwords. A quarter of sites disallow spaces in passwords, preventing the use of passphrases (passphrases are recommended~\cite{grassi2017digital}). While the existence of weak policies may not be surprising, they highlight specific policy parameters that require attention, where effort is needed to translate modern recommendations into real-world changes. % (discussed further in Section~\ref{sec:future}).
%In particular, implementing longer passwords and password checks appears to be non-trivial for websites, as significantly fewer sites employ such checks (e.g., BSI 2020 differs from BSI 2019 only in the addition of dictionary word checks, yet only a third of sites satisfying BSI 2019 also satisfy BSI 2020).

\textbf{Improving Software Defaults and Implementation Support}. Our case studies in Section~\ref{sec:parameters} identified that insecure password policy decisions were closely aligned with the default configurations of popular web software (such as WooCommerce and Shopify).
These findings demonstrate the influence of software defaults on web authentication, but also illuminate a potential remediation path: if popular web software implemented recommended password policy configurations by default, many websites could be moved to stronger password policies. 
%This direction is particularly appealing as only a limited number of organizations/entities would need to update their software, and the changes would propagate to websites broadly.
For example, \emph{nearly half} of our sites with password length minimums below the 8 characters recommended~\cite{grassi2017digital, OWASP, NCSC2018, bsi2020, cert2009} use the Shopify platform and its default 5 characters minimum. Thus, if Shopify increases its default length to 8 characters, potentially a third of our sites would become newly aligned with modern guidelines.
We are currently in the process of communicating with platforms identified offering weak default configurations to encourage such changes.

Related to defaults are the feature support by popular web software. We observed in Section~\ref{sec:results_permissive} that only a minority of sites blocked passwords with certain characteristics, which is widely recommended~\cite{grassi2017digital, OWASP, bsi2020, cert2009, NCSC2018}. We hypothesize that this arises partly because many popular web platforms do not provide full support for such blocking, so web developers would need to custom implement such functionality.
For example, both Python's Django library~\footnote{https://docs.djangoproject.com/en/4.2/topics/auth/passwords/} and the WordPress CMS~\footnote{https://www.wpbeginner.com/plugins/how-to-force-strong-password-on-users-in-wordpress} by default do not support all password checks.
%checks against dictionary words, as well as sequential and repeated characters. 
By implementing such features (and enabling by default) for popular web frameworks (many of which are open-source), our community can meaningfully improve web authentication.
%recommended password policies could be more readily adopted.
%Thus, meaningful improvements to web authentication can be driven by contributing implementations of these features to popular web platforms, many of which are open-source.

\textbf{Promoting Modern Password Guideline Adoption.} Our analysis in Section~\ref{sec:compliance} revealed that many sites exhibit policies satisfying password guidelines, but primarily more dated versions. This result provides evidence that password guidelines do generally inform the policy decisions of many websites. However, there must be barriers inhibiting the adoption of more recent recommendations.

A lack of awareness may be one barrier. Here, education and outreach efforts can help inform websites about current guidelines. Prior work on web administrator notifications~\cite{li2016remedying,li2016you, stock2016hey, stock2018didn} demonstrated that such outreach efforts can drive the remediation of security issues at scale. Future work can also investigate the resources available about web authentication, and identify information sources that should be updated with current recommendations.

In addition, in Section~\ref{sec:compliance}, we saw different guidelines from various organizations, with sometimes conflicting recommendations. For example, NIST 2017~\cite{grassi2017digital} and OWASP~\cite{OWASP} guidelines avoid password complexity requirements, unlike BSI 2020~\cite{bsi2020}. A unified password guideline would provide more consistent and clear recommendations to web administrators around the world. We also uncovered that some guidelines (e.g., OWASP, NCSC 2018) are rarely adopted, suggesting that these guidelines are overly strict or lack visibility and incentives to drive adoption.

Even if adopting a new policy, a remaining challenge is the policy update process. How should websites handle passwords created under the old policy? If old passwords are left as is, the new policy's benefits are not realized. Meanwhile, forced password resets are often onerous to users (as seen with the password resets during data breaches).
Future work should investigate effective processes for upgrading password creation policies, and integrate them into existing web software.
Organizations releasing password guidelines also must be cognizant of the high burden imposed upon websites when adopting new policies, and guidelines must be released with care (e.g., BSI released two guidelines only one year apart~\cite{bsi2019, bsi2020}). %frequent guideline updates may disincentivize websites from promptly adopting.

\textbf{Standardizing Password Creation Policies to Promote Usability.} 
In Section~\ref{sec:top_policies}, we observed that websites exhibit wildly diverse policies, with many policies unique to one site. This heterogeneity is likely a usability burden during password creation, where users do not know what constraints are enforced on chosen passwords across different sites. This is especially true as we found that few sites explicitly document their password policies (from Section~\ref{sec:method_alt}). Standardizing password policies would significantly reduce this user friction, providing a unified policy across the web.

Such standardization would benefit password managers as well, as many password managers assist users by automatically generating random and strong passwords. To do so correctly, they must generate a password valid under a site's policy, which is inhibited by the diversity of real-world site policies.
For example, some sites disallow long passwords or require certain character compositions (from Section~\ref{sec:Results}), which may not be satisfied by a password manager's randomly generated password.
We note that even with the absence of standardization, our results  help inform password managers of the common policy constraints enforced by most sites. For example, we found that passwords of length 12--16 are the most likely to be accepted, permitted by 96-98\% of sites. Our measurement dataset can also be inputted directly to password managers to provide the specific constraints on the sites that we analyzed.

\textbf{Future Research Directions.}
\ccschangestwo{
\label{sec:future}
Our study highlights avenues for future investigation. One direction is in improving upon our measurement techniques. While our collected dataset is significantly larger than those of prior work~\cite{furnell2007assessment, furnell2011assessing, mannan2008security, kuhn2009survey, florencio2010security, mayer2017second, wang2015emperor, bonneau2010password, preibusch2010password, seitz2017differences, lee2022password}, we still successfully analyzed only a minority of sites with account signups. Expanding measurement coverage would allow for more generalizable findings and more extensive analysis of authentication policies across different site characteristics. Similarly, longitudinal measurements could afford insights into policy evolution. Future work could also investigate which website characteristics correlate with secure and usable password policies, such as website categories, geographic regions, and languages.}

\section{Acknowledgements}
We thank the anonymous reviewers for their constructive feedback. The first author was supported by the Kuwait University Scholarship. This work was also supported in part by the National Science Foundation award CNS-2055549. The opinions expressed in this paper do not necessarily reflect those of the research sponsors.

\bibliographystyle{ACM-Reference-Format}
\bibliography{bibliography}


\begin{thebibliography}{57}

%%% the Web address.

\ifx \showCODEN    \undefined \def \showCODEN     #1{\unskip}     \fi
\ifx \showDOI      \undefined \def \showDOI       #1{#1}\fi
\ifx \showISBNx    \undefined \def \showISBNx     #1{\unskip}     \fi
\ifx \showISBNxiii \undefined \def \showISBNxiii  #1{\unskip}     \fi
\ifx \showISSN     \undefined \def \showISSN      #1{\unskip}     \fi
\ifx \showLCCN     \undefined \def \showLCCN      #1{\unskip}     \fi
\ifx \shownote     \undefined \def \shownote      #1{#1}          \fi
\ifx \showarticletitle \undefined \def \showarticletitle #1{#1}   \fi
\ifx \showURL      \undefined \def \showURL       {\relax}        \fi
% invisible to TeX
\providecommand\bibfield[2]{#2}
\providecommand\bibinfo[2]{#2}
\providecommand\natexlab[1]{#1}
\providecommand\showeprint[2][]{arXiv:#2}

\bibitem[Sec(2020)]%
        {SecLists}
 \bibinfo{year}{2020}\natexlab{}.
\newblock \bibinfo{title}{SecLists / Passwords /Common-Credentials}.
\newblock
\newblock
\urldef\tempurl%
\url{https://github.com/danielmiessler/SecLists/tree/master/Passwords/Common-Credentials}
\showURL{%
\tempurl}


\bibitem[azc(2023)]%
        {azcaptcha}
 \bibinfo{year}{2023}\natexlab{}.
\newblock \bibinfo{booktitle}{\emph{Auto Captcha Solver Service and Cheap
  Captcha Bypass Service Provider - AZcaptchas}}.
\newblock
\urldef\tempurl%
\url{https://azcaptcha.com/}
\showURL{%
\tempurl}


\bibitem[scr(2023)]%
        {scraperapi}
 \bibinfo{year}{2023}\natexlab{}.
\newblock \bibinfo{booktitle}{\emph{The Proxy API For Web Scraping}}.
\newblock
\urldef\tempurl%
\url{https://www.scraperapi.com/}
\showURL{%
\tempurl}


\bibitem[sel(2023)]%
        {selenium}
 \bibinfo{year}{2023}\natexlab{}.
\newblock \bibinfo{booktitle}{\emph{Selenium}}.
\newblock
\urldef\tempurl%
\url{https://www.selenium.dev/}
\showURL{%
\tempurl}


\bibitem[fak(2023)]%
        {faker}
 \bibinfo{year}{2023}\natexlab{}.
\newblock \bibinfo{booktitle}{\emph{Welcome to Faker's documentation!}}
\newblock
\urldef\tempurl%
\url{https://faker.readthedocs.io/}
\showURL{%
\tempurl}


\bibitem[Alliance(2022)]%
        {fido}
\bibfield{author}{\bibinfo{person}{FIDO (Fast IDentity~Online) Alliance}.}
  \bibinfo{year}{2022}\natexlab{}.
\newblock \bibinfo{booktitle}{\emph{FIDO Alliance study reveals global password
  usage is down - yet its continued dominance is proving costly}}.
\newblock
\urldef\tempurl%
\url{https://fidoalliance.org/barometer-2022/}
\showURL{%
\tempurl}


\bibitem[Alroomi and Li(2023)]%
        {ourpaper}
\bibfield{author}{\bibinfo{person}{Suood Alroomi} {and} \bibinfo{person}{Frank
  Li}.} \bibinfo{year}{2023}\natexlab{}.
\newblock \showarticletitle{{Measuring Website Password Creation Policies At
  Scale}}. In \bibinfo{booktitle}{\emph{ACM Conference on Computer and
  Communications Security (CCS)}}.
\newblock


\bibitem[Bonneau and Preibusch(2010)]%
        {bonneau2010password}
\bibfield{author}{\bibinfo{person}{Joseph Bonneau} {and}
  \bibinfo{person}{S{\"o}ren Preibusch}.} \bibinfo{year}{2010}\natexlab{}.
\newblock \showarticletitle{The Password Thicket: Technical and Market Failures
  in Human Authentication on the Web.}. In \bibinfo{booktitle}{\emph{Workshop
  on the Economics of Information Security (WEIS)}}.
\newblock


\bibitem[Burnett(2015)]%
        {xato}
\bibfield{author}{\bibinfo{person}{Mark Burnett}.}
  \bibinfo{year}{2015}\natexlab{}.
\newblock \bibinfo{booktitle}{\emph{Today I Am Releasing Ten Million
  Passwords}}.
\newblock
\urldef\tempurl%
\url{https://xato.net/today-i-am-releasing-ten-million-passwords-b6278bbe7495}
\showURL{%
\tempurl}


\bibitem[Burr et~al\mbox{.}(2014)]%
        {burr2004electronic}
\bibfield{author}{\bibinfo{person}{William Burr}, \bibinfo{person}{Donna
  Dodson}, {and} \bibinfo{person}{W.~Timothy Polk}.}
  \bibinfo{year}{2014}\natexlab{}.
\newblock \showarticletitle{Electronic Authentication Guidelines}.
\newblock \bibinfo{journal}{\emph{NIST Special Publication}}
  \bibinfo{volume}{800} (\bibinfo{year}{2014}), \bibinfo{pages}{63--1}.
\newblock


\bibitem[Cybersecurity and (CISA)(2009)]%
        {cert2009}
\bibfield{author}{\bibinfo{person}{Cybersecurity} {and}
  \bibinfo{person}{Infrastructure Security~Agency (CISA)}.}
  \bibinfo{year}{2009}\natexlab{}.
\newblock \showarticletitle{Security Tip (ST04-002) Choosing and Protecting
  Passwords}.
\newblock  (\bibinfo{year}{2009}).
\newblock


\bibitem[DeBlasio et~al\mbox{.}(2017)]%
        {deblasio2017tripwire}
\bibfield{author}{\bibinfo{person}{Joe DeBlasio}, \bibinfo{person}{Stefan
  Savage}, \bibinfo{person}{Geoffrey~M Voelker}, {and} \bibinfo{person}{Alex~C
  Snoeren}.} \bibinfo{year}{2017}\natexlab{}.
\newblock \showarticletitle{Tripwire: Inferring internet site compromise}. In
  \bibinfo{booktitle}{\emph{ACM Internet Measurement Conference (IMC)}}.
\newblock


\bibitem[Drakonakis et~al\mbox{.}(2020)]%
        {drakonakis2020cookie}
\bibfield{author}{\bibinfo{person}{Kostas Drakonakis}, \bibinfo{person}{Sotiris
  Ioannidis}, {and} \bibinfo{person}{Jason Polakis}.}
  \bibinfo{year}{2020}\natexlab{}.
\newblock \showarticletitle{The cookie hunter: Automated black-box auditing for
  web authentication and authorization flaws}. In \bibinfo{booktitle}{\emph{ACM
  Conference on Computer and Communications Security (CCS)}}.
\newblock


\bibitem[Flor{\^e}ncio and Herley(2010)]%
        {florencio2010security}
\bibfield{author}{\bibinfo{person}{Dinei Flor{\^e}ncio} {and}
  \bibinfo{person}{Cormac Herley}.} \bibinfo{year}{2010}\natexlab{}.
\newblock \showarticletitle{Where do security policies come from?}. In
  \bibinfo{booktitle}{\emph{Symposium on Usable Privacy and Security (SOUPS)}}.
\newblock


\bibitem[for United States Department~of Defense~(DoD)(2014)]%
        {disa2014}
\bibfield{author}{\bibinfo{person}{Defense Information Systems Agency~(DISA)
  for United States Department~of Defense~(DoD)}.}
  \bibinfo{year}{2014}\natexlab{}.
\newblock \showarticletitle{Application Security and Development Security
  Technical Implementation Guide}.
\newblock  (\bibinfo{year}{2014}).
\newblock


\bibitem[Furnell(2007)]%
        {furnell2007assessment}
\bibfield{author}{\bibinfo{person}{Steven Furnell}.}
  \bibinfo{year}{2007}\natexlab{}.
\newblock \showarticletitle{An assessment of website password practices}.
\newblock \bibinfo{journal}{\emph{Computers \& Security}} \bibinfo{volume}{26},
  \bibinfo{number}{7-8} (\bibinfo{year}{2007}), \bibinfo{pages}{445--451}.
\newblock


\bibitem[Furnell(2011)]%
        {furnell2011assessing}
\bibfield{author}{\bibinfo{person}{Steven Furnell}.}
  \bibinfo{year}{2011}\natexlab{}.
\newblock \showarticletitle{Assessing password guidance and enforcement on
  leading websites}.
\newblock \bibinfo{journal}{\emph{Computer Fraud \& Security}}
  \bibinfo{volume}{2011}, \bibinfo{number}{12} (\bibinfo{year}{2011}),
  \bibinfo{pages}{10--18}.
\newblock


\bibitem[für Sicherheit in~der Informationstechnik~(BSI)(2005)]%
        {bsi2005}
\bibfield{author}{\bibinfo{person}{Bundesamt für Sicherheit in~der
  Informationstechnik~(BSI)}.} \bibinfo{year}{2005}\natexlab{}.
\newblock \showarticletitle{BSI-Standard 100-2. IT-Grundschutz Methodology.}
\newblock  (\bibinfo{year}{2005}).
\newblock


\bibitem[für Sicherheit in~der Informationstechnik~(BSI)(2019)]%
        {bsi2019}
\bibfield{author}{\bibinfo{person}{Bundesamt für Sicherheit in~der
  Informationstechnik~(BSI)}.} \bibinfo{year}{2019}\natexlab{}.
\newblock \showarticletitle{IT-Grundschutz Compendium}.
\newblock  (\bibinfo{year}{2019}).
\newblock


\bibitem[für Sicherheit in~der Informationstechnik~(BSI)(2020)]%
        {bsi2020}
\bibfield{author}{\bibinfo{person}{Bundesamt für Sicherheit in~der
  Informationstechnik~(BSI)}.} \bibinfo{year}{2020}\natexlab{}.
\newblock \showarticletitle{IT-Grundschutz-Kompendium}.
\newblock  (\bibinfo{year}{2020}).
\newblock


\bibitem[Grassi et~al\mbox{.}(2017)]%
        {grassi2017digital}
\bibfield{author}{\bibinfo{person}{P Grassi}, \bibinfo{person}{Michael~E
  Garcia}, {and} \bibinfo{person}{James~L Fenton}.}
  \bibinfo{year}{2017}\natexlab{}.
\newblock \showarticletitle{Digital identity guidelines}.
\newblock \bibinfo{journal}{\emph{NIST Special Publication}}
  \bibinfo{volume}{800} (\bibinfo{year}{2017}), \bibinfo{pages}{63--3}.
\newblock


\bibitem[Inglesant and Sasse(2010)]%
        {inglesant2010true}
\bibfield{author}{\bibinfo{person}{Philip~G Inglesant} {and}
  \bibinfo{person}{M~Angela Sasse}.} \bibinfo{year}{2010}\natexlab{}.
\newblock \showarticletitle{The true cost of unusable password policies:
  password use in the wild}. In \bibinfo{booktitle}{\emph{SIGCHI Conference on
  Human Factors in Computing Systems (CHI)}}.
\newblock


\bibitem[Jones(1972)]%
        {jones1972statistical}
\bibfield{author}{\bibinfo{person}{Karen~Sparck Jones}.}
  \bibinfo{year}{1972}\natexlab{}.
\newblock \showarticletitle{A statistical interpretation of term specificity
  and its application in retrieval}.
\newblock \bibinfo{journal}{\emph{Journal of documentation}}
  (\bibinfo{year}{1972}).
\newblock


\bibitem[Komanduri et~al\mbox{.}(2011)]%
        {komanduri2011passwords}
\bibfield{author}{\bibinfo{person}{Saranga Komanduri}, \bibinfo{person}{Richard
  Shay}, \bibinfo{person}{Patrick~Gage Kelley}, \bibinfo{person}{Michelle~L
  Mazurek}, \bibinfo{person}{Lujo Bauer}, \bibinfo{person}{Nicolas Christin},
  \bibinfo{person}{Lorrie~Faith Cranor}, {and} \bibinfo{person}{Serge
  Egelman}.} \bibinfo{year}{2011}\natexlab{}.
\newblock \showarticletitle{Of passwords and people: measuring the effect of
  password-composition policies}. In \bibinfo{booktitle}{\emph{SIGCHI
  Conference on Human Factors in Computing Systems (CHI)}}.
\newblock


\bibitem[Kuhn and Garrison(2009)]%
        {kuhn2009survey}
\bibfield{author}{\bibinfo{person}{Bryan~Thomas Kuhn} {and}
  \bibinfo{person}{Chlotia Garrison}.} \bibinfo{year}{2009}\natexlab{}.
\newblock \showarticletitle{A survey of passwords from 2007 to 2009}. In
  \bibinfo{booktitle}{\emph{Information Security Curriculum Development
  Conference}}.
\newblock


\bibitem[Lee et~al\mbox{.}(2022)]%
        {lee2022password}
\bibfield{author}{\bibinfo{person}{Kevin Lee}, \bibinfo{person}{Sten
  Sj{\"o}berg}, {and} \bibinfo{person}{Arvind Narayanan}.}
  \bibinfo{year}{2022}\natexlab{}.
\newblock \showarticletitle{Password policies of most top websites fail to
  follow best practices}. In \bibinfo{booktitle}{\emph{Eighteenth Symposium on
  Usable Privacy and Security (SOUPS 2022)}}. \bibinfo{publisher}{USENIX
  Association}, \bibinfo{address}{Boston, MA}, \bibinfo{pages}{561--580}.
\newblock
\showISBNx{978-1-939133-30-4}
\urldef\tempurl%
\url{https://www.usenix.org/conference/soups2022/presentation/lee}
\showURL{%
\tempurl}


\bibitem[Li et~al\mbox{.}(2016a)]%
        {li2016you}
\bibfield{author}{\bibinfo{person}{Frank Li}, \bibinfo{person}{Zakir
  Durumeric}, \bibinfo{person}{Jakub Czyz}, \bibinfo{person}{Mohammad Karami},
  \bibinfo{person}{Michael Bailey}, \bibinfo{person}{Damon McCoy},
  \bibinfo{person}{Stefan Savage}, {and} \bibinfo{person}{Vern Paxson}.}
  \bibinfo{year}{2016}\natexlab{a}.
\newblock \showarticletitle{{You've Got Vulnerability: Exploring Effective
  Vulnerability Notifications}}. In \bibinfo{booktitle}{\emph{USENIX Security
  Symposium}}.
\newblock


\bibitem[Li et~al\mbox{.}(2016b)]%
        {li2016remedying}
\bibfield{author}{\bibinfo{person}{Frank Li}, \bibinfo{person}{Grant Ho},
  \bibinfo{person}{Eric Kuan}, \bibinfo{person}{Yuan Niu},
  \bibinfo{person}{Lucas Ballard}, \bibinfo{person}{Kurt Thomas},
  \bibinfo{person}{Elie Bursztein}, {and} \bibinfo{person}{Vern Paxson}.}
  \bibinfo{year}{2016}\natexlab{b}.
\newblock \showarticletitle{{Remedying Web Hijacking: Notification
  Effectiveness and Webmaster Comprehension}}. In
  \bibinfo{booktitle}{\emph{International World Wide Web Conference (WWW)}}.
\newblock


\bibitem[Mannan and Van~Oorschot(2008)]%
        {mannan2008security}
\bibfield{author}{\bibinfo{person}{Mohammad Mannan} {and}
  \bibinfo{person}{Paul~C Van~Oorschot}.} \bibinfo{year}{2008}\natexlab{}.
\newblock \showarticletitle{Security and usability: the gap in real-world
  online banking}. In \bibinfo{booktitle}{\emph{Workshop on New Security
  Paradigms}}.
\newblock


\bibitem[Mayer et~al\mbox{.}(2017)]%
        {mayer2017second}
\bibfield{author}{\bibinfo{person}{Peter Mayer}, \bibinfo{person}{Jan
  Kirchner}, {and} \bibinfo{person}{Melanie Volkamer}.}
  \bibinfo{year}{2017}\natexlab{}.
\newblock \showarticletitle{A second look at password composition policies in
  the wild: Comparing samples from 2010 and 2016}. In
  \bibinfo{booktitle}{\emph{Symposium on Usable Privacy and Security (SOUPS)}}.
\newblock


\bibitem[Mihalcea and Tarau(2004)]%
        {mihalcea2004textrank}
\bibfield{author}{\bibinfo{person}{Rada Mihalcea} {and} \bibinfo{person}{Paul
  Tarau}.} \bibinfo{year}{2004}\natexlab{}.
\newblock \showarticletitle{Textrank: Bringing order into text}. In
  \bibinfo{booktitle}{\emph{Conference on Empirical Methods in Natural Language
  Processing}}.
\newblock


\bibitem[Mitre(2023)]%
        {spraying}
\bibfield{author}{\bibinfo{person}{Mitre}.} \bibinfo{year}{2023}\natexlab{}.
\newblock \bibinfo{booktitle}{\emph{Brute Force: Password Spraying}}.
\newblock
\urldef\tempurl%
\url{https://attack.mitre.org/techniques/T1110/003/}
\showURL{%
\tempurl}


\bibitem[Motoyama et~al\mbox{.}(2010)]%
        {captcha_economics}
\bibfield{author}{\bibinfo{person}{Marti Motoyama}, \bibinfo{person}{Kirill
  Levchenko}, \bibinfo{person}{Chris Kanich}, \bibinfo{person}{Damon McCoy},
  \bibinfo{person}{Geoffrey~M. Voelker}, {and} \bibinfo{person}{Stefan
  Savage}.} \bibinfo{year}{2010}\natexlab{}.
\newblock \showarticletitle{Re: {CAPTCHAs{\textemdash}Understanding}
  {CAPTCHA-Solving} Services in an Economic Context}. In
  \bibinfo{booktitle}{\emph{USENIX Security Symposium}}.
\newblock


\bibitem[(NCSC)(2018)]%
        {NCSC2018}
\bibfield{author}{\bibinfo{person}{National Cyber Security~Centre (NCSC)}.}
  \bibinfo{year}{2018}\natexlab{}.
\newblock \showarticletitle{Password administration for system owners}.
\newblock \bibinfo{journal}{\emph{NIST Special Publication}}
  (\bibinfo{year}{2018}).
\newblock


\bibitem[Norvig(2012)]%
        {Norvig}
\bibfield{author}{\bibinfo{person}{Peter Norvig}.}
  \bibinfo{year}{2012}\natexlab{}.
\newblock \bibinfo{booktitle}{\emph{English Letter Frequency Counts: Mayzner
  Revisited or ETAOIN SRHLDCU}}.
\newblock
\urldef\tempurl%
\url{http://norvig.com/mayzner.html}
\showURL{%
\tempurl}


\bibitem[of~Standards(1985)]%
        {NIST1985}
\bibfield{author}{\bibinfo{person}{National~Bureau of Standards}.}
  \bibinfo{year}{1985}\natexlab{}.
\newblock \showarticletitle{“Password Usage” Guidelines}.
\newblock  (\bibinfo{year}{1985}).
\newblock


\bibitem[Pacuit(2019)]%
        {pacuit2019voting}
\bibfield{author}{\bibinfo{person}{Eric Pacuit}.}
  \bibinfo{year}{2019}\natexlab{}.
\newblock \showarticletitle{Voting methods}.
\newblock \bibinfo{journal}{\emph{Stanford Encyclopedia of Philosophy}}
  (\bibinfo{year}{2019}).
\newblock


\bibitem[Pochat et~al\mbox{.}(2019)]%
        {pochat2018tranco}
\bibfield{author}{\bibinfo{person}{Victor~Le Pochat}, \bibinfo{person}{Tom
  Van~Goethem}, \bibinfo{person}{Samaneh Tajalizadehkhoob},
  \bibinfo{person}{Maciej Korczy{\'n}ski}, {and} \bibinfo{person}{Wouter
  Joosen}.} \bibinfo{year}{2019}\natexlab{}.
\newblock \showarticletitle{Tranco: A research-oriented top sites ranking
  hardened against manipulation}. In \bibinfo{booktitle}{\emph{Network and
  Distributed Systems Security Symposium (NDSS)}}.
\newblock


\bibitem[Preibusch and Bonneau(2010)]%
        {preibusch2010password}
\bibfield{author}{\bibinfo{person}{S{\"o}ren Preibusch} {and}
  \bibinfo{person}{Joseph Bonneau}.} \bibinfo{year}{2010}\natexlab{}.
\newblock \showarticletitle{The password game: Negative externalities from weak
  password practices}. In \bibinfo{booktitle}{\emph{International Conference on
  Decision and Game Theory for Security}}.
\newblock


\bibitem[Press(2011)]%
        {oxford}
\bibfield{author}{\bibinfo{person}{Oxford~University Press}.}
  \bibinfo{year}{2011}\natexlab{}.
\newblock \bibinfo{booktitle}{\emph{The Oxford English Corpus: Facts about the
  language}}.
\newblock
\urldef\tempurl%
\url{https://web.archive.org/web/20111226085859/http:/oxforddictionaries.com/words/the-oec-facts-about-the-language}
\showURL{%
\tempurl}


\bibitem[Project(2023)]%
        {OWASP}
\bibfield{author}{\bibinfo{person}{The Open Web Application~Security Project}.}
  \bibinfo{year}{2023}\natexlab{}.
\newblock  (\bibinfo{year}{2023}).
\newblock
\urldef\tempurl%
\url{https://cheatsheetseries.owasp.org/}
\showURL{%
\tempurl}


\bibitem[Rose et~al\mbox{.}(2010)]%
        {rose2010automatic}
\bibfield{author}{\bibinfo{person}{Stuart Rose}, \bibinfo{person}{Dave Engel},
  \bibinfo{person}{Nick Cramer}, {and} \bibinfo{person}{Wendy Cowley}.}
  \bibinfo{year}{2010}\natexlab{}.
\newblock \showarticletitle{Automatic keyword extraction from individual
  documents}.
\newblock \bibinfo{journal}{\emph{Text mining: applications and theory}}
  \bibinfo{volume}{1} (\bibinfo{year}{2010}), \bibinfo{pages}{1--20}.
\newblock


\bibitem[Scikit-learn(2023)]%
        {Scikit}
\bibfield{author}{\bibinfo{person}{Scikit-learn}.}
  \bibinfo{year}{2023}\natexlab{}.
\newblock \bibinfo{booktitle}{\emph{sklearn.svm.SVC'}}.
\newblock
\urldef\tempurl%
\url{https://scikit-learn.org/stable/modules/generated/sklearn.svm.SVC.html}
\showURL{%
\tempurl}


\bibitem[Seitz et~al\mbox{.}(2017)]%
        {seitz2017differences}
\bibfield{author}{\bibinfo{person}{Tobias Seitz}, \bibinfo{person}{Manuel
  Hartmann}, \bibinfo{person}{Jakob Pfab}, {and} \bibinfo{person}{Samuel
  Souque}.} \bibinfo{year}{2017}\natexlab{}.
\newblock \showarticletitle{Do differences in password policies prevent
  password reuse?}. In \bibinfo{booktitle}{\emph{SIGCHI Conference Extended
  Abstracts on Human Factors in Computing Systems}}.
\newblock


\bibitem[Shay et~al\mbox{.}(2015)]%
        {shay2015spoonful}
\bibfield{author}{\bibinfo{person}{Richard Shay}, \bibinfo{person}{Lujo Bauer},
  \bibinfo{person}{Nicolas Christin}, \bibinfo{person}{Lorrie~Faith Cranor},
  \bibinfo{person}{Alain Forget}, \bibinfo{person}{Saranga Komanduri},
  \bibinfo{person}{Michelle~L Mazurek}, \bibinfo{person}{William Melicher},
  \bibinfo{person}{Sean~M Segreti}, {and} \bibinfo{person}{Blase Ur}.}
  \bibinfo{year}{2015}\natexlab{}.
\newblock \showarticletitle{A spoonful of sugar? The impact of guidance and
  feedback on password-creation behavior}. In \bibinfo{booktitle}{\emph{SIGCHI
  Conference on Human Factors in Computing Systems (CHI)}}.
\newblock


\bibitem[Shay et~al\mbox{.}(2014)]%
        {shay2014can}
\bibfield{author}{\bibinfo{person}{Richard Shay}, \bibinfo{person}{Saranga
  Komanduri}, \bibinfo{person}{Adam~L Durity}, \bibinfo{person}{Phillip Huh},
  \bibinfo{person}{Michelle~L Mazurek}, \bibinfo{person}{Sean~M Segreti},
  \bibinfo{person}{Blase Ur}, \bibinfo{person}{Lujo Bauer},
  \bibinfo{person}{Nicolas Christin}, {and} \bibinfo{person}{Lorrie~Faith
  Cranor}.} \bibinfo{year}{2014}\natexlab{}.
\newblock \showarticletitle{Can long passwords be secure and usable?}. In
  \bibinfo{booktitle}{\emph{SIGCHI Conference on Human Factors in Computing
  Systems (CHI)}}.
\newblock


\bibitem[Shay et~al\mbox{.}(2016)]%
        {shay2016designing}
\bibfield{author}{\bibinfo{person}{Richard Shay}, \bibinfo{person}{Saranga
  Komanduri}, \bibinfo{person}{Adam~L Durity}, \bibinfo{person}{Phillip Huh},
  \bibinfo{person}{Michelle~L Mazurek}, \bibinfo{person}{Sean~M Segreti},
  \bibinfo{person}{Blase Ur}, \bibinfo{person}{Lujo Bauer},
  \bibinfo{person}{Nicolas Christin}, {and} \bibinfo{person}{Lorrie~Faith
  Cranor}.} \bibinfo{year}{2016}\natexlab{}.
\newblock \showarticletitle{Designing password policies for strength and
  usability}.
\newblock \bibinfo{journal}{\emph{ACM Transactions on Information and System
  Security (TISSEC)}} \bibinfo{volume}{18}, \bibinfo{number}{4}
  (\bibinfo{year}{2016}), \bibinfo{pages}{1--34}.
\newblock


\bibitem[Shay et~al\mbox{.}(2010)]%
        {shay2010encountering}
\bibfield{author}{\bibinfo{person}{Richard Shay}, \bibinfo{person}{Saranga
  Komanduri}, \bibinfo{person}{Patrick~Gage Kelley},
  \bibinfo{person}{Pedro~Giovanni Leon}, \bibinfo{person}{Michelle~L Mazurek},
  \bibinfo{person}{Lujo Bauer}, \bibinfo{person}{Nicolas Christin}, {and}
  \bibinfo{person}{Lorrie~Faith Cranor}.} \bibinfo{year}{2010}\natexlab{}.
\newblock \showarticletitle{Encountering stronger password requirements: user
  attitudes and behaviors}. In \bibinfo{booktitle}{\emph{Symposium on Usable
  Privacy and Security (SOUPS)}}.
\newblock


\bibitem[Stock et~al\mbox{.}(2018)]%
        {stock2018didn}
\bibfield{author}{\bibinfo{person}{Ben Stock}, \bibinfo{person}{Giancarlo
  Pellegrino}, \bibinfo{person}{Frank Li}, \bibinfo{person}{Michael Backes},
  {and} \bibinfo{person}{Christian Rossow}.} \bibinfo{year}{2018}\natexlab{}.
\newblock \showarticletitle{{Didn't You Hear Me? Towards More Successful Web
  Vulnerability Notifications}}. In \bibinfo{booktitle}{\emph{Network and
  Distributed System Security Symposium (NDSS)}}.
\newblock


\bibitem[Stock et~al\mbox{.}(2016)]%
        {stock2016hey}
\bibfield{author}{\bibinfo{person}{Ben Stock}, \bibinfo{person}{Giancarlo
  Pellegrino}, \bibinfo{person}{Christian Rossow}, \bibinfo{person}{Martin
  Johns}, {and} \bibinfo{person}{Michael Backes}.}
  \bibinfo{year}{2016}\natexlab{}.
\newblock \showarticletitle{{Hey, You Have a Problem: On the Feasibility of
  Large-Scale Web Vulnerability Notification}}. In
  \bibinfo{booktitle}{\emph{USENIX Security Symposium}}.
\newblock


\bibitem[Ur et~al\mbox{.}(2016)]%
        {ur2016users}
\bibfield{author}{\bibinfo{person}{Blase Ur}, \bibinfo{person}{Jonathan Bees},
  \bibinfo{person}{Sean~M Segreti}, \bibinfo{person}{Lujo Bauer},
  \bibinfo{person}{Nicolas Christin}, {and} \bibinfo{person}{Lorrie~Faith
  Cranor}.} \bibinfo{year}{2016}\natexlab{}.
\newblock \showarticletitle{{Do Users' Perceptions of Password Security Match
  Reality?}}. In \bibinfo{booktitle}{\emph{ACM CHI Conference on Human Factors
  in Computing Systems (CHI)}}.
\newblock


\bibitem[Ur et~al\mbox{.}(2012)]%
        {ur2012does}
\bibfield{author}{\bibinfo{person}{Blase Ur}, \bibinfo{person}{Patrick~Gage
  Kelley}, \bibinfo{person}{Saranga Komanduri}, \bibinfo{person}{Joel Lee},
  \bibinfo{person}{Michael Maass}, \bibinfo{person}{Michelle~L Mazurek},
  \bibinfo{person}{Timothy Passaro}, \bibinfo{person}{Richard Shay},
  \bibinfo{person}{Timothy Vidas}, \bibinfo{person}{Lujo Bauer},
  {et~al\mbox{.}}} \bibinfo{year}{2012}\natexlab{}.
\newblock \showarticletitle{How does your password measure up? The effect of
  strength meters on password creation}. In \bibinfo{booktitle}{\emph{USENIX
  Security Symposium}}.
\newblock


\bibitem[Ur et~al\mbox{.}(2015)]%
        {ur2015added}
\bibfield{author}{\bibinfo{person}{Blase Ur}, \bibinfo{person}{Fumiko Noma},
  \bibinfo{person}{Jonathan Bees}, \bibinfo{person}{Sean~M Segreti},
  \bibinfo{person}{Richard Shay}, \bibinfo{person}{Lujo Bauer},
  \bibinfo{person}{Nicolas Christin}, {and} \bibinfo{person}{Lorrie~Faith
  Cranor}.} \bibinfo{year}{2015}\natexlab{}.
\newblock \showarticletitle{I Added '!' at the End to Make It Secure":
  Observing Password Creation in the Lab}. In
  \bibinfo{booktitle}{\emph{Symposium on Usable Privacy and Security (SOUPS)}}.
\newblock


\bibitem[Wang and Wang(2015)]%
        {wang2015emperor}
\bibfield{author}{\bibinfo{person}{Ding Wang} {and} \bibinfo{person}{Ping
  Wang}.} \bibinfo{year}{2015}\natexlab{}.
\newblock \showarticletitle{The emperor’s new password creation policies}. In
  \bibinfo{booktitle}{\emph{European Symposium on Research in Computer Security
  (ESORICS)}}.
\newblock


\bibitem[Wash et~al\mbox{.}(2016)]%
        {wash2016understanding}
\bibfield{author}{\bibinfo{person}{Rick Wash}, \bibinfo{person}{Emilee Rader},
  \bibinfo{person}{Ruthie Berman}, {and} \bibinfo{person}{Zac Wellmer}.}
  \bibinfo{year}{2016}\natexlab{}.
\newblock \showarticletitle{Understanding password choices: How frequently
  entered passwords are re-used across websites}. In
  \bibinfo{booktitle}{\emph{Symposium on Usable Privacy and Security (SOUPS)}}.
\newblock


\bibitem[Weir et~al\mbox{.}(2010)]%
        {weir2010testing}
\bibfield{author}{\bibinfo{person}{Matt Weir}, \bibinfo{person}{Sudhir
  Aggarwal}, \bibinfo{person}{Michael Collins}, {and} \bibinfo{person}{Henry
  Stern}.} \bibinfo{year}{2010}\natexlab{}.
\newblock \showarticletitle{Testing metrics for password creation policies by
  attacking large sets of revealed passwords}. In \bibinfo{booktitle}{\emph{ACM
  Conference on Computer and Communications Security (CCS)}}.
\newblock


\bibitem[Wheeler(2016)]%
        {wheeler2016zxcvbn}
\bibfield{author}{\bibinfo{person}{Daniel~Lowe Wheeler}.}
  \bibinfo{year}{2016}\natexlab{}.
\newblock \showarticletitle{zxcvbn: Low-Budget Password Strength Estimation}.
  In \bibinfo{booktitle}{\emph{USENIX Security Symposium}}.
\newblock


\end{thebibliography}

\appendix

\begin{table}[t]
\centering

\begin{tabular}{|l|c|c|}

\hline
\textbf{Paper} & \textbf{Year} & \textbf{\# Websites} \\ \hline
S. Furnell~\cite{furnell2007assessment}              & 2007                      & 10                 \\ \hline
Mannan and Van Oorschot~\cite{mannan2008security}              & 2007                      & 5                  \\ \hline
Kuhn et al.\cite{kuhn2009survey}              & 2009                      & 69                 \\ \hline

Florencio et al.~\cite{florencio2010security}              & 2010                      & 75                 \\ \hline
Bonneau et al.~\cite{bonneau2010password}              & 2010                      & 150                \\ \hline
Preibusch and Bonneau~\cite{preibusch2010password}              & 2010                      & 150                \\ \hline
S. Furnell~\cite{furnell2011assessing}              & 2011                      & 10                 \\ \hline
Wang et al.~\cite{wang2015emperor}              & 2015                      & 50                 \\ \hline
Seitz et al.~\cite{seitz2017differences}              & 2017                      & 83                 \\ \hline
Mayer et al.~\cite{mayer2017second}              & 2017                      & 137                \\ \hline
Lee et al.~\cite{lee2022password}              & 2022                      & 120                \\ \hline
%R. Dudheria~\cite{dudheria2019assessing}              & 2019                      & 50 (mobile apps)                \\ \hline

%Beno et al.~\cite{beno2020hacking}              & 2020                      & 14                 \\ \hline
\end{tabular}

\caption{Summary of prior work investigating real-world website password policies. We list the paper with its publication year and the number of websites analyzed.}
\label{tab:datasize_short}
\end{table}

\begin{table*}[]
\centering

\resizebox{18cm}{!}{
\begin{tabular}{|c|c|c|c|}

\hline
\textbf{Reference} &\textbf{Year}&
  \textbf{Country} &
  \textbf{Guidelines and Recommendations} \\ \hline

  NIST~\cite{NIST1985} &
  1985 &
  USA &
  \begin{tabular}[c]{@{}c@{}}\textbf{Low Security Level:} Min Length\textgreater{}=4, Requires Digits\\ \textbf{Medium:} Min Length\textgreater{}=4, Requires Uppercase, Lowercase, and Digits\\ \textbf{High:} Min Length\textgreater{}=6, Requires Uppercase, Lowercase, Digits, and Symbol\end{tabular} \\ \hline

NIST~\cite{burr2004electronic} &2004
&
  USA &
  \begin{tabular}[c]{@{}c@{}}\textbf{Level 1:} Min Length\textgreater{}=6, Allows Special Character \\ \textbf{Level 2:} Min Length\textgreater{}=8, Allows Special Character, Dictionary (Common) Check,  Has composition rules\end{tabular} \\ \hline
BSI~\cite{bsi2005} &2005&
  DE &
  Min Length\textgreater{}=8, Requires a digit or a symbol, dictionary (Common) Check \\ \hline
US-CERT~\cite{cert2009} &2009&
  USA &
  \begin{tabular}[c]{@{}c@{}}Min Length\textgreater{}=8, Max Length\textgreater{}=64, Dictionary (words) Check, No Personal Information, \\ Requires Uppercase, Lowercase, Digits, and Symbol\end{tabular} \\ \hline
DISA~\cite{disa2014} &2014&
  USA &
  \begin{tabular}[c]{@{}c@{}}\textbf{High Severity:} Min Length\textgreater{}=15 \\ \textbf{Medium Severity:} Requires Uppercase, Lowercase, Digits, and Symbol\end{tabular} \\ \hline
NIST~\cite{grassi2017digital} &2017&
  USA &
  \begin{tabular}[c]{@{}c@{}}\textbf{Shall:} Min Length\textgreater{}=8, Dictionary (Breach) Check, Dictionary (Words) Check, No Repetitive, No Sequential \\ \textbf{Should:} Max Length\textgreater{}=64, Accepts Space,  all Printable ASCII, Unicode including Emoji, No other Composition Rules\end{tabular} \\ \hline

NCSC~\cite{NCSC2018} &2018&
  UK &
  Dictionary (Common) Check, No Complexity Requirements, No Short Passwords* \\ \hline
BSI~\cite{bsi2019} &2019&
  DE &
  Minimal length = Sufficient*, Complexity = Sufficient* \\ \hline
BSI~\cite{bsi2020} &2020&
  DE &
  Minimal length = Sufficient*, Complexity = Sufficient*, Dictionary (Common) Check \\ \hline
OWASP~\cite{OWASP} &2022&
  %\begin{tabular}[c]{@{}c@{}}Online \\ Community\end{tabular} 
  - &
  Min length\textgreater{}=8, Max length\textgreater{}=64, Allow Unicode, Allow Space, Dictionary (Breach) Check \\ \hline

\end{tabular}
}
\caption{Summary of the password policy recommendations provided by multiple organizations. *We interpreted short length by NCSC to be less than 8 characters, sufficient length by BSI 2019 and BSI 2020 to be at least 8 characters long, and sufficient complexity to have at least 2 character classes present.}
\label{tab:guideline_descriptions}
\end{table*}

\section{Keywords Extraction Details}

%\chapter{\textbf{Keywords Extraction Details}}
\label{app:keywords}

\begin{figure}[t]
   \centering
  \includegraphics[scale=0.21]{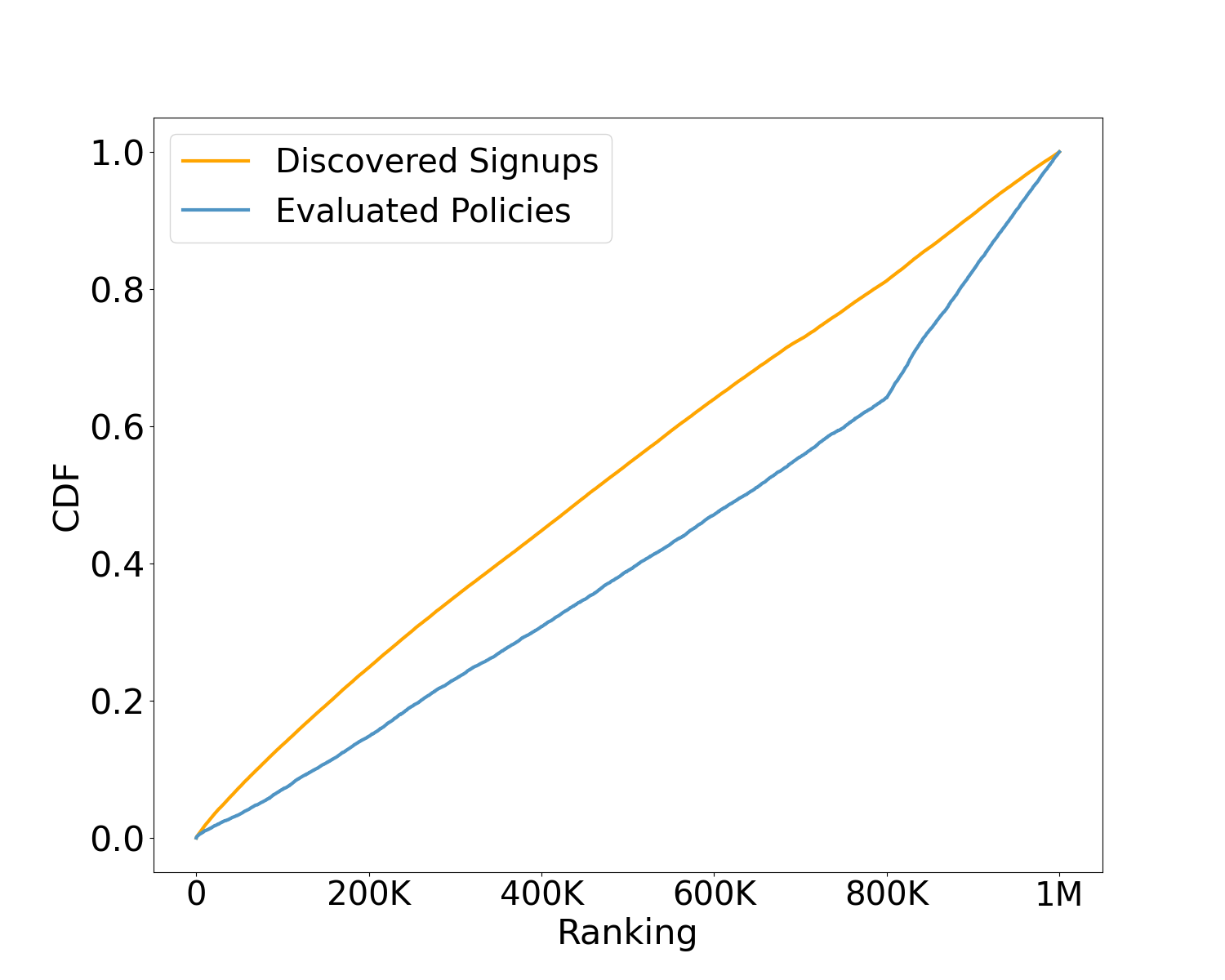}
  \caption{CDF of the rankings of sites for which our measurement method discovered signups pages (141K) and fully evaluated password policies (20K).}
 \label{fig:rankings}
\end{figure}

To identify relevant keywords for our method, we manually analyzed a random sample of domains and independently applied three common keyword ranking algorithms: Term Frequency-Inverse Document Frequency (TF-IDF) \cite{jones1972statistical}, Rapid Automatic Keyword Extraction (RAKE) \cite{rose2010automatic}, and Text Rank \cite{mihalcea2004textrank}. Since each method ranks the importance of keywords/phrases differently, we merge the rankings produced by the three methods in two ways. We applied Borda count voting~\cite{pacuit2019voting}, and calculated their Euclidean distance in a 3-dimensional space from the origin (point's axes values are its ranks across the three rankings). We then manually investigate both merged lists (which we note are typically nearly identical) and select all top-ranked words and phrases, regardless of language. Thus, our keyword selection was language-agnostic, although we note that keywords were primarily English due to the prevalence of English on our sampled sites as well as in HTML webpages (even for non-English sites).

\section{Password Inference Details}
\label{Appendix:PasswordInferenceDetails}
\changes{Here, we expand on the details of our password inference algorithm, as discussed in Section~\ref{sec:inference_alg_method}.}
%detailed description of the logic behind our password inference algorithm with examples of the pseudocode. % The full logic presented in~\ref{Pseudocode} %It begins with a special section, which explains the notions of \emph{Safe Bets} and \emph{Safe Sets}. Then a section for each of the four main steps of the algorithm.

\textbf{Step 1. Finding an Admissible Password and $R_{no\_a\_sps}$:}
As discussed in Step 1 of Section~\ref{sec:inference_alg_method}, we first find one admissible password by attempting passwords of different lengths and configurations (listed in Table~\ref{tab:safesets}).

\changes{
\emph{Correctness.} Here, we prove that \emph{one} of our tested passwords must be admissible given the policy parameters investigated and the assumed bounds. }

\changes{Given our conservative assumed bounds ($L_{min} \in [0,32]$ and $L_{max} \in [6, 128]$), a password with $l \in [6, 32]$ must be admissible. We test passwords across this length range, one of which will necessarily satisfy the length requirements. For an acceptable length, our tested passwords must also cover any configuration of the restrictive requirements. %\frank{What order of lengths do we do for admissible passwords?} \roomi{We relied on previously reported length range limits to optimize our order~\cite{furnell2007assessment,mannan2008security,furnell2011assessing,wang2015emperor,seitz2017differences,dudheria2019assessing}. Specifically, we tried 16, 12, 32, 10, 8, and 20, and then the rest were filled from highest to lowest. (we note that this only affects the average performance but not the worst case).}
}

\begin{itemize}[nosep, leftmargin=*]

\item \changes{$LOW_{min}$, $UPP_{min}$, $DIG_{min}$, $SPS_{min}$: For each length, there is at least one password that satisfies any possible combination of $LOW_{min}$, $UPP_{min}$, $DIG_{min}$, and $SPS_{min}$ values (see Table~\ref{tab:safesets}).
Note that if a password has $N$ characters of a class, it inherently satisfies any policy that requires $N$ or fewer characters of that class. For lengths 6 and 7, the number of safe set passwords is larger than for other lengths because a single password cannot satisfy all minimum class character requirements simultaneously.} 

\item \changes{$R_{cmb23}$, $R_{cmb24}$, and $R_{cmb34}$: The evaluated passwords satisfy all class combination requirements by containing at least three character classes.}

\item \changes{$R_{2word}$: We assumed that if $R_{2word}=True$, $L_{max} \geq 10$ (see Section~\ref{sec:inference_alg_method}). All 
our evaluated passwords of length 10 or greater contain a two-words structure.}

\item \changes{$R_{no\_a\_sps}$: For each length, we test a password that does not contain any special characters, and thus would satisfy this restriction. Note that if this parameter is true, implicitly $SPS_{min}=0$, and all tested passwords also satisfy $LOW_{min}=UPP_{min}=DIG_{min}=2$.}

\item \changes{$R_{lstart}$: All evaluated passwords begin with a letter.}

\end{itemize}

\changes{Note that as tested passwords longer than length 10 are padded versions of length 10 passwords, which already satisfies any restrictive requirements, these longer passwords also remain in adherence to restrictive requirements. Thus, for any acceptable length, there is at least one password that satisfies all restrictive parameters. Permissive parameters are simultaneously satisfied as none of these passwords contain dictionary words, sequential/repeated characters, only digits, personal identifiers, certain sequences and characters, or are breached passwords. Thus, one of the safe set passwords must satisfy all policy constraints and is admissible.}

\changes{\emph{Efficiency:} As seen in Table~\ref{tab:safesets} (and with two safe set passwords for lengths 11--32), the total number of passwords is 65. Thus, signup attempts range from 1--65, until one admissible password is found (exactly one signup success).}

%%%%%%%%%%%%%%%%

\textbf{Step 2: Restrictive Parameters.}
Given an admissible password, we next determine each restrictive parameter's value by modifying the admissible password to violate just that parameter's requirements and observing whether the password is still accepted (while still satisfying the other restrictions previously satisfied by the original admissible password).
%Determining restrictive parameters relies on violating a particular parameter (or lowering its value) and checking if the new password with the modified parameter is still admissible. The modification is carefully constructed not to violate more than the tested parameter. 
%
Note, $R_{no\_a\_sps}$ was already determined in Step 1.
We determine the restrictive parameters as described in Step 2 of Section~\ref{sec:inference_alg_method}. Here, we discuss the correctness and efficiency of our evaluation for each restrictive parameter.

$R_{2word}$: We tested this parameter by violating the two-word structure in the admissible password, if present.
%If the admissible password is less than length 10,  $R_{2word}=False$ as the admissible password will not contain a two-word structure (see Table~\ref{tab:safesets}). Otherwise, the admissible password always contains the two-word structure.
%o determine $R_{2word}$ in this case, we modify the admissible password to violate the two-word structure, by moving the digit delimiter to the end of the two-word structure (e.g., ``\textbf{MxT7zcS}4t1'' is modified to ``\textbf{MxTzcS7}4t1''). If this new password is accepted, $R_{2word}=False$, otherwise $R_{2word}=True$. 

\begin{itemize}[nosep, leftmargin=*]

\item \emph{Correctness:} This move does not affect the admissible password's length, other restrictive parameters (characters do not change and the first character is not affected, as related to $R_{lstart}$), nor any permissive parameters (as the new six-letter string does not introduce a repeated/sequential/word/username substring). Thus, only $R_{2word}$ is affected, and whether the modified password is accepted reveals its value.

\item \emph{Efficiency:} Up to one password/signup is attempted, which may not succeed (0 or 1 signup success).

\end{itemize}

$R_{lstart}$: This parameter is evaluated by violating the letter-start structure of the admissible password.

%To determine the value of $R_{lstart}$, we first utilize \textsc{ConstructPassword} (discussed in the next step) to create an admissible password of length $L_{max}$. The first non-letter within the password is then moved to the beginning, and the password is attempted. If non-exist, the last character (least priority) of the password is removed, the password is checked if it still satisfies all the restrictive and length parameters (otherwise, return False), and if so, a number (zero) is added at the beginning of the password to test $R_{lstart}$. In cases of $R_{2word}$ being $True$, we preserve the first 7 letter substring to keep the 2-word structure in the password.

\begin{itemize}[nosep, leftmargin=*]
\item \emph{Correctness:} The modified password does not change the password's length or set of characters. It also preserves existing two-word structures, if true, and thus does not affect other restrictive parameters. Permissive requirements are also not affected as the reordering does not introduce potentially disallowed strings. Thus, only the $R_{lstart}$ restrictive parameter is affected.

\item \emph{Efficiency:} One password is tested, potentially resulting in one successful signup.

\end{itemize}

%as we either move the position of a non-letter or replace the last character with another of different class. The procedure of moving a non-letter or adding one at the end only deals with the substring from position 8 to Lmax when $R_{2word}$ is $True$ to preserve the two-word structure. Further, we only introduce a number as a non-letter to avoid violating $R_{no\_a\_sps}$. When removing the last character and replacing it with a number, we count the classes present in the modified password to ensure it still satisfies the minimum requirement and the $R_{cmb34}$, $R_{cmb23}$, $R_{cmb24}$ if they are true. Also, the modified password does not introduce a sequential, repetitive, or dictionary word to the password.

$LOW_{min}$, $UPP_{min}$, $DIG_{min}$, and $SPS_{min}$: 
We determine the class minimums by modifying the password to have fewer instances of a class, and testing for acceptance.

\begin{itemize}[nosep, leftmargin=*]
\item \emph{Correctness:} Our test password has the same length as the admissible one, so the length parameters are not violated. As the admissible password already satisfies the minimum class count requirements, changing characters from class $C$ to another only affects the restrictive requirement about $C$, and not the other classes. Our use of special characters also remains adherent to $R_{no\_a\_sps}$. % (as we only add to the number of characters of the other classes, and only replace with special characters if allowed by $R_{no\_a\_sps}$).
Our method for replacing class $C$ (Section~\ref{sec:inference_alg_method}) also preserves class combination requirements by ensuring that the modified password always still has three classes, and positional requirements remain satisfied (respecting $R_{lstart}$ and $R_{2word}$). Any permissive parameters remain satisfied, as the modified password still does not contain a repeated/sequential/word/username substring, only digits, or certain characters, and is not a breached password. Thus, only the specific class minimum is affected.

\item \emph{Efficiency:} Each class requires up to two modified passwords attempted, with at most one signup success. Across four classes, this results in up to 8 attempts and 4 successes.

\end{itemize}

$R_{cmb34}$, $R_{cmb23}$, $R_{cmb24}$: 
We determine class combination requirements by modifying the admissible password to have fewer classes, testing for acceptance.

\begin{itemize}[nosep, leftmargin=*]

\item \emph{Correctness:} Evaluated passwords are the same length as the admissible one, the length parameters are not affected. Only non-required classes ($C_{min} = 0$) are considered for replacement when reducing class combinations, maintaining the class minimum requirements. We also only use special characters if required so  $R_{no\_a\_sps}$ is not violated. If $R_{lstart}=True$, then a letter class is required, and all evaluated passwords start with a letter in such cases. If $R_{2word}=True$, a letter class and a non-letter class are already required and a two-word structure already exists in the admissible password. Our replacement strategy replaces letters of one class with letters of another class (likewise between digits and special characters), so $R_{2word}$ is preserve. Our modified passwords do not contain repeated/sequential/word/username substrings, relevant permissive characters, and is not a top breached password. It is all-digit only if digits are required when evaluating $R_{cmb24}$. If the policy accepts this all-digit password, $R_{cmb24}=False$ regardless of $P_{shortd}$ and $P_{longd}$. Thus, permissive parameters remain unaffected, and the class combination parameters are evaluated in isolation.

\item \emph{Efficiency:} Evaluating each of the three class combination parameters requires one password tested, which may result in a successful signup. Together, this evaluation may make up to three signup attempts and up to three successes.

\end{itemize}

\textbf{Step 3: Length Parameters.}
We determine the password length ranges using binary search while testing passwords of varying lengths.
%
%\IEEEchange{We utilize binary search to identify the bounds of $L_{min} \in [0,32]$ and $L_{max} \in [6, 128]$. Effectively, we construct and attempt passwords of ascending and descending lengths to conclude these bounds. For example, to evaluate $L_{max}$ with an admissible password of length 8, we first construct and test a password of length 68 (halfway point), the status of the attempt then guide us to either test the upper portion of the range ($[69-128]$) or the lower one ($[9-68]$). Similarly done but in reverse direction to find the value of $L_{min}$}.
%
%
For each iterations of the search, we generate a password of a specified length $l$ that satisfies all determined restrictive requirements. Our password generation method (\emph{ConstructPassword}) works as follows:

\begin{enumerate}[nosep, leftmargin=*]
    \item \changes{If $R_{lstart}=True$, we place a letter as the first character (choosing case based on the class requirements, arbitrarily selecting one if neither differ in requirements).}
    %whether $LOW_{min}>0$ or $UPP_{min}>0$, 
    
    \item If  $R_{2word}=True$ (which only occurs for $l \ge 10$), we add the two-word structure. Here, if $R_{lstart}=True$, we use the first character, a letter, as the start of the first word in the two-word structure. 
    %If $R_{lstart}=False$, we start the password with a full two-word structure. 
    During two-word construction, for each letter, we use a case that remains needed in the password based on lowercase and uppercase requirements. Similarly for the delimiter for digit and special symbol requirements, and also the value of $R_{no\_a\_sps}$, selecting a digit if the requirements do not differ.
    %$LOW_{min}$ and $UPP_{min}$. 
    
    %whether $DIG_{min}>0$, $SPS_{min}>0$, and $R_{no\_a\_sps}$, selecting a digit if the requirements do not differ.
    
    \item \changes{The class minimums provide a set of  characters that must be in the password (with counts reduced by the characters already placed in the password due to $R_{lstart}$ and $R_{2word}$), which we then append to the password in arbitrary order..}
    
    \item \changes{ If there is a class combination restriction that is not yet satisfied, we append a character of each missing class, using special symbols only if allowed by $R_{no\_a\_sps}$. }
    
    \item \changes{Our password is now as short as possible while satisfying all restrictions. If the length exceeds $l$, we deem $l$ as infeasible. If it is shorter than $l$, then we pad the password to length $l$ with arbitrary alternating letters and digits.}
\end{enumerate}

\changes{
\emph{Correctness:} When we construct an initial password that satisfies all restrictive requirements, we do so with the shortest password possible. A two-word structure requires adding 7 characters in all cases. If a letter-start is required, then we overlap this with the two-word structure to avoid adding an additional character. If only the letter start is required, but not the two-word structure, then one initial letter is required. During starting letter and two-word structure selection, we draw down the required characters as much as possible. The remaining required characters (first from class minimums and then from class combinations) must be added somewhere in the password in all cases (and special symbols are only added if required, keeping the password adherent to $R_{no\_a\_sps}$). If this password already exceeds $l$, there cannot be a shorter password that satisfies the restrictive parameters. Otherwise, the password can be padded to length $l$ while remaining adherent to the restrictions, as the addition of only letters and digits does not affect any of the already-satisfied restrictive parameters. The $l$-length password also does not affect the permissive parameters as it does not contain a repeated/sequential/dictionary word/username substring, and is not an all-digit string or a top breached password. Thus, if a policy accepts passwords of length $l$, the selected password must be admissible under the policy constraints. Binary search then allows us to determine which range of password lengths are accepted.
}

\changes{
\emph{Efficiency:} Binary search for $L_{min}$ requires $log_2(32)=5$ signup attempts and potential successes. $L_{max}$ requires $log_2(128)=7$ signup attempts and potential successes.
}

\textbf{Step 4: Permissive Parameters.}
To evaluate each permissive parameter, we take an admissible password and modify it as related to the parameter, and test whether that parameter is accepted.
%\roomi{Each of the parameters is handled by a sub-procedure that makes a single attempt with specifically constructed for the case password. All such passwords use the greatest number of characters possible ($L_{max}$), so that there are enough characters in the string to construct and/or adjust specific configurations.}
$P_{longd}$, $P_{shortd}$: We construct and test all-digit passwords of length $L_{min}$ and $L_{max}$.

\begin{itemize}[nosep, leftmargin=*]

\item  \emph{Correctness:} All-digit passwords satisfy length constraints, and are not constructed with repeated/sequential substrings and are not breached passwords, so are not affected by other permissive parameters. Restrictive parameters are ignored here to understand if a policy has special exceptions for all-digit passwords. 
%Checking for digit-only passwords (parameters $P_{longd}$ and $P_{shortd}$) is done by constructing strings\ that are made only of digits. An all-digit string with length $L_{max}$ is used for $P_{longd}$, while one with length $L_{min}$ is used for $P_{shortd}$. For both, we construct a digit pattern that is non-repetitive and non-sequential. Here we we are not concerned with satisfying restrictive parameters as we are specifically looking for the permission of all digits.

\item \emph{Efficiency:} Each parameter requires one signup attempt and potential success.

\end{itemize}

$P_{rep}$, $P_{seq}$, $P_{dict}$, $P_{id}$: To evaluate these parameters, we construct and test a password with the evaluated substring/sequence.
For $P_{rep}$, the sequence is three repeating consecutive characters of the same class (i.e., ``AAA''). For $P_{seq}$, the sequence is either ``abc'', ``ABC'', or ``123''. For $P_{dict}$, the sequence is the most popular dictionary word (according to the Oxford English Corpus~\cite{oxford}) of a substring length that fits within the password, ranging from 3 to 8 letters (specifically \{"one", "time", "world", "number", "company", "sentence"\}). For $P_{id}$, the usernames we choose are 3-letter names followed by 5 random digits, and the sequence is the 3-letter portion of the username.

\begin{itemize}[nosep, leftmargin=*]

\item \emph{Correctness:} Before inserting the evaluated sequence, we start with the shortest partial password that satisfies all restrictive requirements, created using \emph{ConstructPassword} (whose correctness was demonstrated in Step 3). 
If $L_{max}$ allows the evaluated sequence to be appended to this partial password, then the augmented password remains adherent to all restrictive and length requirements (letters and digits are added).
For the case where appending the evaluated sequence is not possible due to length constraints, we only replace characters in the partial password with different characters of the same class. If adding new characters are required (i.e., there are fewer than three characters of a class to form a sequence), we create a sequence using the most frequent character class in the partial password, to minimize added characters.
We handle positional constraints specially, as discussed in Section~\ref{sec:inference_alg_method}, where again characters are replaced only with those of the same class (and no characters are added).
Thus, the restrictive requirements remain satisfied, and the constructed password is within the allowed length range (if feasible).

%When $R_{2word}=True$, the constructed password is identical to Step 3's except the padded characters follow the second word in the two-word structure, rather than at the password end, which does not affect any restrictive requirements. When $R_{2word}=False$, the character order is irrelevant (except when $R_{lstart}=True$, which we satisfy by using a letter as a first character), so the restrictive requirements remain satisfied with the character reordering and padding.
Beyond the evaluated sequence (e.g., repeated characters), the constructed password does not contain substrings for the other permissive parameters (e.g., sequential/word/username sequences), and is not a breached or all-digit password. Thus, other permissive parameters are not affected.

\item \emph{Efficiency:} Each parameter may require one password attempted, potentially resulting in a signup success.

\end{itemize}

\changes{
$P_{space}$, $P_{unicd}$, $P_{emoji}$, $P_{spn1}-P_{spn4}$: To evaluate these permissive parameters, we construct and test a password that includes the associated character. %For evaluating whether a certain character type $T$ is permitted in the password, we use the same \emph{ConstructPassword($L_{max}$, $R$)} as in Step 3 to generate an $L_{max}$-length admissible password. We replace the last character of the admissible password with the character of class $T$. %\IEEEchange{(as shown in Line 2-3 of DETERMINEPEMOJI)}. 
%The one exception is for $P_{space}$, where instead we remove the last character of the admissible password and place the whitespace character in the middle of the password (where if $R_{2word}=True$, we ensure that the space is after the two-word structure). If this modified password (with the character of type $T$) is accepted, the permissive parameter for type $T$ is true.
}

%For parameter $P_{space}$, we construct a password of length Lmax (using construct password) and then remove the last character (least priority). The new password is then checked for compliance to the restrictive parameters and then a space is added to the middle index and attempted. For $R_{2word}$ the 2word substring is preserved when possible and space is added afterwards, otherwise we replace "MxT7zcS" with "MxT7 zcS". Parameters $P_{unicd}$, $P_{emoji}$, and $P_{spn1}$ - $P_{spn4}$ are handled in a similar fashion: Replace one optional character from the password (usually if possible, the one at the end of it) with a character from the group, which is featured by the parameter. E.g. for $P_{unicd}$ the new character is a letter from a non-Latin (or non-English) alphabet, for $P_{emoji}$ the new character is an emoji, and for any of $P_{spn1}$ to $P_{spn4}$ the new character is the special symbol which is represented by the current parameter.  

\begin{itemize}[nosep, leftmargin=*]
\item \emph{Correctness:} We start with an $L_{max}$-length password that satisfies all restrictive and permissive requirements using \emph{ConstructPassword} (correctness in Step 3). An optional character (i.e., one not involved with satisfying any restrictive parameters) is then replaced with the evaluated character. If there is no optional character, the parameter's value is inherently false. Otherwise, the modified password remains adherent to all length, restrictive, and permissive parameters (besides the permissive parameter). % If not, then the parameter value is inherently false. For, $P_{space}$, the whitespace character is added in the middle of the password and is after the second word if $R_{2word}=True$.

%We find one optional character that is not required by any restrictive requirements, including positional constraints, and replace it 
%\changes{
%Recall that the order of stages in \emph{ConstructPassword} ensures that characters required by the restrictive parameters are placed at the beginning of the password, and optional characters (if any) are placed at the end. Thus, the last character must be optional under the restrictive requirements, if any characters are optional, and can be replaced/removed without violating restrictions (if not, then the restrictive parameters do not permit any additional characters of type $T$ anyways). In the special case of $P_{space}$, we ensure that the placement of the whitespace in the middle of the password does not affect any restrictive parameters (namely the two-word structure, if required). Thus, no restrictive parameters are affected.

%\changes{
%Finally, the replacement of one optional character with a character of type $T$ will not introduce sequential/repeated substrings, dictionary words, usernames, other character types, or make the password a breached password. Thus no other permissive parameters are affected, and only the parameter for character type $T$ is evaluated.
%
\item \emph{Efficiency:} Evaluating each character type may require one password attempt, with a potentially successful signup.

\end{itemize}

\changes{
$P_{br}$: Given all other policy parameters have now been determined, we test the highest-ranked breached password that is admissible under all other constraints, where ranking is based on a popular breached password dataset~\cite{xato}. }

\begin{itemize}[nosep, leftmargin=*]
\item \emph{Correctness:} The breached password selected satisfies all other constraints and should be admissible. We select the highest-ranked password to increase the likelihood it is blocked if the website does disallow breached passwords.
%is the last parameter tested as it considers the values of all policy parameters evaluated. Here we search a list of the most common breach passwords~\cite{xato} for the highest ranked password the compiles with the inferred policy (e.g. it satisfies the policy and should be accepted), when such a password is found, an attempt is made with it to determine the value for $P_{br}$. \roomi{If no breach password found that satisfy the policy, we assume the parameter is false by default}. 

\item \emph{Efficiency:} One password is attempted, with a potentially successful signup.
\end{itemize}

\changes{
\textbf{Step 5: Sanity Check.}
As our last step, we attempt one account signup using a password that must violate the inferred policy, based on the parameters inferred. If this password seems to be accepted, we assume an error when evaluating this site and can subsequently manually investigate or filter such sites.
\begin{itemize}[nosep, leftmargin=*]
\item \emph{Efficiency:} One password is attempted, which should not lead to a successful signup (but could).
\end{itemize}
}

\subsection{\textbf{Policy Evaluation Example}}
\label{app:example}
Here, we describe a full example execution of our policy inference algorithm from Section~\ref{sec:PasswordPolicyInferenceAlgorithm}, to infer an example policy. Considering our policy parameters (from Section~\ref{sec:parameters}), the example policy is:
%\emph{complex 8} password policy often discussed in the password literature~\cite{shay2014can, shay2016designing, ur2012does, seitz2017differences}. This policy largely derived from older NIST guidelines that encourage passwords to include different classes. Considering our policy parameters (from Section~\ref{sec:parameters}), the complex 8 policy precisely is:

\begin{itemize}[nosep, leftmargin=*]

\item  Length: $L_{min}=8$; $L_{max}=20$ 

\item Restrictive: $DIG_{min}=UPP_{min}=LOW_{min}=SPS_{min} = 1$; $R_{cmb23}$, $R_{cmb24}$, $R_{cmb34}$, $R_{lstart}$ are all true (other restrictive parameters are false)

\item Permissive: $P_{dict}$, $P_{seq}$, $P_{rep}$, $P_{id}$, $P_{sp1}$-$P_{sp4}$, $P_{br}$) all are true (other permissive parameters are false)

\end{itemize}

The first step is finding an admissible password. We start attempting the safe set passwords %(M-7S4@ Mx-cS@ x-7c4@ M-7c4@ Mx-7c@ Mx7c4@ M-7cS@ M-7cS4 Mx7cS@ Mx7-cS4 Mx-cS4@, M7-cS4@ Mx7-c4@) 
(see Table~\ref{tab:safesets}) in increasing length order, starting with length 6 passwords. These will all fail until we reach length 8's safe set passwords. The next attempt is the length 8 password Mx7-cS4@, which will be accepted as it satisfies all the restrictive and length requirements without violating the permissive ones. $R_{no\_a\_sps}$ is evaluated as false since the accepted password includes a special character. Also, since the accepted password length is less than 10, we evaluate $R_{2word}$ to be false. 

Next, we modify the admissible password by moving a non-letter character to its start, attempting the password 7Mx-cS4@. This will be rejected, so $R_{lstart}$ is true. 

For character class minimums, we first evaluate $LOW_{min}$. We attempt MX7-CS4@, which fails (does not have a lowercase character). Then we test Mx7-CS4@, which succeeds, making $LOW_{min}=1$.

We do similarly for the other character classes, testing the following pairs: (mx7-cs4@, Mx7-cs4@), (Mx\&-cS\&@, Mx7-cSw\@), and (Mx75cS41, Mx7-cS41), for $UPP$, $DIG$ and $SPS$, respectively. The first password in each pair will be rejected (no character of the tested class is present) and the second will succeed (one character of the class is present). Thus, $UPP_{min}$, $DIG_{min}$, and $SPS_{min}$ are all 1. This directly implies that $R_{cmb23}=R_{cmb24}=R_{cmb34}=True$, without needing explicit evaluation. 

For length evaluation, our admissible password search already indicates that $L_{min}=8$. For $L_{max}$, we do binary search, first constructing a password of length 68 (the halfway length between 8 and 128), which starts with Mx7- (the shortest password satisfying all restrictive parameters), and then we pad the rest with arbitrary letters and digits up to the length required (e.g., Mx7-k6d1e2o9k0w5f7b....). This password will be rejected, so we continue binary search until determining that $L_{max}= 20$. % (e.g., the attempted password of length 20 is Mx7-k6d1e2o9k0w5f7b6). We then ascend from that point to test and fail passwords of lengths 29,24,22, and 21 to find our $L_{max}$ to be 20. Similarly done in reverse fashion to find $L_{min}$. We attempt and fail passwords of lengths 5 (halfway point),6 and 7 (e.g., a password of length 5 would be Mx7-w) to find our $L_{min}=8$. 

For the permissive parameters, we construct a password of length 20 (maximum length), which starts with Mx7- padded by arbitrary letters and digits (e.g., Mx7-k6d1e2o9k0w5f7b6). The password has 16 non-required characters (the padded letters/numbers), which can be used to include the permissive character(s) to test. This can either be the last character (e.g., $P_{emoji}$), the middle character ($P_{space}$), the last three characters ($P_{rep}$, $P_{seq}$), or the last 3-8 characters ($P_{dict}$). For example, to test for $P_{rep}$, we  modify the length 20 password to be (Mx7-k6d1e2o9k0w5faaa), which will be accepted, revealing $P_{rep}=True$. For $P_{space}$, we include a space in the middle (e.g., Mx7-k6d1e2 o9k0w5f7b), which will be rejected, revealing $P_{space}=False$. Finally, the highest-ranked common password from our breach dataset~\cite{xato} satisfying all other policy parameters is P@ssw0rd. %, ranked 15,584 in~\cite{xato}.
That password is accepted, making $P_{br}=True$.

\section{Measurement Performance}

\subsection{Error Cases}
\label{appendix:Challenges}

\begin{figure}[t]
   \centering
  \includegraphics[scale=0.2]{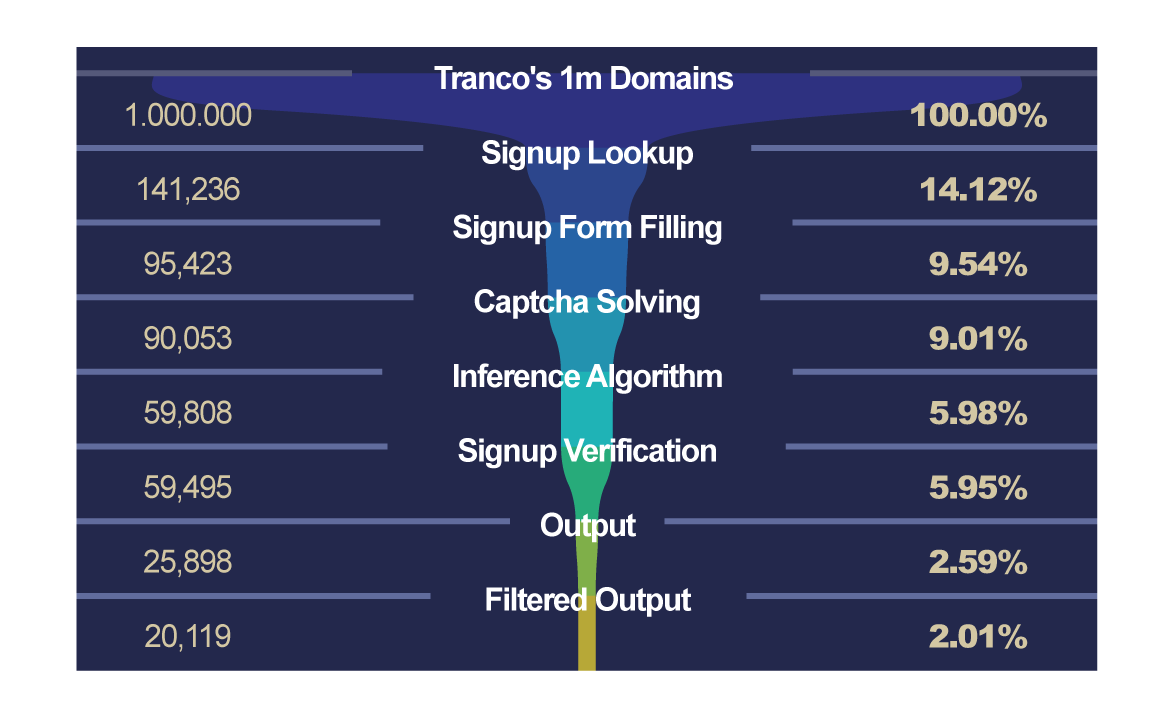}
  \caption{Funnel chart of the one million domains as they flow through the main stages of the framework}
 \label{fig:funnel}
\end{figure}

Here, we list the biggest challenges we found in the different stages of our measurement. Figure~\ref{fig:funnel} shows the funnel diagram of our site population as we proceed through the stages of our method. 
%
\iffalse
The following are the main errors we faced:

Signup Lookup:
\begin{itemize}
  \item Unresponsive websites.
  \item Unidentified keywords in the signup URL or form.
  \item Signup pages deep within a site (i.e., more than 2 link hops away from the landing page).
  \item Signup pages not in our Google search query results.
  \item Complex signup workflow (e.g., multi-page, required user actions)
   \item Domains with registration fees or requiring offline membership.
\end{itemize}

Form Filling:
\begin{itemize}
  \item Inaccessible forms (e.g., requiring user interaction).
  \item Incorrect field data submitted.
  \item  Timeouts.
   \item Locating form submission button.
   \item  Accessing form submission button (e.g., requiring user interaction).
\end{itemize}

CAPTCHA Solving:
\begin{itemize}
  \item Unrecognized CAPTCHA element. 
    \item Unrecognized CAPTCHA types.
    \item Error from the CAPTCHA solver.
    \item Incorrect solutions.
\end{itemize}

\fi
%
The Signup Lookup stage experienced a significant drop in the number of domains. This also observed in our manual analysis of domains, where only 26\% had signup URLs (thus much of the drop is expected). Domains where we did not find signup pages include:

\begin{itemize}[nosep, leftmargin=*]
  \item Unresponsive domains
  \item Unidentified keywords in the signup URL or form.
  \item Signup pages that were deep within websites (multiple hops away from the landing page) or not in Google search query results
  \item Domains with complex signup workflow (e.g., multi-page signup processes)
   \item Domains with registration fees or requiring offline membership
\end{itemize}

\changes{In the Form Filling stage, a portion of sites with signup pages are no longer evaluated due to form filling errors, which include inaccessible forms (e.g., those requiring user interaction), incorrect field data being provided, timeouts, and anti-bot defenses.} %We noticed that domains of lower ranks are highly volatile as their status and response time vary considerably, even with several retries. 
%Some domains intentionally implement invisible fields as a mechanism to detect bots when those fields are auto-filled. We attempted to avoid filling such fields by only filling out fields we could classify and generate .
%
When solving CAPTCHAs, we noted that their appearance was inconsistent and often partway through evaluating a site (likely as an anti-bot mechanism). Using AZcaptcha, we succeeded in solving the majority of the CAPTCHAs encountered throughout our measurements (94\%). When AZCaptcha failed, we investigated the errors and identified that most arose due to unrecognized CAPTCHA types (often custom CAPTCHA implementations),  unrecognized CAPTCHA fields, or generating incorrect solutions. %The average AZCaptcha execution is about 75 seconds, with reCAPTCHA and hCaptcha taking on average 95 seconds. Thus, CAPTCHA-solving adds a significant overhead.
Other issues during this stage included diverse modes of interacting with the form submission button (some of which were not enabled by default, changing state in response to certain user actions). We also found other HTML buttons other than the submission button in some forms, requiring distinguishing between button functionalities.

Our inference algorithm's correctness was presented in Appendix~\ref{Appendix:PasswordInferenceDetails}. However, our inference depends on accurate signup verification. While our trained signup verification classifier exhibits high accuracy (as discussed in Section~\ref{sec:SignupSuccessVerification}), it can provide false positive or negative classifications. However, these classifications are typically consistent on a given site across multiple signup attempts; thus we can detect and filter out sites with these errors as they exhibit uniform signup success or failure across all attempts, including the sanity check attempt (infeasible for a realistic password policy).

\subsection{Performance Comparison to Prior Work}
\label{appendix:comparisontocookiehunter}
The prior large-scale automated account creation method~\cite{drakonakis2020cookie} attempted to create accounts on 1.5M domains, successfully creating one account for analysis on 25K domains (a 1.6\% success rate). In comparison, our work fully analyzed 20K domains across only 1M domains (a 2\% success rate, 20\% higher than~\cite{drakonakis2020cookie}). 
Successfully evaluating a site in our context is a significantly more challenging task though, as we require multiple successful account creations, whereas the prior work only required one. We observed that our method was successful at creating at least one account on 30.3\% of websites with signup pages (4.3\% of all sites), more than twice the rate of the prior work. Furthermore, our approach exhibits higher rates for website signup discovery (on 14.1\% of sites vs. 9\%). Thus, while our method shares similarities with~\cite{drakonakis2020cookie}, our ground-up design and implementation yielded a larger population to analyze.

%%%%%%% REMOVING FIGURE %%%%%%%%%%%

\subsection{Result Generalizability}
\label{Generalizability}
\begin{table}[t]
\centering

\begin{tabular}{|c|c|c|}
\hline
\textbf{}                       & \textbf{\begin{tabular}[c]{@{}c@{}}Manually \\ Analyzed\end{tabular}} & \textbf{\begin{tabular}[c]{@{}c@{}}Our \\ Method\end{tabular}} \\ \hline
\textbf{Restrictive Parameters} &                        &                               \\ \hline
Requires 2 Words                & 2                      & 2.4                           \\ \hline
Requires No Arbitrary Special   & 1                      & 1.8                           \\ \hline
Any 3 of 4 Classes              & 12                     & 7.4                           \\ \hline
Any 2 of 4 Classes              & 15                     & 9.1                           \\ \hline
Any 2 of 3 General Classes      & 14                     & 9.3                           \\ \hline
Starting With a Letter          & 5                      & 2.9                           \\ \hline
\textbf{Permissive Parameters}  &                        &                               \\ \hline
Dictionary Words                & 76                     & 72.0                            \\ \hline
Sequential Characters           & 78                     & 71.7                          \\ \hline
Repeated Characters             & 77                     & 71.1                          \\ \hline
Short Digit-only                & 80                     & 78.0                            \\ \hline
Long Digit-Only                 & 72                     & 78.2                          \\ \hline
Personal Identifier             & 76                     & 71.4                          \\ \hline
Space                           & 77                     & 69.0                            \\ \hline
Unicode                         & 82                     & 67.7                          \\ \hline
Emoji                           & 72                     & 64.4                          \\ \hline
Breach Password                 & 93                     & 88.2                          \\ \hline
1st Popular Special = .         & 80                     & 70.0                            \\ \hline
2nd Popular Special = !         & 81                     & 69.6                          \\ \hline
3rd Popular Special = \_        & 80                     & 69.7                          \\ \hline
4th Popular Special = \#        & 82                     & 69.4                          \\ \hline
\end{tabular}

\caption{\ccschangestwo{Policy parameter values for our study's large-scale population (produced through our automated method) compared to a random sample of 100 sites that our method did not successfully evaluate, which we manually analyzed.}}
\label{tab:parameter_vals_sample}

\end{table}

\begin{table}[t]
\centering

\begin{tabular}{|cc|c|c|c|c|}
\hline
\multicolumn{2}{|l|}{\textbf{}} & \textbf{Lower} & \textbf{Upper} & \textbf{Digit} & \textbf{Special} \\ \hline
\multicolumn{1}{|c|}{\multirow{2}{*}{0}} & Manually Analyzed     & 91   & 86   & 82 & 86   \\ \cline{2-6} 
\multicolumn{1}{|c|}{}                   & Our Method & 84.1 & 83.7 & 82.0 & 86.3 \\ \hline
\multicolumn{1}{|c|}{\multirow{2}{*}{1}} & Manually Analyzed     & 7    & 11   & 11 & 10   \\ \cline{2-6} 
\multicolumn{1}{|c|}{}                   & Our Method & 8.3  & 8.7  & 10.0 & 7.0    \\ \hline
\multicolumn{1}{|c|}{\multirow{2}{*}{2}} & Manually Analyzed     & 2    & 3    & 7  & 4    \\ \cline{2-6} 
\multicolumn{1}{|c|}{}                   & Our Method & 7.5  & 7.6  & 8.0  & 6.8  \\ \hline
\end{tabular}

\caption{\ccschangestwo{Character class minimums for our study's large-scale population (produced through our automated method) compared to a random sample of 100 sites that our method did not successfully evaluate, which we manually analyzed.}}
\label{tab:restrictive_sample}

\end{table}

\ccschangestwo{Our study's results represent the domains that we could automatically evaluate with our measurement method (see Section~\ref{limitation} for the discussion of our method's limitations). Here, we investigate the generalizability of our findings by comparing them to those for domains we could not evaluate automatically. To do so, we randomly sampled 100 domains (within the same Tranco top 1M list) that our method did not automatically analyze, and manually analyzed their password policy parameters. Overall, while we observe some variations in parameter distributions between the manually analyzed sample and our large-scale population (produced by our automated method), we find similar password policy trends between both groups, indicating that our core findings are broadly applicable even to the types of sites that our method did not handle.}

 \ccschangestwo{\emph{Restrictive Parameters:} As shown in Tables~\ref{tab:parameter_vals_sample} and~\ref{tab:restrictive_sample}, our manually analyzed sample exhibited similar restrictive characteristics as our large-scale population, although the manually analyzed sites were slightly more restrictive with character class combinations and slightly less restrictive with character class minimums.}

 \ccschangestwo{\emph{Permissive Parameters:} As shown in Table~\ref{tab:parameter_vals_sample}, our manually analyzed sample is similar to our large-scale population in terms of permissive parameters, although the manually analyzed sites are  more accepting of certain characters and sequences. For example, our manually analyzed domains were more likely to accept Unicode, emojis, whitespaces, and special characters.} 

 \ccschangestwo{\emph{Length Parameters:} For password length requirements, we observe similar results for our manually analyzed sample as with our large-scale population. In both groups, a non-trivial fraction of sites (7\% in the manually analyzed sample, 12\% in our large-scale population) allow single-character passwords, and less than a third of sites required passwords of 8 characters or longer.  The median password length minimum in our manually analyzed sample was 6, compared to 5 for our large-scale population. For password length maximums, we found about 76\% of our manually analyzed sample allowed passwords exceeding 64 characters, compared to 60\% in our large-scale population. }

%\ccschangestwo{Overall, our sample exhibits similar policy parameter prevalences as in our large-scale results, signaling our results are largely generalizable. Small variations was observed in the restrictive, permissive, and length parameters. Particularly, the sample was found to be slightly more restrictive and allowed more permissive parameters. For length, the minimum limits were found to be similar (with a median difference of one), and the maximum length was found to be less restricting. Thus, while our study is biased to those domains we can measure (see Section~\ref{limitation}), our results and conclusions extend to the overall domains of the web.}

\end{document}